\documentclass[showpacs,showkeys,aps,prd,twocolumn,floatfix,byrevtex]{revtex4-1}
\usepackage{CJK,upgreek,fancyhdr}
\usepackage{graphicx}
\usepackage{booktabs}
\usepackage{amssymb,bm,mathrsfs,bbm,amscd}
\usepackage[tbtags]{amsmath}
\usepackage{lastpage}
\usepackage{color}
\usepackage{placeins}

\def \jp {J/\psi} \def \psip {\psi(3686)} \def \ccbar {c\bar c} \def
\lundcharm {{\small\rm LUNDCHARM~}}

\newcommand{\be}{\begin{enumerate}}
\newcommand{\ee}{\end{enumerate}}

\newcommand{\etal}{\it et al.\rm}
\newcommand{\red}{\color{black}}
\newcommand{\blue}{\color{black}}
\newcommand\T{\rule{0pt}{2.6ex}}       
\newcommand\B{\rule[-1.2ex]{0pt}{0pt}} 

\begin{document}
\begin{CJK*}{UTF8}{gkai}


\title{\boldmath Inclusive charged and neutral particle multiplicity
  distributions in $\chi_{cJ}$ and $J/\psi$ decays}

\author{
\begin{small}
\begin{center}
M.~Ablikim(麦迪娜)$^{1}$, M.~N.~Achasov$^{10,e}$, P.~Adlarson$^{63}$, S. ~Ahmed$^{15}$, M.~Albrecht$^{4}$, A.~Amoroso$^{62A,62C}$, Q.~An(安琪)$^{47,59}$, ~Anita$^{21}$, Y.~Bai(白羽)$^{46}$, O.~Bakina$^{28}$
% ...
\\
\vspace{0.2cm}
(BESIII Collaboration)\\
\vspace{0.2cm} {\it
$^{1}$ Institute of High Energy Physics, Beijing 100049, People's Republic of China\\
% ...
$^{4}$ Bochum Ruhr-University, D-44780 Bochum, Germany\\
% ...
$^{10}$ G.I. Budker Institute of Nuclear Physics SB RAS (BINP), Novosibirsk 630090, Russia\\
% ...
$^{15}$ Helmholtz Institute Mainz, Johann-Joachim-Becher-Weg 45, D-55099 Mainz, Germany\\
% ...
$^{21}$ Indian Institute of Technology Madras, Chennai 600036, India\\
% ...
$^{28}$ Joint Institute for Nuclear Research, 141980 Dubna, Moscow region, Russia\\
% ...
$^{46}$ Southeast University, Nanjing 211100, People's Republic of China\\
$^{47}$ State Key Laboratory of Particle Detection and Electronics, Beijing 100049, Hefei 230026, People's Republic of China\\
% ...
$^{59}$ University of Science and Technology of China, Hefei 230026, People's Republic of China\\
% ...
$^{62}$ (A)University of Turin, I-10125, Turin, Italy; (B)University of Eastern Piedmont, I-15121, Alessandria, Italy; (C)INFN, I-10125, Turin, Italy\\
$^{63}$ Uppsala University, Box 516, SE-75120 Uppsala, Sweden\\
% ...
$^{e}$ Also at the Novosibirsk State University, Novosibirsk, 630090, Russia\\
% ...
}
\end{center}
\vspace{0.2cm}
\end{small}}

\affiliation{}
\date{\today}

\begin {abstract}
  Using a sample of 106 million $\psi(3686)$ decays, $\psi(3686) \to
  \gamma \chi_{cJ} \: {\red (J = 0, 1, 2)}$ and $\psi(3686) \to \gamma
  \chi_{cJ}, \chi_{cJ} \to \gamma J/\psi$ ${\red (J = 1, 2)}$ events are
  utilized to study inclusive $\chi_{cJ} \to$ anything, $\chi_{cJ}
  \to$ hadrons, and $J/\psi \to$ anything distributions, including
  distributions of the number of charged tracks, electromagnetic
  calorimeter showers, and $\pi^0$s, and to compare them with
  distributions obtained from the BESIII Monte Carlo simulation.
  Information from each Monte Carlo simulated decay event is used to
  construct matrices connecting the detected distributions to the
  input predetection ``produced'' distributions. Assuming these
  matrices also apply to data, they are used to predict the analogous
  produced distributions of the decay events. Using these, the charged
  particle multiplicities are compared with results from MARK I.
  Further, comparison of the distributions of the number of photons in
  data with those in Monte Carlo simulation indicates that G-parity
  conservation should be taken into consideration in the simulation.

\end{abstract}



\maketitle
\end{CJK*}


\section{Introduction}

The multiplicity distributions of charged hadrons, which can be
characterized by their means and dispersions, are an important observable
in high energy collisions and an input to models of multihadron
production.  Charged particle means from below 2 GeV to LEP energies
have been fit as a function of energy with a variety of models {\red in}
Ref.~\cite{opal}, and a review of theoretical understanding can be
found in Ref.~\cite{carruthers}.

The study of $\chi_{cJ} \: {\red (J = 0, 1, 2)}$ decays is important since they
are expected to be an important source of glueballs, and future
studies require both more data and better simulation of generic
$\chi_{cJ}$ decays.  Also since $\chi_{cJ}$ decays make up
approximately 30\% of $\psi(3686)$ decays, better understanding of
$\chi_{cJ}$ decays improves that of $\psi(3686)$ {\red decays}.

The branching fractions of $\psi(3686) \to \gamma \chi_{cJ}$ and
$\chi_{cJ} \to \gamma J/\psi$ were measured previously by BESIII using
a sample of 106 million $\psi(3686)$ decays~\cite{bam248}.  The
accuracy of these measurements depends critically on the ability of
the Monte Carlo (MC) simulation to model data well. Since a large fraction of
$\chi_{cJ}$ hadronic decay modes are still unmeasured~\cite{PDG}, it
is particularly important to verify the modeling of their inclusive
decays, where we rely heavily on the \lundcharm model~\cite{lundcharm}
to simulate these events.

In this paper, which is based on the analysis performed in
Ref.~\cite{bam248}, we report on the ``detected'' distributions: the
efficiency-corrected charged particle multiplicity distributions, as
well as the efficiency-corrected distributions of the number of
electromagnetic calorimeter showers and $\pi^0$s for $\chi_{cJ}$ and
$J/\psi$ decays. Our detected distributions are compared with MC
simulation, and the results can be used to improve the \lundcharm
model simulation, in particular for $\chi_{cJ}$ hadronic decays.  

Information from each {\red MC} simulation decay event is used to
construct matrices connecting the detected charged particle and photon multiplicity distributions to the input predetection
distributions. Assuming the matrices also apply to data, they are used
to predict the analogous ``produced'' distributions of the decay
events.  Produced charged particles and photons correspond to
those coming directly from the $\chi_{cJ}$ or $J/\psi$ decays or the
decays of their daughter particles.  The means of the charged particle
multiplicity distribution are compared with those of MARK I, which
measured the mean charged particle multiplicity for $e^+ e^- \to$
hadrons as a function of center-of-mass energy from 2.6 to 7.8
GeV~\cite{marki}.

In Ref.~\cite{bam248}, an electromagnetic calorimeter {\red (EMC)} shower
(EMCSH) was labeled a ``photon'', but as described in
Section~{\ref{EMC_energy_deposits}}, showers include hadronic
interactions in the EMC crystals and electronic noise, so here we will
explicitly refer to them as EMCSHs.  The comparison of data and
inclusive $\psi(3686)$ MC simulation showed good agreement for charged
track distributions and most EMCSH energy ($E_{\rm sh}$)
distributions, however{\red ,} there was some difference in the distribution
of the number of $\pi^0$s~\cite{bam248}.  Here, we explore the
agreement for $\chi_{cJ} \to$ anything and $\chi_{cJ} \to$ hadrons via
$\psi(3686) \to \gamma \chi_{cJ}$ and $J/\psi \to$ anything via
$\chi_{cJ} \to \gamma J/\psi$.  Recently BESIII observed
electromagnetic Dalitz decays $\chi_{cJ} \to l^+ l^- J/\psi \: {\red
  (l\: = \: e
\: {\rm or} \: \mu)} $~\cite{lljp}, so our $\chi_{cJ} \to$ hadron
distributions also include $\chi_{cJ} \to l^+ l^- J/\psi$.  However,
the branching fractions for these decays are very small, on the order
of $10^{-4}$, which are negligible compared with those of $\chi_{cJ}
\to$ hadrons.  Below we will continue to refer to these distributions
as $\chi_{cJ} \to$ hadrons.  {\red ``Hadrons'' is used very loosely and
includes all processes except $\chi_{cJ} \to \gamma J/\psi$, such as
other $\chi_{cJ}$ radiative decays  and $\chi_{cJ} \to \gamma \gamma$.}

This analysis is based on the 106 million $\psi(3686)$ event sample
gathered in 2009, the corresponding continuum sample with integrated
luminosity of 44 pb$^{-1}$ at $\sqrt{s} = 3.65$~GeV~\cite{Npsip}, and
a 106 million $\psi(3686)$ {\red event} inclusive MC sample.

The paper is organized as follows: In Section~{\ref{Lundcharm}}, the
\lundcharm model is described. In Sections~{\ref{nch_results}} -
{\ref{npi0_results}}, the distributions of the number of detected
charged tracks, EMCSHs, and $\pi^0$s, respectively, are determined and
compared with MC simulation. Section~{\ref{produced_section}} presents
the produced distributions.  Section~{\ref{systematics}} discusses
systematic uncertainties, while Section~{\ref{summary}} provides a
summary.  Additional EMCSH and $\pi^0$ tables are included in an appendix.

\section{LUNDCHARM model}
\label{Lundcharm}
The \lundcharm model is an event generator to produce events for
charmonium decaying inclusively to anything{\red ~\cite{lundcharm}}. This
model, which was inspired by QCD theory, was developed at BESII and
migrated to the BESIII experiment.  In this model, $\jp$ or $\psip$
decaying into light hadrons is described as $\ccbar$ quark
annihilation into one photon, three gluons or one photon plus two
gluons, followed by the photon and gluons transforming into light
quarks and further materializing into final light hadron states. To
leading order accuracy, the $\ccbar$ quark annihilations are modeled
by perturbative QCD~\cite{pqcd}, while the hadronization of light
quark fragmentation is described with the Lund model~\cite{pythia6.4}
using a set of parameters to describe the baryon/meson ratio,
strangeness and $\{\eta,\eta'\}$ suppression, and the distribution of
orbital angular momentum, etc.

\begin{table}[bth]
\begin{center}
  \caption{Fractions of charmonium unmeasured
    decays~\cite{PDG}. \label{missingdecays}}
\begin{footnotesize}
\begin{tabular}{c|c} \hline
Charmonium & fraction of unmeasured decays\\\hline
$\psip$    & 0.1656\\
$\chi_{c0}$  & 0.8547\\
$\chi_{c1}$  & 0.5725\\
$\chi_{c2}$  & 0.7208\\
$\jp$      & 0.5456 \\
$\eta_c$   & 0.7094\\ \hline
\end{tabular}
\end{footnotesize}
\end{center}
\end{table}

\begin{figure}[htbp]
\begin{center}
\includegraphics*[width=0.3\textwidth]{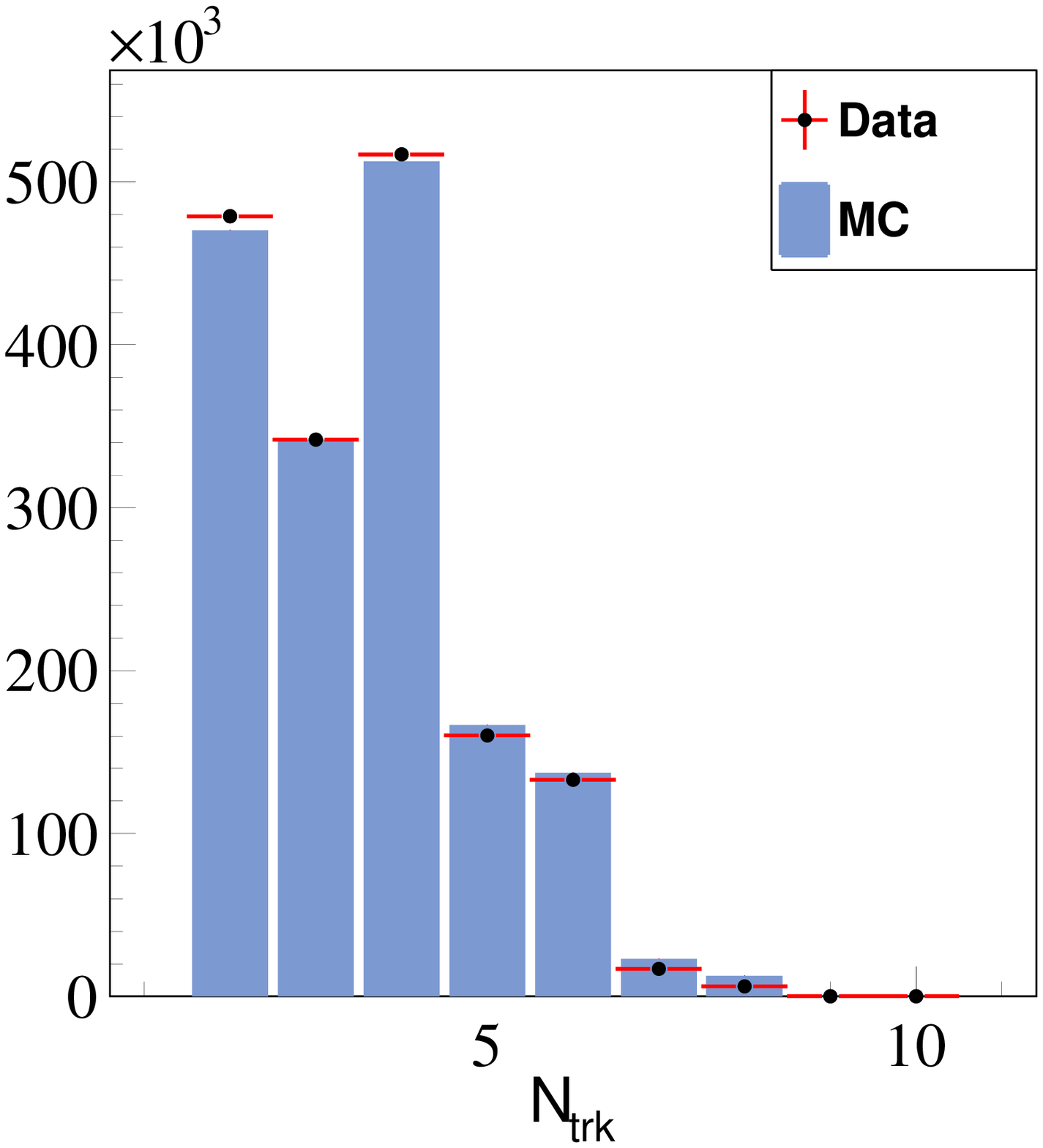} 
\put(-70,140){\bf \large (a)}

\includegraphics*[width=0.3\textwidth]{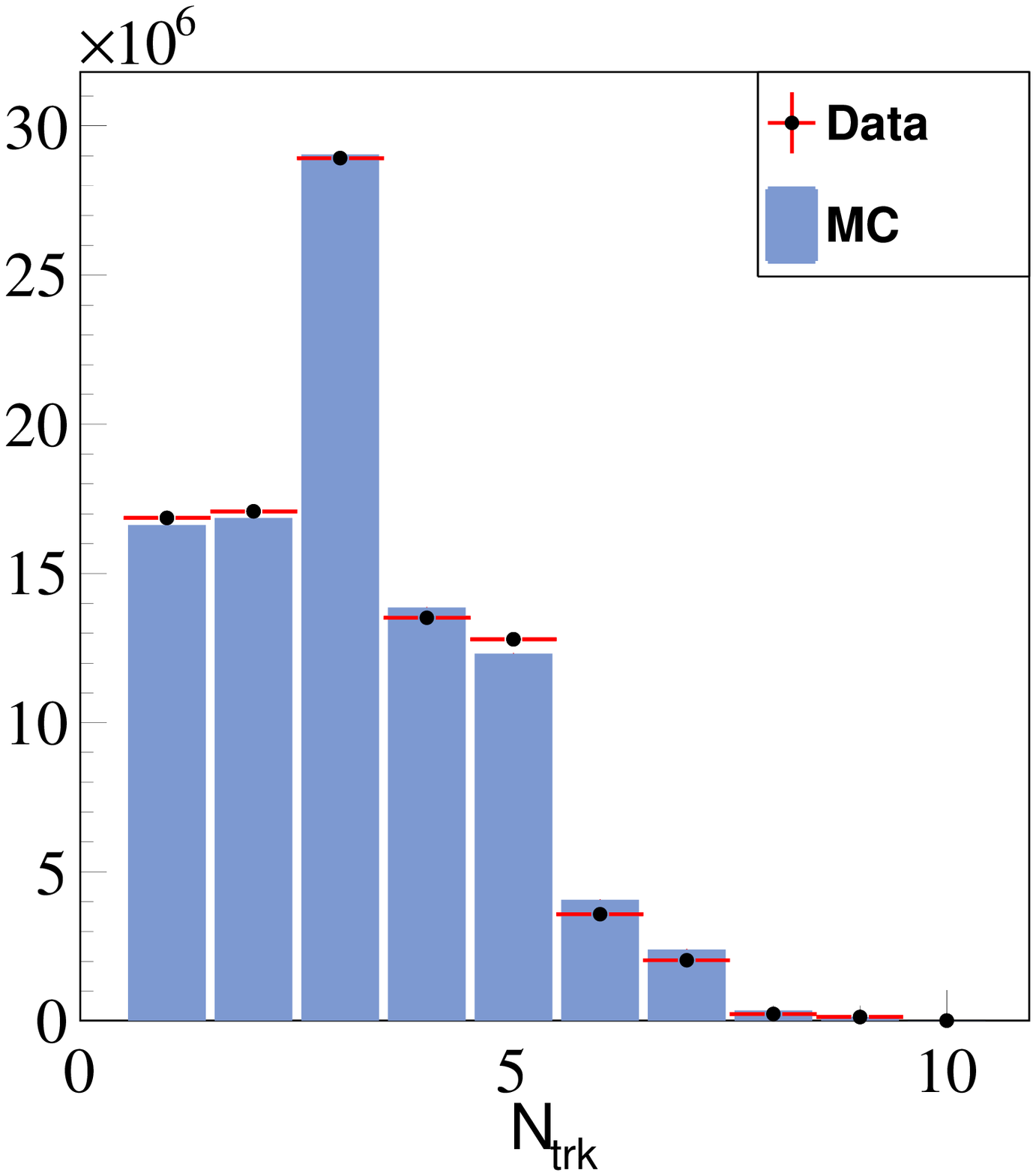}
\put(-70,140){\bf \large (b)}
\caption{The multiplicity distributions of detected charged tracks,
  (a) $\jp$ {\red decays} and (b) $\psip$ {\red decays}, where
  black histograms are from data and the shaded histograms are
  produced from the inclusive $\psi(3686)$ MC sample with tuned
  \lundcharm model parameters.}
  \label{chargedTrack}
\end{center}
\end{figure}

The \lundcharm model is used to generate the unmeasured charmonium
decays, while the established decays are exclusively generated with
their appropriate \rm{BesEvtGen} models~\cite{besevtgen} using
branching fractions from the Particle Data Group~\cite{PDG}. The
fraction of unmeasured decays for each charmonium state is given in
Table~\ref{missingdecays}~\cite{PDG}. Since the fractions are quite large for
$\chi_{cJ}$ decays, the \lundcharm model is very important for the
simulation of these decays.  The parameters of the \lundcharm model
are optimized using 20 million $\jp$ decays accumulated at the BESIII
experiment~\cite{tuning}.  Figure~\ref{chargedTrack} shows the
comparison between data and MC simulation of the multiplicity of
detected charged tracks for $\jp$ and $\psip$ decays.  More
comparisons of data and MC simulation for $J/\psi$ decays can be found
in Ref.~\cite{tuning} and for $\psi(3686)$ decays in
Refs.~\cite{tuning,bam248}.

\section{\boldmath Multiplicity distribution of charged tracks}
\label{nch_results}
\subsection{Method}
The basic approach is the same as in Ref.~\cite{bam248}.  Charged
tracks must be in the active region of the drift chamber and have
their points of closest approach consistent with the run-by-run
interaction point.  Neutral tracks must be in the active regions of
the barrel {\red EMC} or end-cap {\red EMC}, satisfy minimum and
maximum energy requirements and a time requirement.  The basic
$\psi(3686)$ event selection requires at least one charged track
({\red except for the study of the events with no charged tracks, where this
requirement is dropped}), at least one neutral track, and a minimum
event energy. A background filter removes non-$\psi(3686)$ events, and
events consistent with being a $\psi(3686) \to \pi \pi J/\psi$ decay
are removed{\blue ~\cite{bam248}}.  Following this, the $E_{\rm sh}$ distribution is
constructed for the remaining events, where the EMCSH must be in the
barrel {\red EMC}, not originate from a charged track ($\delta >
14^{\circ}$, where $\delta$ is the angle between the shower and the
nearest charged track), and not be a photon from a $\pi^0$ decay.
Fitting the peaks in the $E_{\rm sh}$ distribution due to $\psi(3686)
\to \gamma \chi_{cJ}$ and $\chi_{cJ} \to \gamma J/\psi$, as shown in
Fig.~\ref{egam_fit_0prongs}, allows the determination of the number of
the inclusive decays and the final branching fractions.  Please refer
to Ref.~\cite{bam248} for many important details.

To determine the distribution of the number of charged tracks,
$N_{\rm ch}$, ten $E_{\rm sh}$ distributions are constructed for $N_{\rm ch}$
ranging from 0 to 9.  These distributions are then fitted to determine
the numbers of $\chi_{cJ} \to$ anything and $\chi_{cJ} \to \gamma
J/\psi,\:J/\psi \to$ anything events, and these numbers determine the
$N_{\rm ch}$ distributions for $\chi_{cJ} \to$ anything and $J/\psi \to$
anything.  

In Ref.~\cite{bam248}, simultaneous fitting of inclusive and exclusive
$E_{\rm sh}$ distributions was performed, but this is not done here, except for
the $N_{\rm ch} = 0$ case, because there are no exclusive $E_{\rm sh}$
distributions versus $N_{\rm ch}$ to be used in such a fit.  Another
change is that events with $N_{\rm ch} = 0$ have additional requirements
in order to reduce background in the $E_{\rm sh}$ distributions.

\subsection{\boldmath $N_{\rm ch} = 0$ event selection and fit of $E_{\rm sh}$
 distributions}
Events with $N_{\rm ch} = 0$ were selected in our previous analysis only
to determine the systematic uncertainty associated with the $N_{\rm ch} > 0$
requirement. The photon time requirement was removed since without
charged tracks, the event time is not well determined. Although other
selection requirements were tightened, the events still had much
background~\cite{bam248}.

\begin{figure}[htbp]
\begin{center}
\rotatebox{-90}
{\includegraphics*[width=2.75in]{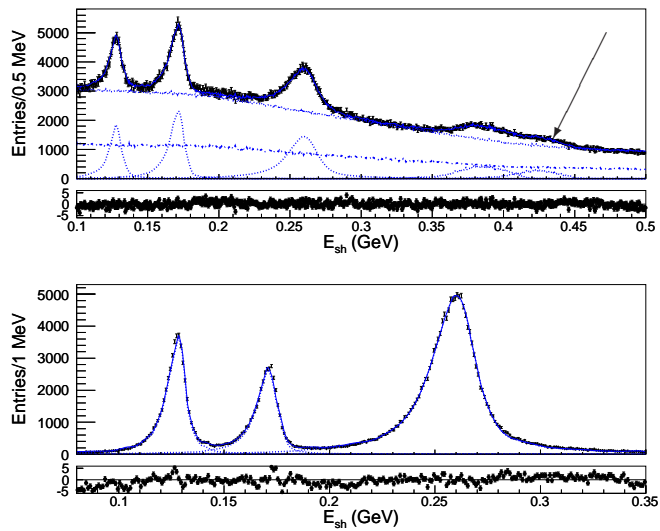}}
\put(-20,-10){\vector(-1,-2){20}}
\caption{\label{egam_fit_0prongs} Simultaneous fits
  to the $E_{\rm sh}$ distributions of data for $N_{\rm ch}=0$. (Top set)
  Inclusive $N_{\rm ch} = 0$ distribution fit and corresponding pulls, and
  (Bottom set) exclusive distribution fit and pull distribution.  The
  five peaks from left to right in the top figure correspond to
  $\psi(3686) \to \gamma \chi_{c2}$, $\gamma \chi_{c1}$, $\gamma
  \chi_{c0}$, $\chi_{c1} \to \gamma J/\psi$ and the small $\chi_{c2}
  \to \gamma J/\psi$ contribution (see arrow). The exclusive modes include $\psi(3668) \to
  \gamma \chi_{cJ}, \chi_{cJ} \to 2$ and $4$ charged track events,
  selected with requirements on the invariant mass of the charged
  tracks and the angle between the direction of the radiative photon
  and the recoil momentum from the charged tracks.  Here the wide
  $\chi_{cJ} \to \gamma J/\psi$ shapes are described by the inclusive
  MC shapes, while the narrow $\psi(3686) \to \gamma \chi_{cJ}$ shapes
  are inclusive MC shapes convolved with bifurcated Gaussians.  The
  smooth curves in the two plots are the fit results. The dash-dotted
  and dotted curves in the top plot are the background distribution
  from the inclusive $\psi(3686)$ MC with radiative photons removed
  and the total background, respectively, where the total is the sum
  of the MC background and a second order polynomial.}
\end{center}
\end{figure}

For the current analysis, events with $|(P_x)_{\rm {\red neu}}| > 1.0$
GeV/$c$ and $|(P_y)_{\rm neu}| > 1.0$ GeV/$c$ are removed, since
these regions contain much background according to MC simulation.
$(P_x)_{\rm neu}$ and $(P_y)_{\rm neu}$ are the sum of the momenta of
all neutrals in the $x$ and $y$ directions, respectively, where $x$
and $y$ are orthogonal axes perpendicular to the axis of the detector.
The $E_{\rm sh}$ distribution with the additional requirements is much
cleaner and easily fitted, as shown in Fig.~\ref{egam_fit_0prongs}.  A
simultaneous fit {\red with inclusive and exclusive events} was used for the
previous $N_{\rm ch}=0$ systematic uncertainty study since the signal
to background ratio was so low, and the same fitting method is used
here, as shown in Fig.~\ref{egam_fit_0prongs}.  The $\chi^2/ndf$ for
the fit to data is 1.3{\red , where $ndf$ is the number of degrees of
freedom}.


\subsection{\boldmath $N_{\rm ch} >0$ selection and fitting}

\begin{figure*}[htbp]
\begin{center}
\rotatebox{-90}{
  \includegraphics*[width=4.4in]{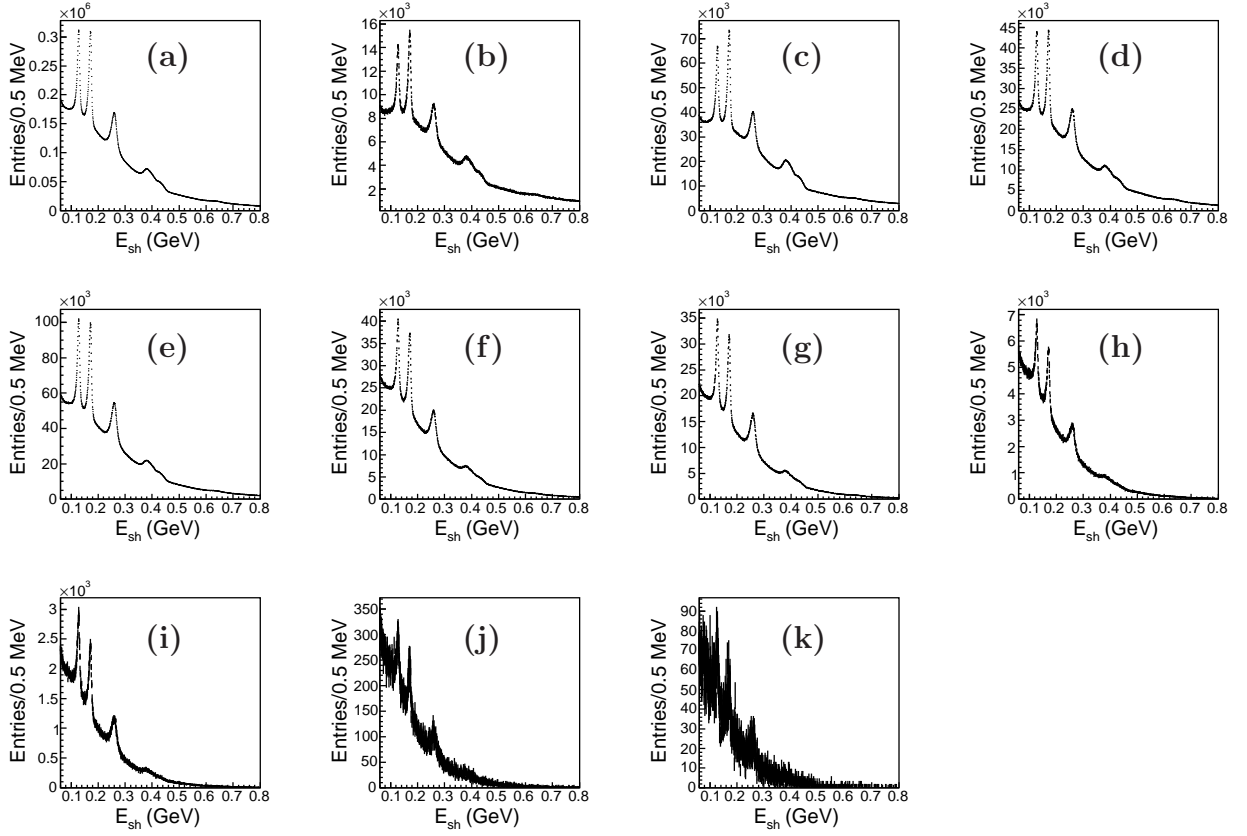}}
\put(-410,-30){\bf \large (a)} \put(-290,-30){\bf \large (b)}
\put(-170,-30){\bf \large (c)} \put(-50,-30){\bf \large (d)}
\put(-410,-140){\bf \large (e)} \put(-290,-140){\bf \large (f)}
\put(-170,-140){\bf \large (g)} \put(-50,-140){\bf \large (h)}
\put(-410,-250){\bf \large (i)} \put(-290,-250){\bf \large (j)}
\put(-170,-250){\bf \large (k)}
\caption{\label{Nchgt0} The distributions of $E_{\rm sh}$ of data for
  (a) all $N_{\rm ch}$ and (b)-(k) $N_{\rm ch} = 1 - 10$. For $N_{\rm ch}=10$, the
  signal is negligible, and this distribution is not fitted.}
\end{center}
\end{figure*}

Figure~\ref{Nchgt0} shows
the $E_{\rm sh}$ distributions for all $N_{\rm ch}$ and for individual
values of $N_{\rm ch} > 0$ for data.  $E_{\rm sh}$ distributions for
different values of $N_{\rm ch}$ for MC simulation and continuum
background are constructed similarly.

\begin{figure}[htbp]
\begin{center}
  \rotatebox{-90}
  {\includegraphics*[width=1.42in]{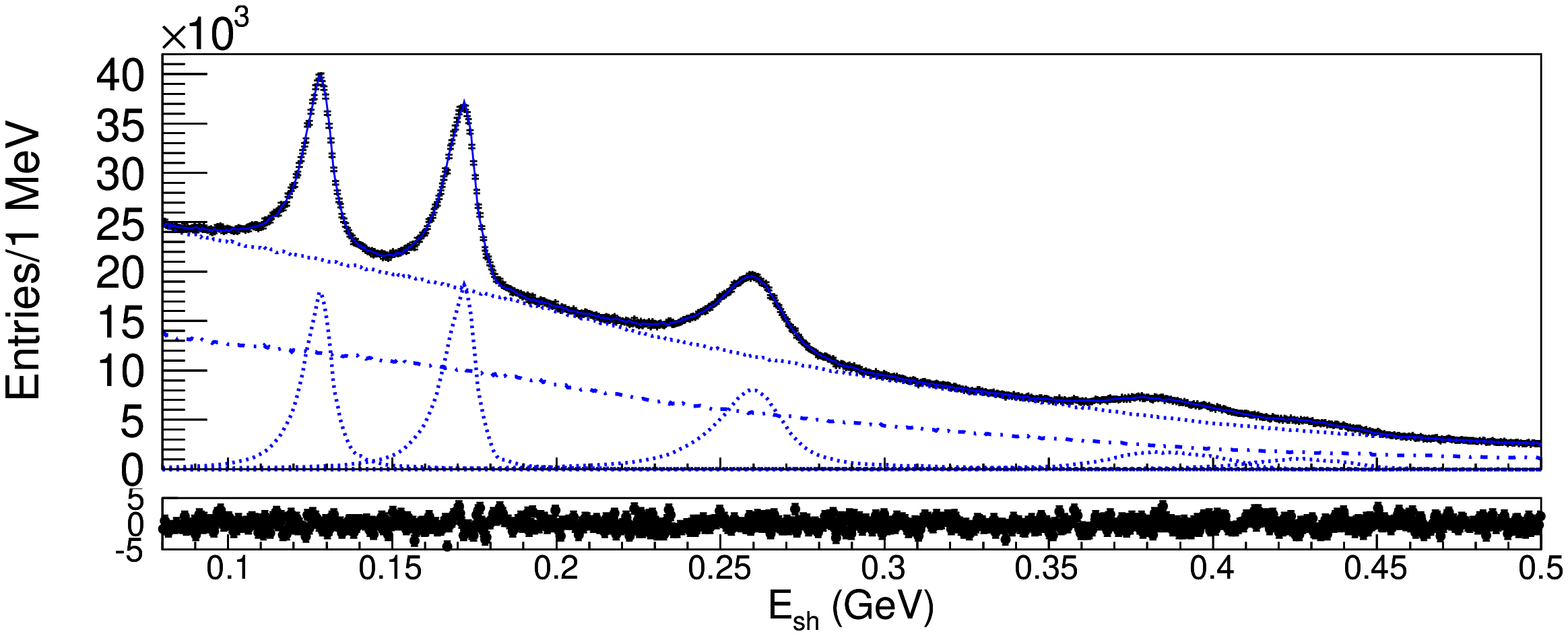}}
  \caption{\label{fitofNcheq5} Fit to the $E_{\rm sh}$
    distribution of data and pulls for $N_{\rm ch}=5$.  See
    Fig.~\ref{egam_fit_0prongs} (Top set) for the plot description.
    Here the MC simulation and background distributions are also for
    $N_{\rm ch} = 5$.}
\end{center}
\end{figure}

Signal shapes and background shapes used in the fit depend on the
value of $N_{\rm ch}$. In fitting the distributions for $N_{\rm ch} > 7$,
because of the small sample sizes, the signal shapes and background
shapes for $N_{\rm ch} = 7$ are used.  The fit result of data for $N_{\rm ch}
= 5$ is shown in Fig.~\ref{fitofNcheq5}, and the $\chi^2/ndf$ is 1.4.  Fit
results for other values of $N_{\rm ch}$ result in similar $\chi^2/ndf$
values.

The MC simulated sample is fitted as a function of $N_{\rm ch}$ in a
similar fashion, but $\psi(3686) \to \gamma \chi_{cJ}$ MC signal
shapes are fitted without convolution.  As described in
Ref.~\cite{bam248}, the MC events are weighted by $wt_{\pi^0} \times
wt_{\rm trans}$, where $wt_{\pi^0}$ accounts for the difference between
data and MC simulation on the number of $\pi^0$s and $wt_{\rm trans}$
accounts for the $E^3_{\gamma}$ energy dependence of the radiative
photon in the electric dipole transitions for $\psi(3686) \to \gamma
\chi_{cJ}$ and $\chi_{cJ} \to \gamma J/\psi$.

\subsection{Results}

The MC simulated sample is analyzed by counting the number of events
versus $N_{\rm ch}$ before applying any selection criteria.  The
efficiency is then the number of events passing all selection criteria
divided by the number of events without imposing any selection versus
$N_{\rm ch}$.  Note that $N_{\rm ch}$ here is the ``detected'' number of
charged tracks.

Using the number of detected data events, $D$, and the MC determined
efficiencies, $\epsilon$, which are dependent on $N_{\rm ch}$, we determine the
distribution of the efficiency-corrected number of events in data for
$\chi_{cJ} \to$ anything and $\chi_{cJ} \to \gamma J/\psi,\:J/\psi \to$
anything.  Results are listed in Table~\ref{result_chic_to_all} for
$\chi_{cJ} \to$ anything and Table~\ref{result_jpsi_to_all} for
$\chi_{c1/2} \to \gamma J/\psi,\:J/\psi \to$ anything.

For comparison, MC simulation numbers, $N^{\rm {\red MC}}$, are also listed in
the tables. $N^{\rm MC}$ corresponds to the $N_{\rm ch}$ distribution before
imposing selection requirements.  Since the branching fractions of MC
simulation are not the same as the measured branching fractions of
Ref.~\cite{bam248}, the MC numbers are scaled by $B_{\rm {\red
    BESIII}}/B_{\rm {\red MC}}$,
where $B_{\rm {\red BESIII}}$ and $B_{\rm {\red MC}}$ are the BESIII branching
fractions~\cite{bam248} and those used by the MC, respectively, and
the $N^{\rm MC}$ in Tables~\ref{result_chic_to_all} and
\ref{result_jpsi_to_all} are the scaled MC numbers.

\begin{table*}[bth]
\begin{center}
\caption{Detected data events, $D$, efficiencies, $\epsilon$, efficiency corrected events, $N$, and number of scaled simulated events $N^{\rm MC}$ for $\chi_{cJ} \to$ anything. 
\label{result_chic_to_all}}
\begin{footnotesize}
\begin{tabular}{l|rrrr|rrrr|rrrr} \hline 
$N_{\rm ch}$ & $D_{\chi_{c0}}$ & $\epsilon_{\chi_{c0}}$ &
  $N_{\chi_{c0}}$  &  $N^{\rm MC}_{\chi_{c0}}$ & $D_{\chi_{c1}}$ &
  $\epsilon_{\chi_{c1}}$ &  $N_{\chi_{c1}}$  &  $N^{\rm MC}_{\chi_{c1}}$  & $D_{\chi_{c2}}$ & $\epsilon_{\chi_{c2}}$ &  $N_{\chi_{c2}}$ &  $N^{\rm MC}_{\chi_{c2}}$\\
  & & (\%) & & & & (\%) & & & & (\%) & &  \\ \hline
  0  &   95664  & 30.7 &  311124  & 207332 &  73922  & 24.1 &  307213
  & 218503 &  51006  & 21.1 &  241455 & 189395\\
  1  &  206872  & 43.7 &  473186  & 450456 & 226613  & 43.6 &  519506
  & 502988 & 165867  & 36.2 &  457732 & 446984 \\
  2  & 1003030  & 48.6 & 2065843  & 2041808 & 1210640  & 49.9 &
  2426435 & 2414376 & 887474  & 41.9 & 2118574 & 2078609 \\
  3  &  663550  & 41.6 & 1594227  & 1782415 & 699804  & 41.5 & 1687651
  & 1775014 &  589383  & 35.8 & 1646546 & 1790336\\
  4  & 1602890  & 54.0 & 2969910  & 3100329 &1662640  & 54.4 & 3058982
  & 3031942 & 1459680  & 47.6 & 3064694 & 3073785\\
  5  &  528842  & 47.3 & 1117174  & 1074490 & 566264  & 48.2 & 1173704
  & 1137965 &  499056  & 42.0 & 1186940 & 1166188\\
  6  &  502471  & 44.5 & 1128369  &  991170 & 533755  & 45.6 & 1171074
  & 1046738 &  492290  & 40.0 & 1230654 & 1076283\\
  7  &   70611  & 34.2 &  206487  &  124917 & 79957  & 35.4 &   225920
  &  158769 & 76321  & 31.3 &  243714 & 163899\\
  8  &   36744  & 25.9 &  141685  &  54033  & 38446  & 31.8 &  120915
  & 73010 &   38390  & 27.5 &  139611 & 75074\\
  9  &    2616  & 14.1 &   18570  &   3782  & 3087  & 24.0 &   12843 &
  5478 &   3562  & 30.1 &   11845 & 5879\\ \hline
\end{tabular}
\end{footnotesize}
\end{center}
\end{table*}

\begin{table*}[bth]
\begin{center}
  \caption{Detected data events, $D$, efficiencies, $\epsilon$,
    efficiency corrected events, $N$, and number of scaled simulated
    events $N^{\rm MC}$ for $\chi_{c1/2} \to \gamma J/\psi,$ $J/\psi \to$
    anything. Here and below, $J/\psi_{1/2}$ represents $\chi_{c1/2}
    \to \gamma J/\psi,\:J/\psi \to $anything.
\label{result_jpsi_to_all}}
\begin{footnotesize}
\begin{tabular}{l|rrrr|rrrr} \hline
$N_{\rm ch}$ & $D_{J/\psi_1}$ & $\epsilon_{J/\psi_1}$
  & $N_{J/\psi_1}$  & $N^{\rm MC}_{J/\psi_1}$ &  $D_{J/\psi_2}$ & $\epsilon_{J/\psi_2}$ &  $N_{J/\psi_2}$  &  $N^{\rm MC}_{J/\psi_2}$\\
  & & (\%) & & & & (\%) & &  \\ \hline
  0  &   36983  & 28.9 &  128178  & 119881 &   19705  & 29.1 &   38250 &
  65012\\
  1  &  110869  & 47.2 &  234686  & 212706 &  60555  & 51.5 &  113737
  & 119930 \\
  2  &  633989  & 54.3 & 1167955  & 1158351 & 320064  & 53.2 &  601156 &
  633894 \\
  3  &  252917  & 47.7 &  530595  &  549543 & 136369  & 48.3 &  282565 &
  297953 \\
  4  &  552012  & 59.7 &  925337  &  911111 & 294272  & 60.1 &  489386 &
  516037 \\
  5  &  157700  & 53.1 &  297245  &  305425 & 83325  & 53.9 &  154712 &
  163137 \\
  6  &  135463  & 49.0 &  276515  &  270788 & 73828  & 49.4 &  149512 &
  157654 \\
  7  &   16602  & 36.9 &   44960  &   49716 &  8172  & 37.6 &   21736 &
  22919 \\
  8  &    6724  & 28.4 &   23717  &   23877 &  2927  & 24.3 &   12033 &
  12688 \\
  9  &     241  & 18.6 &    1296  &    1850 &  240  & 16.4 &    1463 & 1543\\
  \hline
\end{tabular}
\end{footnotesize}
\end{center}
\end{table*}

  The efficiency corrected $N_{\rm ch}$ distributions for $\chi_{cJ} \to$
  anything contain the $\chi_{cJ} \to \gamma J/\psi,\:J/\psi \to$
  anything events, as well as the $\chi_{cJ} \to$ hadrons events.  A
  more interesting comparison between data and the simulated MC sample
  is with the $N_{\rm ch}$ distributions for $\chi_{cJ} \to$ hadrons
  directly.  These are obtained by subtracting $N_{\rm ch}$ distributions
  for $\chi_{cJ} \to \gamma J/\psi,\:J/\psi \to$ anything from those
  of $\chi_{cJ} \to$ anything.  Since we do not have the distribution
  from data for $\chi_{c0} \to \gamma J/\psi,\:J/\psi \to $ anything,
  we use the MC distribution for this process.  The branching fraction
  is {\red small, 1.4 \%,} so the change for $\chi_{c0} \to$ anything is small.

\begin{figure*}[htbp]
\begin{center}
\rotatebox{-90}
{\includegraphics*[width=2.0in]{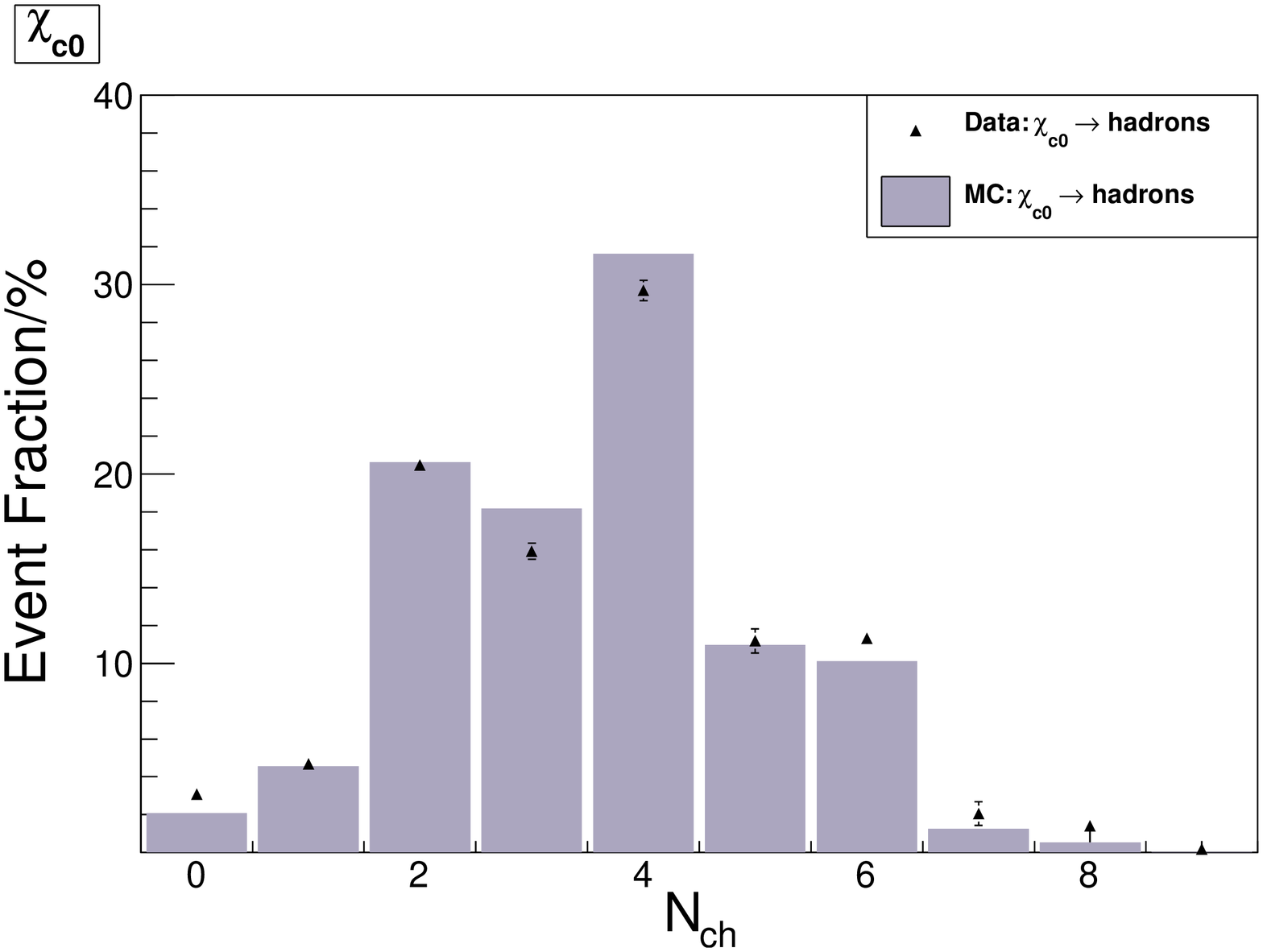}}
\put(-160,-35){\bf \large  {(a)}}
\rotatebox{-90}
{\includegraphics*[width=2.0in]{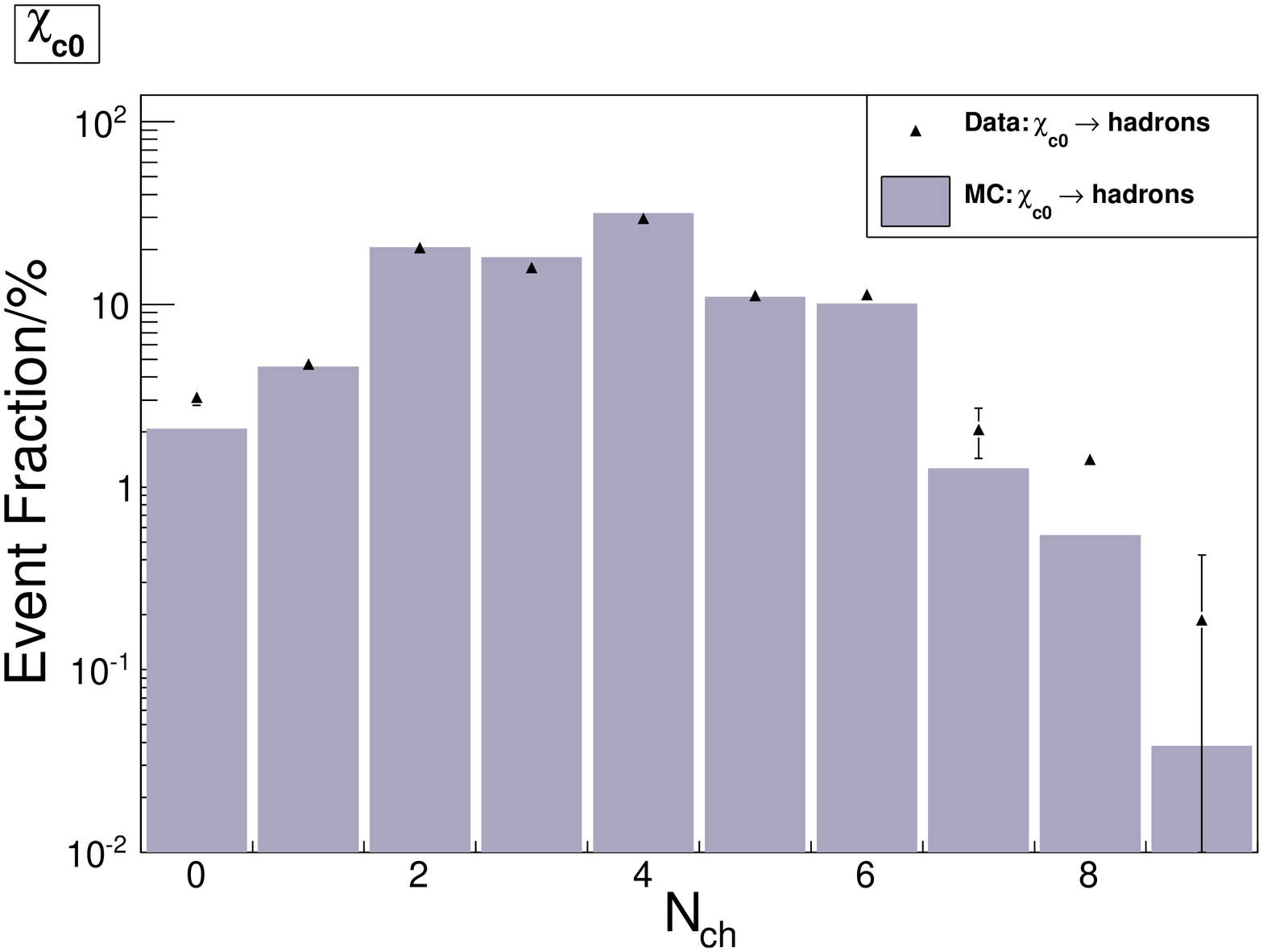}}
\put(-160,-35){\bf \large {(b)}}

\rotatebox{-90}
{\includegraphics*[width=2.0in]{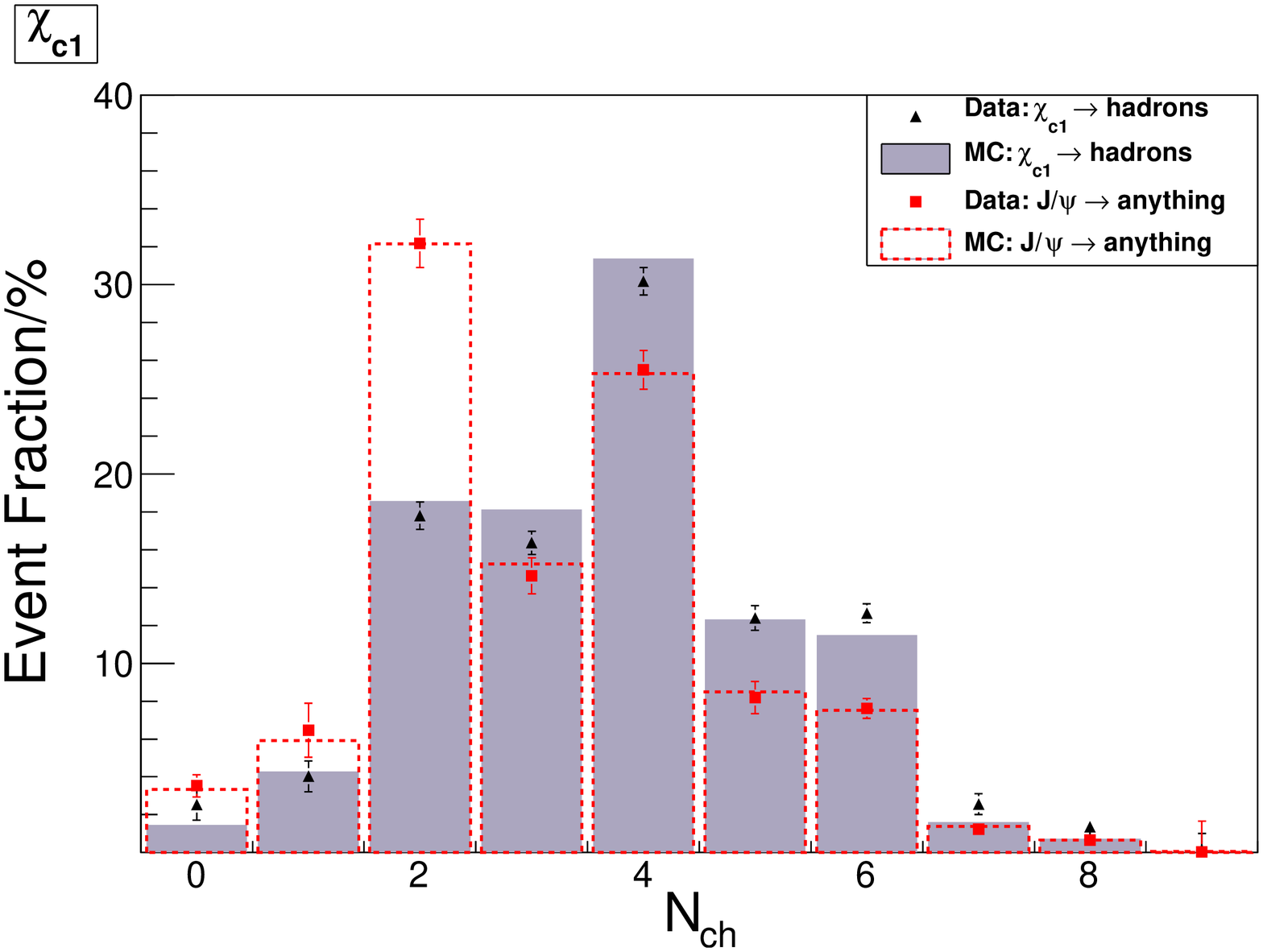}}
\put(-160,-35){\bf \large  {(c)}}
\rotatebox{-90}
{\includegraphics*[width=2.0in]{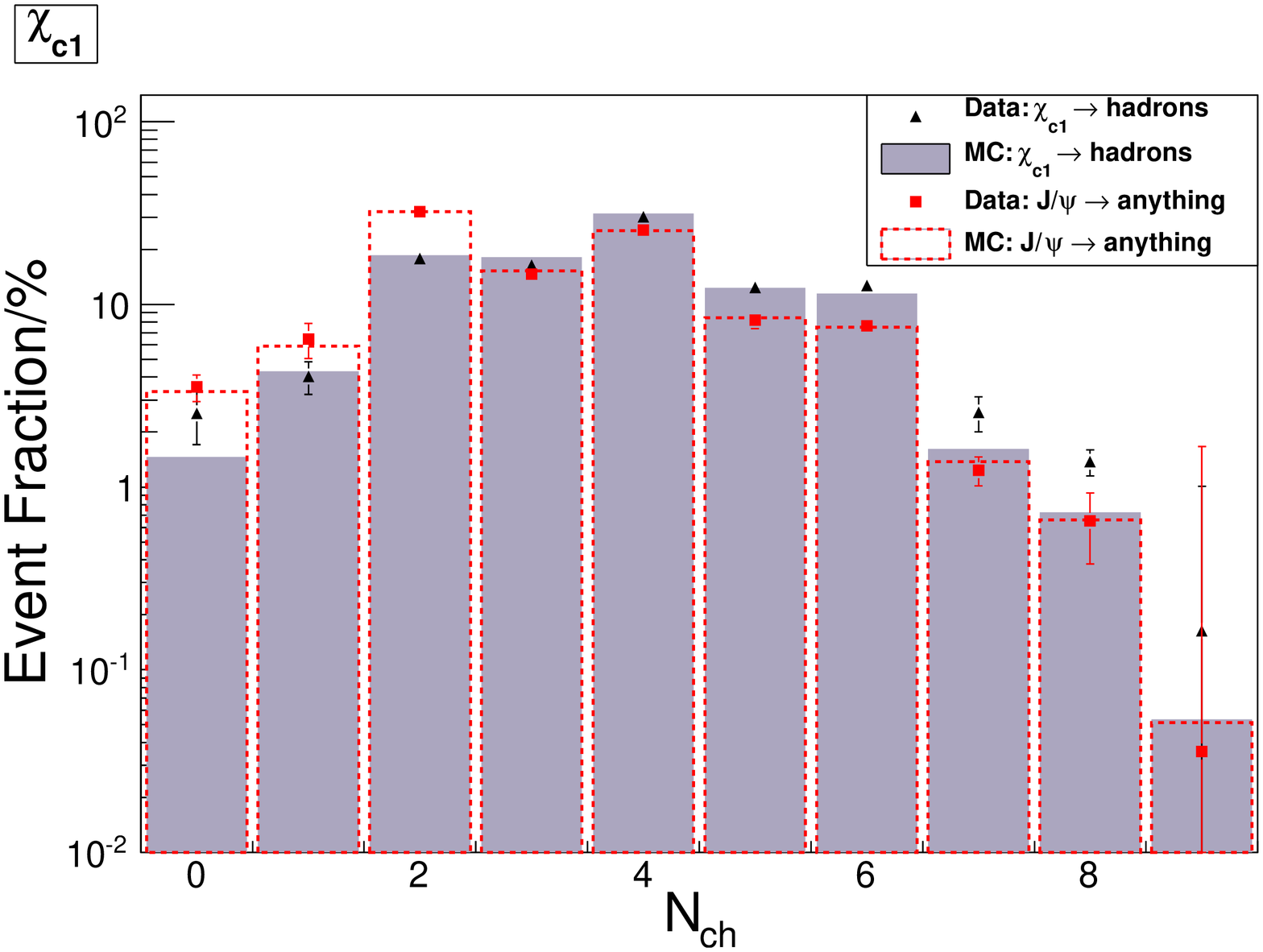}}
\put(-160,-35){\bf \large  {(d)}}

\rotatebox{-90}
{\includegraphics*[width=2.0in]{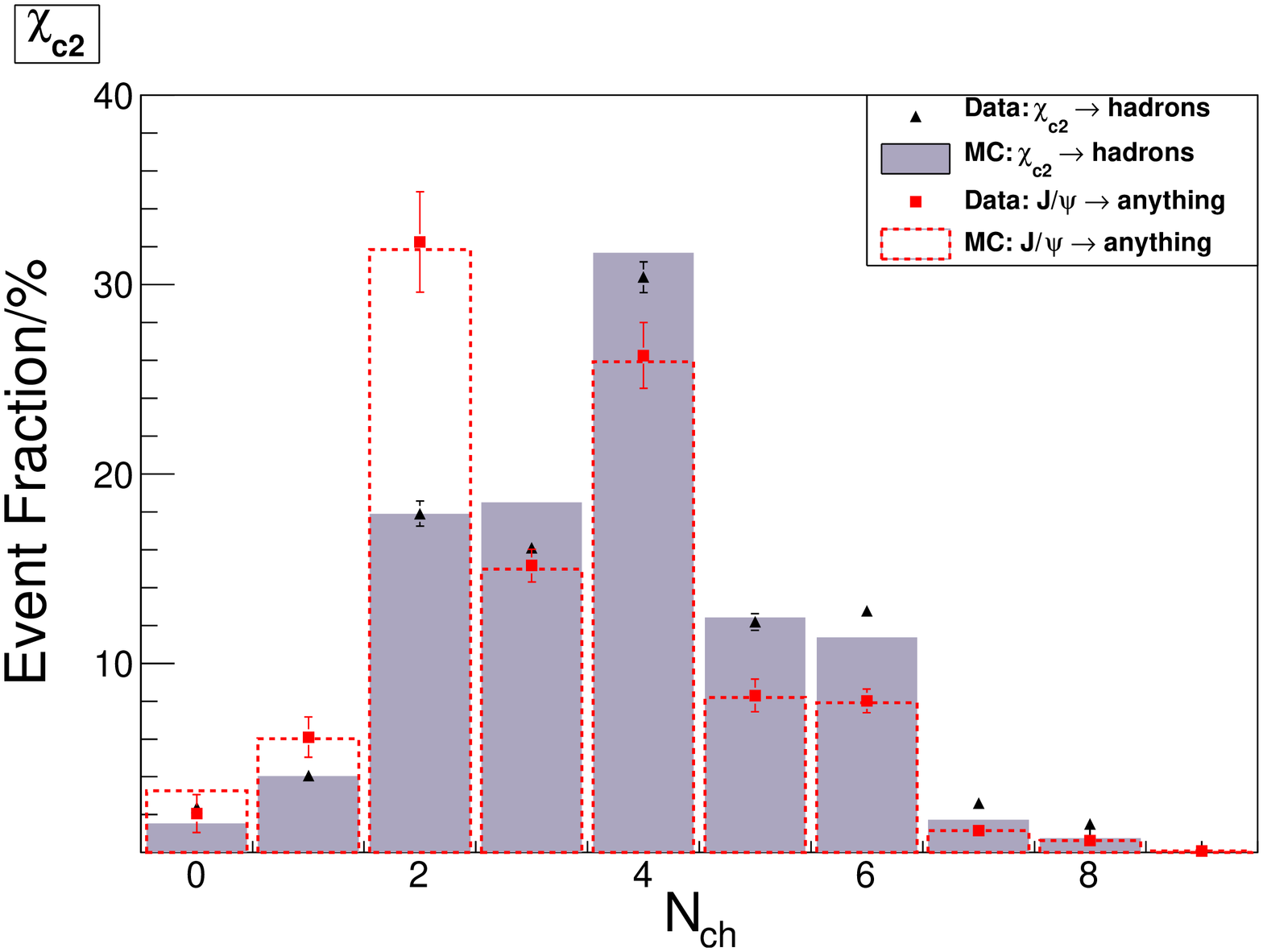}}
\put(-160,-35){\bf \large {(e)}} \rotatebox{-90}
{\includegraphics*[width=2.0in]{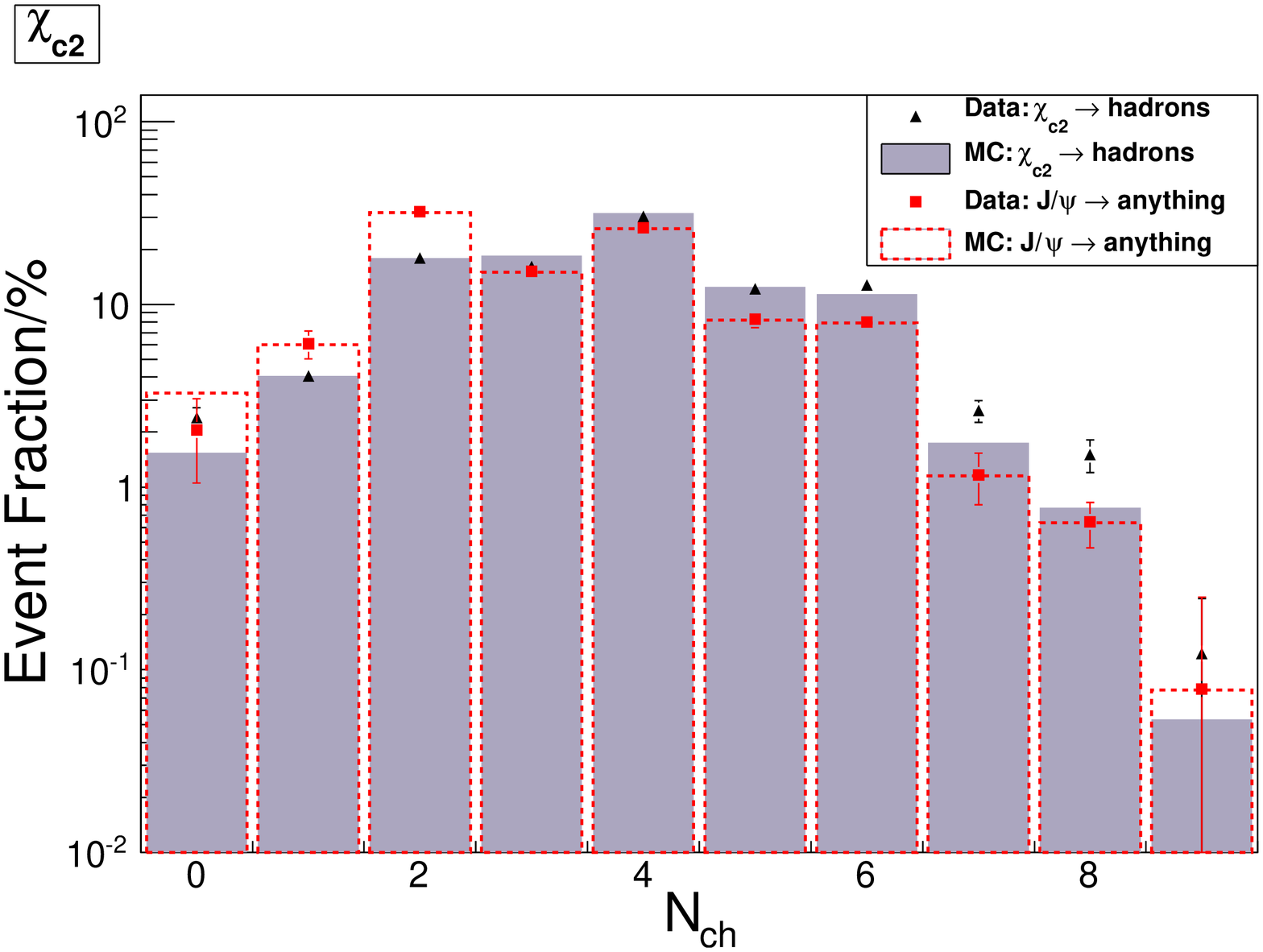}}
\put(-160,-35){\bf \large {(f)}}
\caption{\label{ch_results2} (Color online) Comparisons of the event fractions of
  data and those for scaled MC simulation events versus $N_{\rm ch}$ for
  (a) $\chi_{c0} \to$ hadrons, (c) $\chi_{c1} \to$ hadrons and
  $\chi_{c1} \to \gamma J/\psi,\:J/\psi \to$ anything, and (e)
  $\chi_{c2} \to$ hadrons and $\chi_{c2} \to \gamma J/\psi,\:J/\psi
  \to$ anything, while (b)(d)(f) are the corresponding logarithmic
  plots.  Here and in Fig.~\ref{gam_results2}
  below, the uncertainties shown for MC are the uncertainties from the
  fits to the inclusive $E_{\rm sh}$ distributions, and the uncertainties
  for data are those combined in quadrature with the systematic
  uncertainties, described in Section~\ref{systematics}.}
\end{center}
\end{figure*}

  The $N_{\rm ch}$ fractions, $F$, where the fraction is the number of
  efficiency corrected events with $N_{\rm ch} = j$ ($j$ takes on values
  from 0 to 9) divided by the sum of all $N_{\rm ch}$ events, are
  determined and are listed in Table~\ref{fraction1} for $\chi_{cJ}
  \to$ hadrons and Table~\ref{fraction2} for $\chi_{c1/2} \to \gamma
  J/\psi,\:J/\psi \to$ anything.  For comparison, MC simulation
  numbers, $F^{\rm MC}$, are also listed in the tables.  $F^{\rm MC}$ is
  calculated in an analogous way as was $F$ using the scaled MC
  simulation numbers.  In Figs.~\ref{ch_results2} (a), (c), and (e)
  comparisons of the $N_{\rm ch}$ fractions between data and scaled MC
  simulated sample are shown, while Figs.~\ref{ch_results2} (b), (d), and (f)
  are the corresponding plots in logarithmic scale.

Figure~\ref{ch_results2} shows good agreement between the three
$\chi_{cJ} \to$ anything decay distributions. Data are above MC
simulation for $N_{\rm ch} = 0$ and $N_{\rm ch} >5$ and below for $N_{\rm ch} = 3$
for these distributions. The agreement between data and MC simulation
is good for $J/\psi \to$ anything ($\chi_{c1}$ and $\chi_{c2} \to
\gamma J/\psi$).  Better agreement is expected for those
distributions, since MC tuning was performed on the $J/\psi \to$ anything
events.

\begin{table*}[bth]
\begin{center}
  \caption{Comparison of fraction of events in \% with $N_{\rm ch}$ for
    data and the scaled MC simulated sample for $\chi_{cJ} \to$
    hadrons.  Here and below, the first uncertainties are the
    uncertainties from the fits to the inclusive $E_{\rm sh}$
    distributions and the second ones are systematic, described
    in Section~\ref{systematics}.
\label{fraction1}}
\begin{footnotesize}
\begin{tabular}{l|cc|cc|cc} \hline
\T $N_{\rm ch}$ & $F_{\chi_{c0}}$ & $F^{\rm MC}_{\chi_{c0}}$ & $F_{\chi_{c1}}$ & $F^{\rm MC}_{\chi_{c1}}$ &  $F_{\chi_{c2}}$ & $F^{\rm MC}_{\chi_{c2}}$ \B \\ \hline
0  &  $3.09\pm0.05\pm0.30$ & 2.09 & $2.53\pm0.08\pm0.82$ & 1.46 & $2.40\pm0.06\pm0.31$ & 1.54\\
1  &  $4.70\pm0.05\pm 0.36$ & 4.56 & $4.03\pm0.07\pm0.81$ & 4.29 & $4.06\pm0.06\pm0.32$ & 4.05\\
2  &  $20.45\pm0.06\pm0.40$ & 20.62 & $17.79\pm0.10\pm0.71$ & 18.58 & $17.90\pm0.09\pm0.67$ & 17.89\\
3  &  $15.91\pm0.07\pm0.43$ & 18.17 & $16.36\pm0.09\pm0.60$ & 18.12 & $16.09\pm0.08\pm0.30$ & 18.48\\
4  &  $29.68\pm0.06\pm0.53$ & 31.63 & $30.16\pm0.08\pm0.71$ & 31.37 & $30.38\pm0.07\pm0.81$ & 31.67\\
5  &  $11.18\pm0.06\pm0.64$ & 10.97 & $12.39\pm0.08\pm0.65$ & 12.31 & $12.18\pm0.07\pm0.44$ & 12.42\\
6  &  $11.30\pm0.05\pm0.33$ & 10.12 & $12.65\pm0.08\pm0.50$ & 11.48 & $12.75\pm0.07\pm0.27$ & 11.38\\
7  &   $2.07\pm0.04\pm0.63$ & 1.27 & $2.56\pm0.06\pm0.55$ & 1.61 & $2.62\pm0.05\pm0.36$ & 1.75\\
8  &   $1.42\pm0.04\pm0.08$ & 0.55 & $1.37\pm0.05\pm0.21$ & 0.73 & $1.50\pm0.04\pm0.30$ & 0.77\\
9  &   $0.19\pm0.04\pm0.24$ & 0.04 &  $0.16\pm0.05\pm0.84$ & 0.05 &  $0.12\pm0.04\pm0.12$ & 0.05\\
\hline
\end{tabular}
\end{footnotesize}
\end{center}
\end{table*}

\begin{table}[bth]
\begin{center}
\caption{Comparison of fraction of events in \% with $N_{\rm ch}$ for data
  and the scaled MC simulated sample for $\chi_{c1,2} \to \gamma J/\psi,
  J/\psi \to$ anything. {\red These two sets of measurements describe the
  same distribution.}
\label{fraction2}}
\begin{footnotesize}
\begin{tabular}{l|cc|cc} \hline
\T $N_{\rm ch}$ & $F_{J/\psi_1}$ &  $F^{\rm MC}_{J/\psi_1}$ &  $F_{J/\psi_2}$ &   $F^{\rm MC}_{J/\psi_2}$   \B  \\ \hline
  0  &   $3.53\pm0.11\pm0.58$ &  3.33 &  $2.05\pm0.13\pm0.99$ & 3.27\\
  1  &   $6.46\pm0.11\pm1.42$ &  5.90 &  $6.10\pm0.15\pm1.05$ & 6.02\\
  2  &  $32.17\pm0.12\pm1.27$ & 32.15  & $32.24\pm0.18\pm2.65$ & 31.84\\
  3  &  $14.61\pm0.13\pm0.94$ & 15.25 & $15.15\pm0.18\pm0.84$ & 14.97\\
  4  &  $25.49\pm0.12\pm1.01$ & 25.29 & $26.25\pm0.17\pm1.73$ & 25.92\\
  5  &   $8.19\pm0.10\pm0.84$ &  8.48 &  $8.30\pm0.16\pm0.84$ & 8.19\\
  6  &   $7.62\pm0.10\pm0.51$ &  7.52 &  $8.02\pm0.15\pm0.60$ & 7.92\\
  7  &   $1.24\pm0.08\pm0.21$ &  1.38 &  $1.17\pm0.12\pm0.34$ & 1.15\\
  8  &   $0.65\pm0.07\pm0.26$ &  0.66 &  $0.65\pm0.11\pm0.15$ & 0.64\\
  9  &   $0.04\pm0.07\pm1.63$ &  0.05 &  $0.08\pm0.11\pm0.13$ & 0.08\\
  \hline
\end{tabular}
\end{footnotesize}
\end{center}
\end{table}

\section{\boldmath{Multiplicity distribution of the number of EMC showers}}
\label{ngam_results}
\subsection{MC study of EMC energy deposits} 
\label{EMC_energy_deposits}
The situation for neutral showers is more complicated than for charged
tracks.  Energy deposits in the EMC from $\psi(3686) \to \gamma
\chi_{cJ}$ and $\chi_{cJ} \to \gamma J/\psi$ events are caused by
their radiative photons, photons from the decays of $\pi^0$s from
$\chi_{cJ}$ and $J/\psi$ hadronic decays and their daughter particles,
bremsstrahlung from charged tracks, as well as interactions of hadrons
in the EMC crystals and noise.  The inclusive MC needs
to model all these sources.  We are interested in the number
of photons, $N_{\gamma}$, from the hadronic decays of $\chi_{cJ}$ and
$J/\psi$.  We can use the MC simulation to determine what fraction of
the EMCSHs are due to radiative photons and the photons from the
primary and secondary decays.  We signify the number of EMCSHs by
$N_{\rm sh}$.

\begin{figure}[htbp]
\begin{center}
\centering
\rotatebox{-90}{
\includegraphics*[width=2.in]{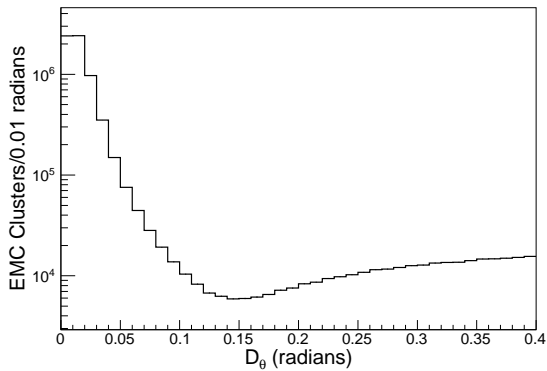}}
\caption{\label{cluster_dtheta} The distribution of $D_{\theta}$,
  which is the minimum difference in angle between an EMCSH and
  the angles of all the MC truth tagged photons.  }
\end{center}
\end{figure}

The MC ``truth'' information tags the radiative photons in the
generator model and photons from the generator final particle decays
in GEANT4~\cite{geant4}, e.g. $\pi^+ \to \mu^+ \nu_{\mu} \gamma$, as
well as final-state radiation photons. {\red  MC truth does not tag the
photons produced from the scattering and/or ionization of generator
final state particles with the detector materials, simulated by
GEANT4.}  The angles of tagged photons can be compared with the angles
of EMCSHs to identify the fraction of showers that are caused by these
photons.  Figure~\ref{cluster_dtheta} shows for a small subsample of
$\psi(3686) \to \gamma \chi_{cJ}$ events the angle $D_{\theta}$, which
is the minimum of the difference in angle between an EMCSH and all the
MC tagged photons.  There is a sharp peak at small $D_{\theta}$
corresponding to good shower matches between the MC predictions and
the EMCSHs. We define showers with $D_{\theta} < 0.1$ radians as a
good shower match.  The efficiency of matching photons in the correct
angular range ($|\cos \theta| < 0.8$) and energy range (0.25 GeV $<
E_{\rm sh} <$ 2 GeV) is 91.2\%.

\begin{figure}[htbp]
\begin{center}
\centering
\rotatebox{-90}{
\includegraphics*[width=2.in]{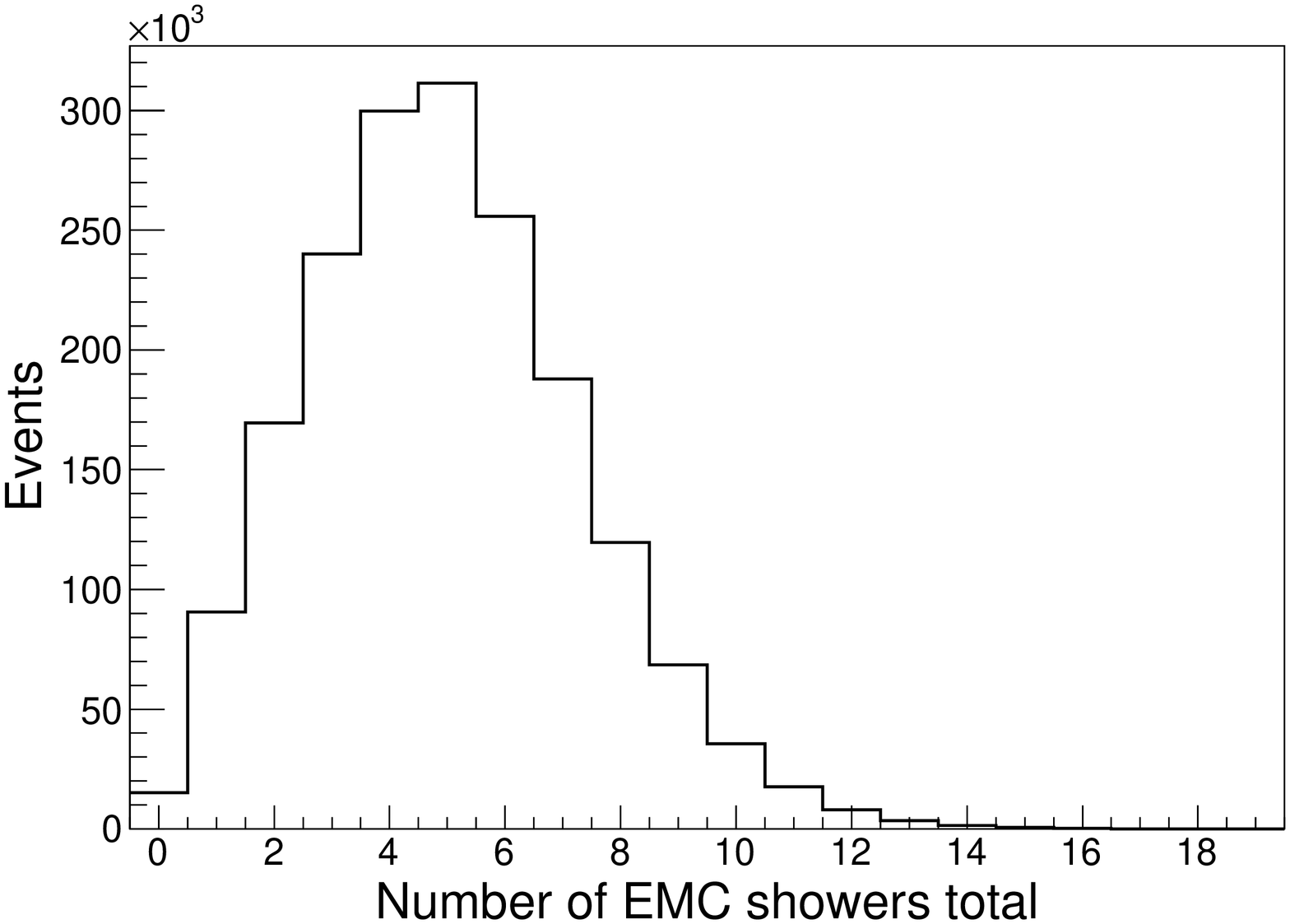}}
\put(-50,-40){\bf \large  {(a)}}

\rotatebox{-90}{
\includegraphics*[width=2.in]{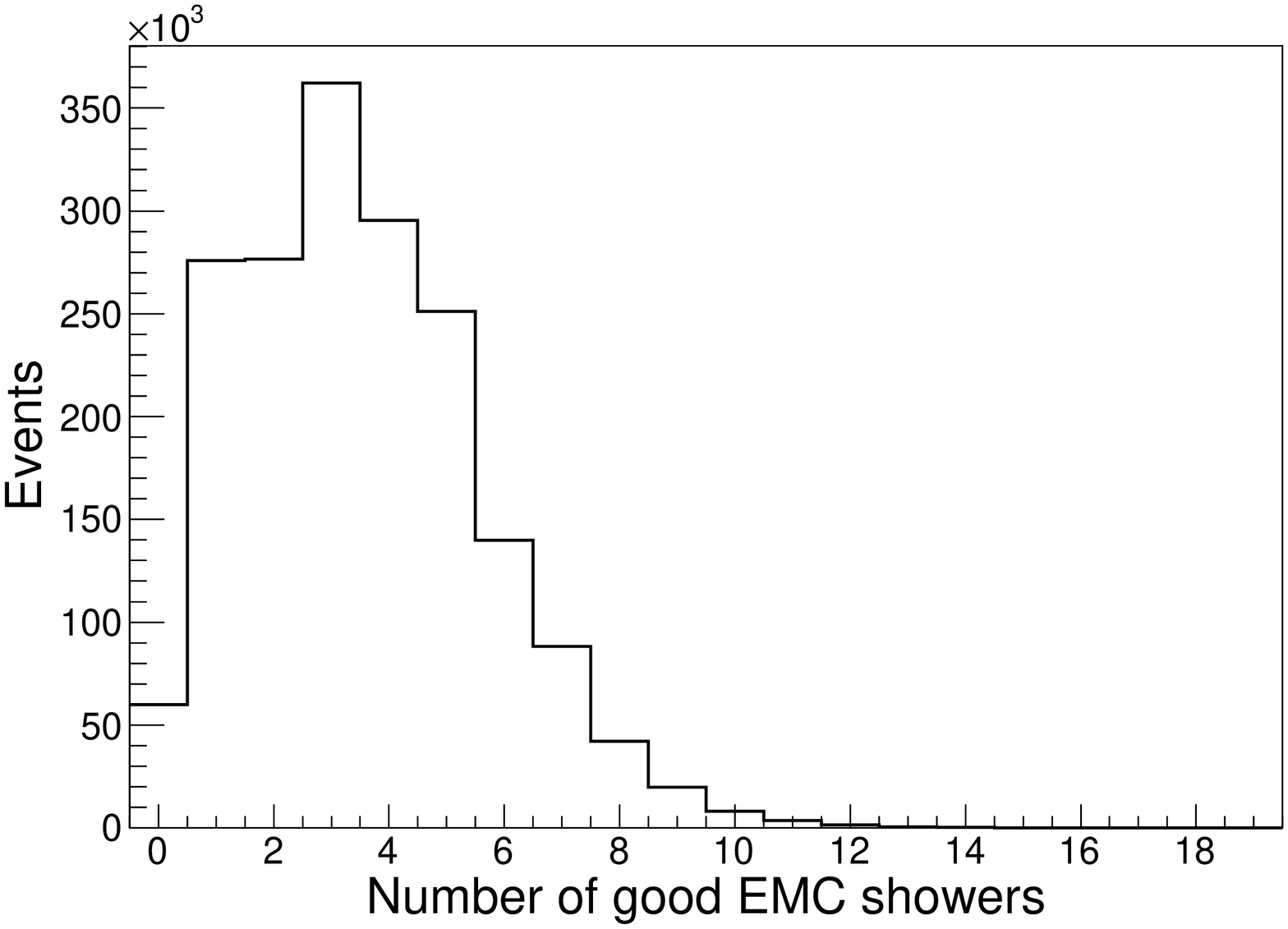}}
\put(-50,-40){\bf \large  {(b)}}
\caption{\label{cluster_number} The number distributions of (a) all
  and (b) good showers.}
\end{center}
\end{figure}

The fraction of good matches varies from 60\% at the lowest energy to
89\% at the highest. Figures~\ref{cluster_number} (a) and (b) show the
number distributions of all and good showers, respectively.  In the
following, we will compare the $N_{\rm sh}$ distributions of data and
MC simulation.

\subsection{\boldmath $N_{\rm sh}$ distribution}

The analysis for the distribution of $N_{\rm sh}$ is similar to that
for $N_{\rm ch}$.  $N_{\rm sh}$ is the number of showers satisfying
requirements on the energy, polar angle, and time, but no requirement
on the angle between the shower and the closest charged track in the
event.  Here 15 energy distributions are constructed for $N_{\rm sh}$
ranging from 1 to $\ge15${\red , where $N_{\rm sh} = 1$ is because at
  least one radiative photon must be detected}. For more direct
comparison of data with MC simulation, MC events are weighted only by
$wt_{\rm trans}$.

As above, using the number of detected data events, $D$, and the MC
determined efficiencies, $\epsilon$, versus $N_{\rm sh}$, we determine the
efficiency corrected $N$ distributions of data for $\chi_{cJ} \to$
anything and $\chi_{cJ} \to \gamma J/\psi,\:J/\psi \to$ anything.
Results are listed in Table~\ref{result_chic_to_all_gam} for
$\chi_{cJ} \to$ anything and Table~\ref{result_jpsi_to_all_gam} for
$\chi_{c1/2} \to \gamma J/\psi,\:J/\psi \to$ anything.  The $N_{\rm sh}$
fractions, $F$, are also determined and are listed in
Table~\ref{fraction1_gam} for $\chi_{cJ} \to$ hadrons and
Table~\ref{fraction2_gam} for $\chi_{c1/2} \to \gamma J/\psi,\:J/\psi
\to$ anything.  For comparison, MC simulation numbers, $N^{\rm MC}$, are
listed in Tables~\ref{result_chic_to_all_gam} and
\ref{result_jpsi_to_all_gam} in the appendix and fractions, $F^{\rm MC}$, are listed in
Tables~\ref{fraction1_gam} and
\ref{fraction2_gam}.  

\begin{table*}[bth]
\begin{center}
\caption{Comparison of fraction of events in \% with $N_{\rm sh}$
  between data and the  scaled MC simulated sample for {\red $\psi(3686) \to
  \gamma \chi_{cJ} \to \gamma$ hadrons}.
\label{fraction1_gam}}
\begin{footnotesize}
\begin{tabular}{l|cc|cc|cc} \hline
\T $N_{\rm sh}$ & $F_{\chi_{c0}}$ & $F^{\rm MC}_{\chi_{c0}}$ & $F_{\chi_{c1}}$ & $F^{\rm MC}_{\chi_{c1}}$ &  $F_{\chi_{c2}}$ & $F^{\rm MC}_{\chi_{c2}}$ \B \\ \hline
1  &   $6.93\pm0.03\pm0.33$  & 6.37 &  $4.77\pm0.06\pm0.46$ &  4.33 &  $5.88\pm0.04\pm0.32$ & 4.75\\
2  &   $9.46\pm0.04\pm0.61$  & 9.51 &  $7.92\pm0.06\pm0.61$ &  8.11 &  $8.53\pm0.05\pm0.58$ & 8.39\\
3  &  $13.29\pm0.05\pm0.29$ & 14.20 & $12.72\pm0.07\pm0.59$ & 13.40 & $12.48\pm0.06\pm0.60$ & 13.49\\
4  &  $16.62\pm0.06\pm0.39$ & 17.28 & $16.70\pm0.07\pm0.75$ & 16.76 & $16.54\pm0.06\pm0.68$ & 16.82\\
5  &  $16.94\pm0.06\pm0.54$ & 17.69 & $17.55\pm0.08\pm0.80$ & 17.86 & $17.42\pm0.07\pm0.95$ & 17.70\\
6  &  $12.34\pm0.06\pm0.57$ & 13.63 & $13.58\pm0.08\pm0.57$ & 14.74 & $13.06\pm0.07\pm0.52$ & 14.42\\
7  &   $9.21\pm0.05\pm0.53$ &  9.48 & $10.10\pm0.08\pm0.63$ & 10.71 & $9.73\pm0.07\pm0.53$ & 10.44\\
8  &   $6.64\pm0.05\pm0.60$ &  5.79 &  $7.10\pm0.07\pm0.70$ &  6.71 &  $6/98\pm0.06\pm0.56$ &  6.63\\
9  &   $4.10\pm0.03\pm1.00$ &  3.20 &  $4.55\pm0.22\pm0.64$ &  3.80 &  $4.32\pm0.05\pm0.57$ &  3.76\\
10 &   $2.14\pm0.03\pm0.90$ &  1.58 &  $2.71\pm0.06\pm0.64$ &  1.95 &  $2.19\pm0.19\pm0.35$ &  1.95\\
11 &   $1.29\pm0.04\pm0.37$ &  0.74 &  $1.09\pm0.06\pm0.38$ &  0.94 &  $1.32\pm0.05\pm0.26$ &  0.94\\
12 &   $0.57\pm0.03\pm0.22$ &  0.32 &  $0.78\pm0.05\pm0.35$ &  0.42 &  $0.73\pm0.05\pm0.46$ &  0.42\\
13 &   $0.27\pm0.03\pm0.16$ &  0.14 &  $0.26\pm0.05\pm0.17$ &  0.17 &  $0.50\pm0.05\pm0.27$ &  0.18\\
14 &   $0.14\pm0.02\pm0.14$ &  0.05 &  $0.11\pm0.04\pm0.12$ &  0.07 &  $0.30\pm0.05\pm0.21$ &  0.08\\
$\ge15$ &   $0.06\pm0.02\pm0.06$ &  0.02 &  $0.06\pm0.03\pm0.04$ &  0.03 &  $0.01\pm0.02\pm0.02$ &  0.03\\
\hline
\end{tabular}
\end{footnotesize}
\end{center}
\end{table*}

\begin{table}[bth]
\begin{center}
\caption{Comparison of fraction of events in \% with $N_{\rm sh}$
  between data and the scaled MC simulated sample for {\red $\chi_{c1,2} \to
  \gamma J/\psi \to \gamma$ anything.} {\red These two sets of measurements
  measure the same distribution and are in agreement within
  uncertainties.}
\label{fraction2_gam}}
\begin{footnotesize}
\begin{tabular}{l|cc|cc} \hline
\T $N_{\rm sh}$ & $F_{J/\psi_1}$ &  $F^{\rm MC}_{J/\psi_1}$ &  $F_{J/\psi_2}$ &   $F^{\rm MC}_{J/\psi_2}$   \B \\ \hline
1  &  $4.33\pm0.10\pm0.54$ &  3.78 &  $2.48\pm0.10\pm0.97$ &  4.12\\
2  & $13.49\pm0.09\pm0.87$ & 12.56 & $10.68\pm0.14\pm1.42$ & 12.21\\
3  & $11.76\pm0.09\pm0.62$ & 11.58 & $11.92\pm0.13\pm1.09$ & 11.77\\
4  & $14.15\pm0.10\pm1.24$ & 15.16 & $14.80\pm0.14\pm1.29$ & 15.03\\
5  & $14.24\pm0.10\pm1.12$ & 15.19 & $15.20\pm0.14\pm2.48$ & 15.06\\
6  & $13.34\pm0.10\pm0.85$ & 13.75 & $14.26\pm0.14\pm0.94$ & 13.75\\
7  & $11.14\pm0.09\pm0.73$ & 10.98 & $11.65\pm0.14\pm1.56$ & 10.96\\
8  &  $7.73\pm0.09\pm0.94$ &  7.65 &  $8.07\pm0.13\pm1.91$ &  7.62\\
9  &  $4.74\pm0.42\pm0.55$ &  4.49 &  $5.06\pm0.09\pm1.00$ &  4.53\\
10 &  $2.43\pm0.06\pm0.67$ &  2.47 &  $3.08\pm0.09\pm0.59$ &  2.50\\
11 &  $1.50\pm0.07\pm0.44$ &  1.28 &  $1.52\pm0.08\pm0.89$ &  1.31 \\
12 &  $0.58\pm0.05\pm0.30$ &  0.63 &  $0.87\pm0.08\pm0.43$ &  0.65 \\
13 &  $0.36\pm0.07\pm0.20$ &  0.30 &  $0.27\pm0.07\pm0.17$ &  0.31\\
14 &  $0.17\pm0.06\pm0.20$ &  0.13 &  $0.14\pm0.08\pm0.32$ &  0.13\\
$\ge15$ &  $0.05\pm0.04\pm0.05$ &  0.05 &  $0.00\pm0.05\pm0.09$ &  0.05\\
\hline
\end{tabular}
\end{footnotesize}
\end{center}
\end{table}


\begin{figure*}[htbp]
\begin{center}
\rotatebox{-90}
{\includegraphics*[width=2.0in]{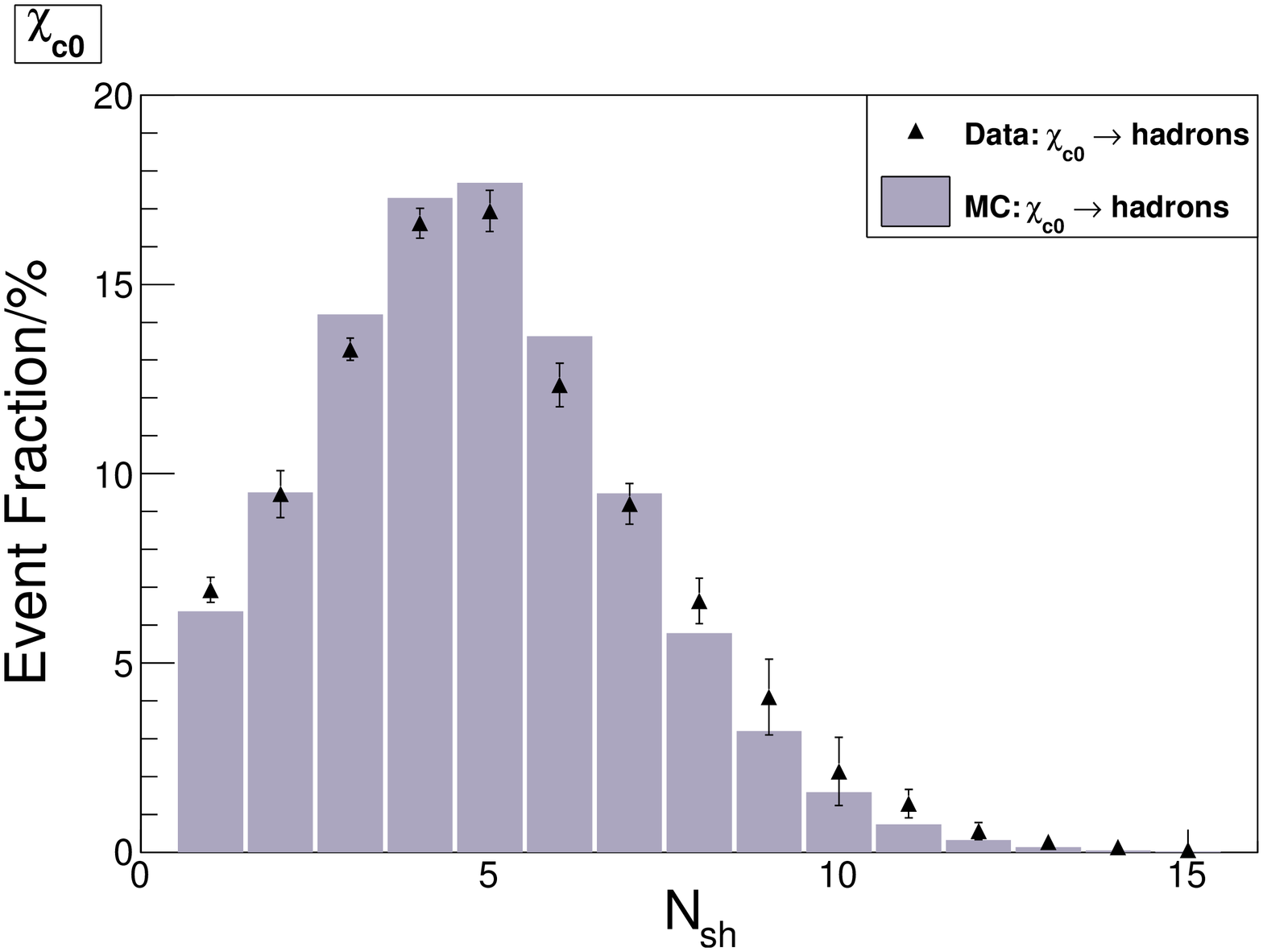}}
\put(-165,-26){\bf \large  {(a)}}
\rotatebox{-90}
{\includegraphics*[width=2.0in]{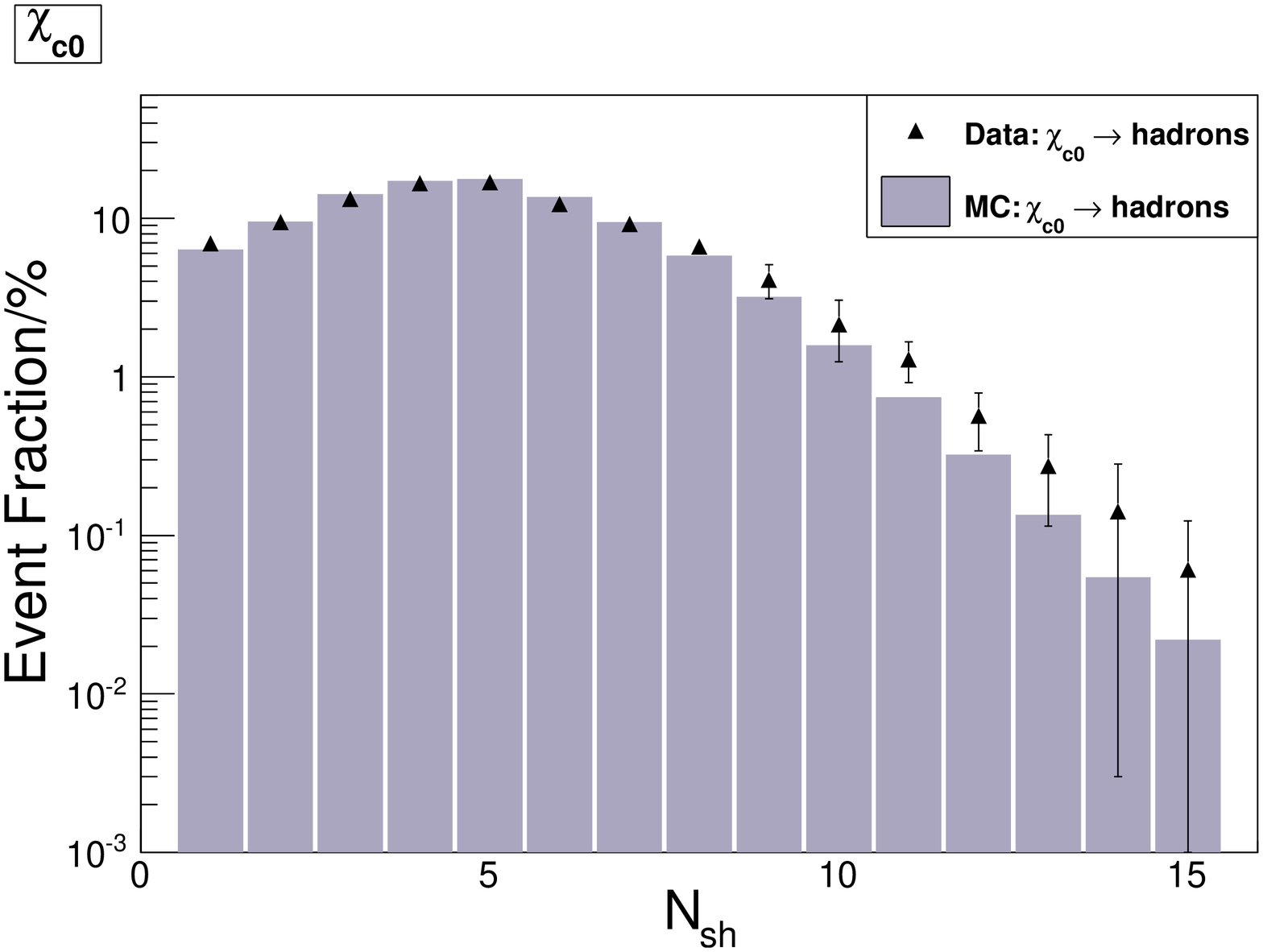}}
\put(-165,-26){\bf \large {(b)}}

\rotatebox{-90}
{\includegraphics*[width=2.0in]{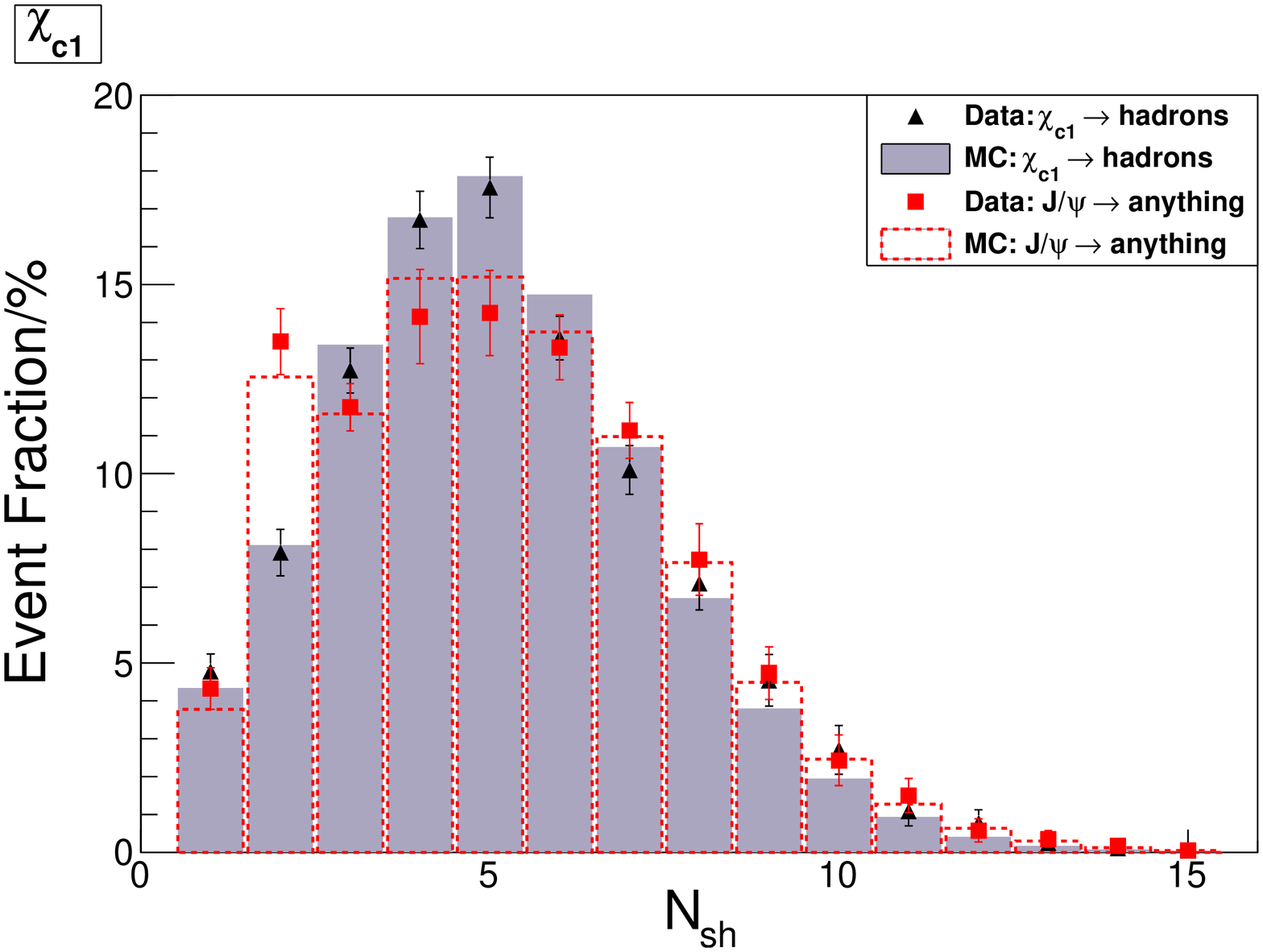}}
\put(-165,-26){\bf \large  {(c)}}
\rotatebox{-90}
{\includegraphics*[width=2.0in]{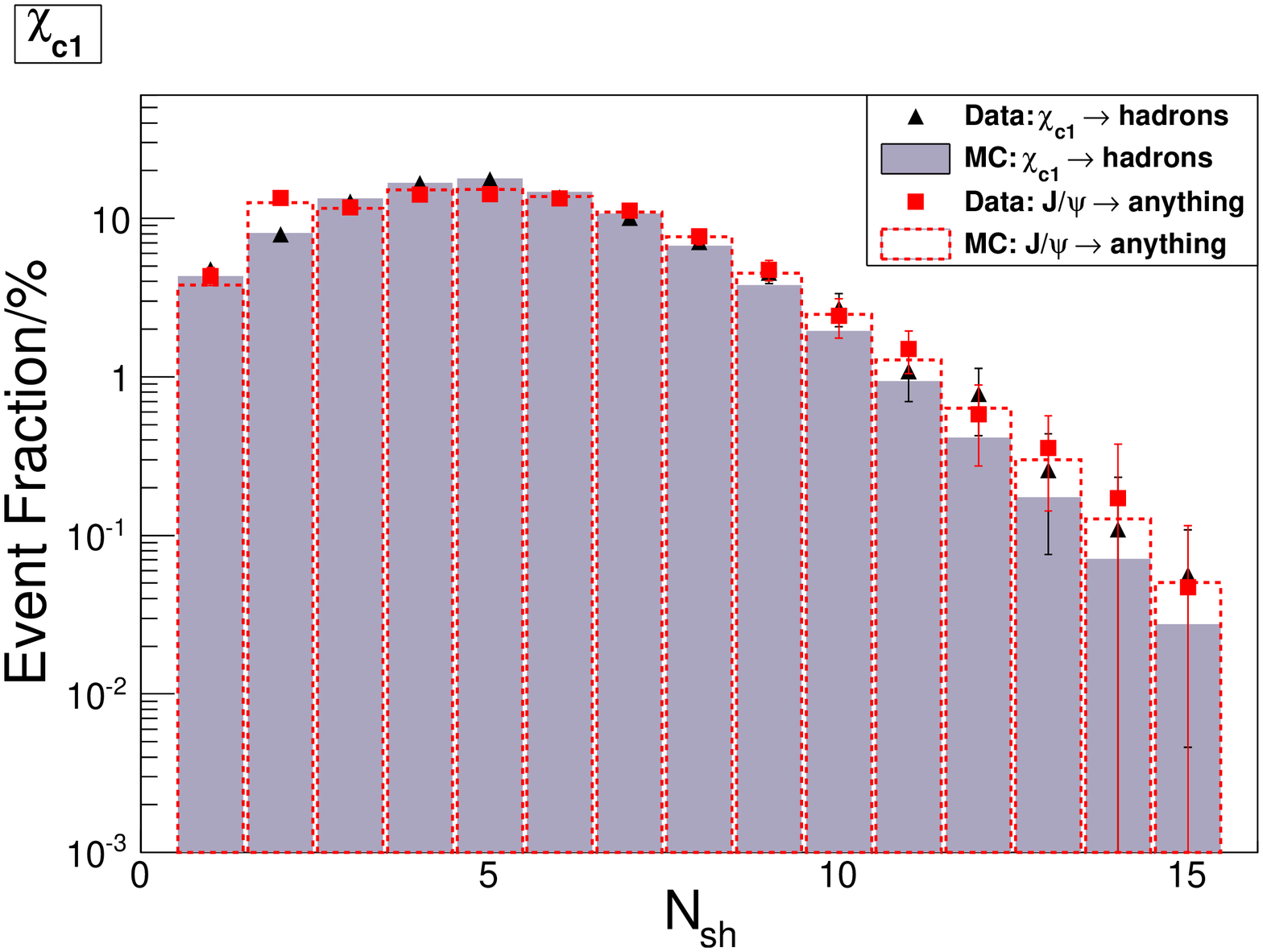}}
\put(-165,-26){\bf \large  {(d)}}

\rotatebox{-90}
{\includegraphics*[width=2.0in]{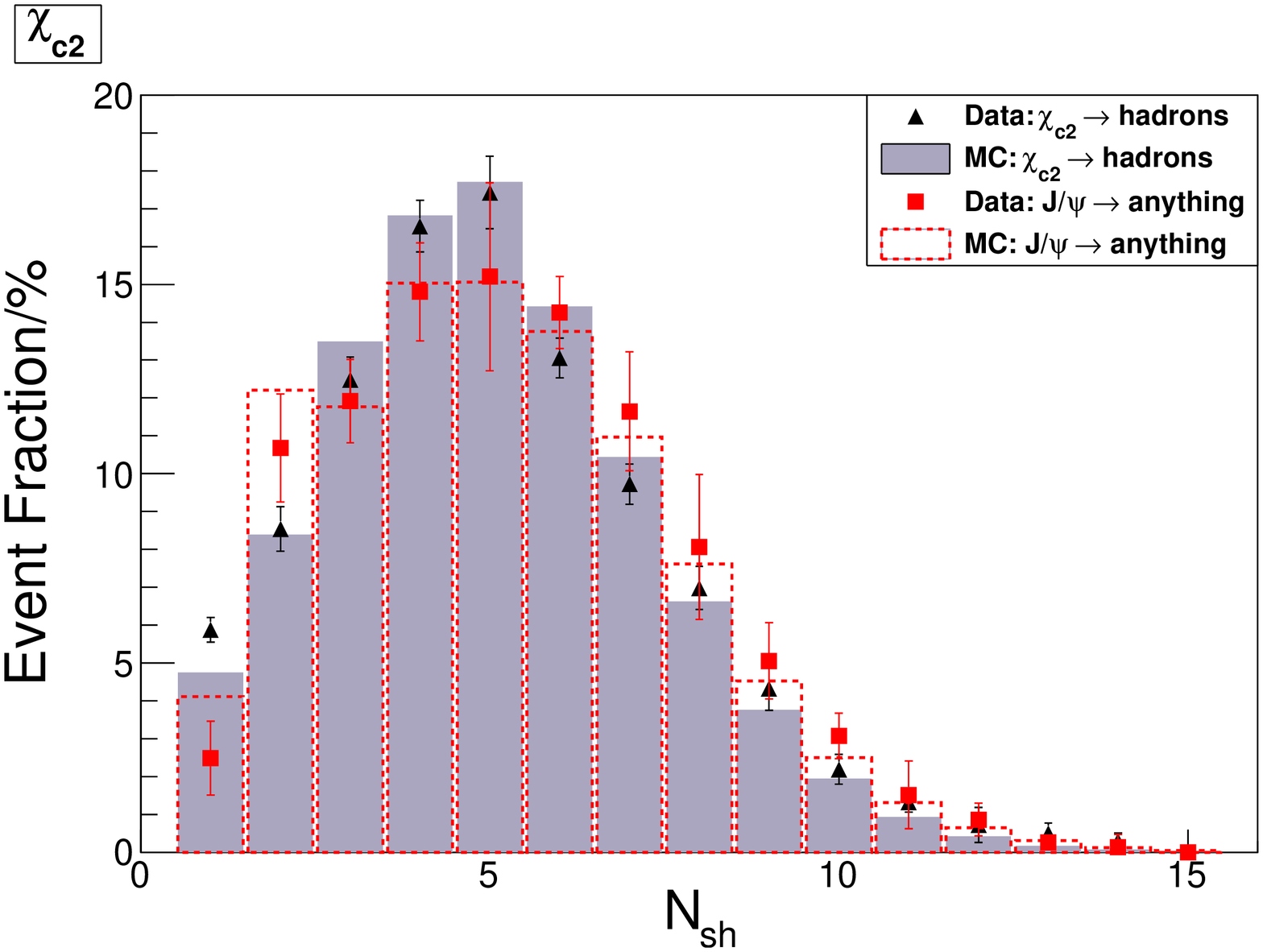}}
\put(-165,-26){\bf \large {(e)}} \rotatebox{-90}
{\includegraphics*[width=2.0in]{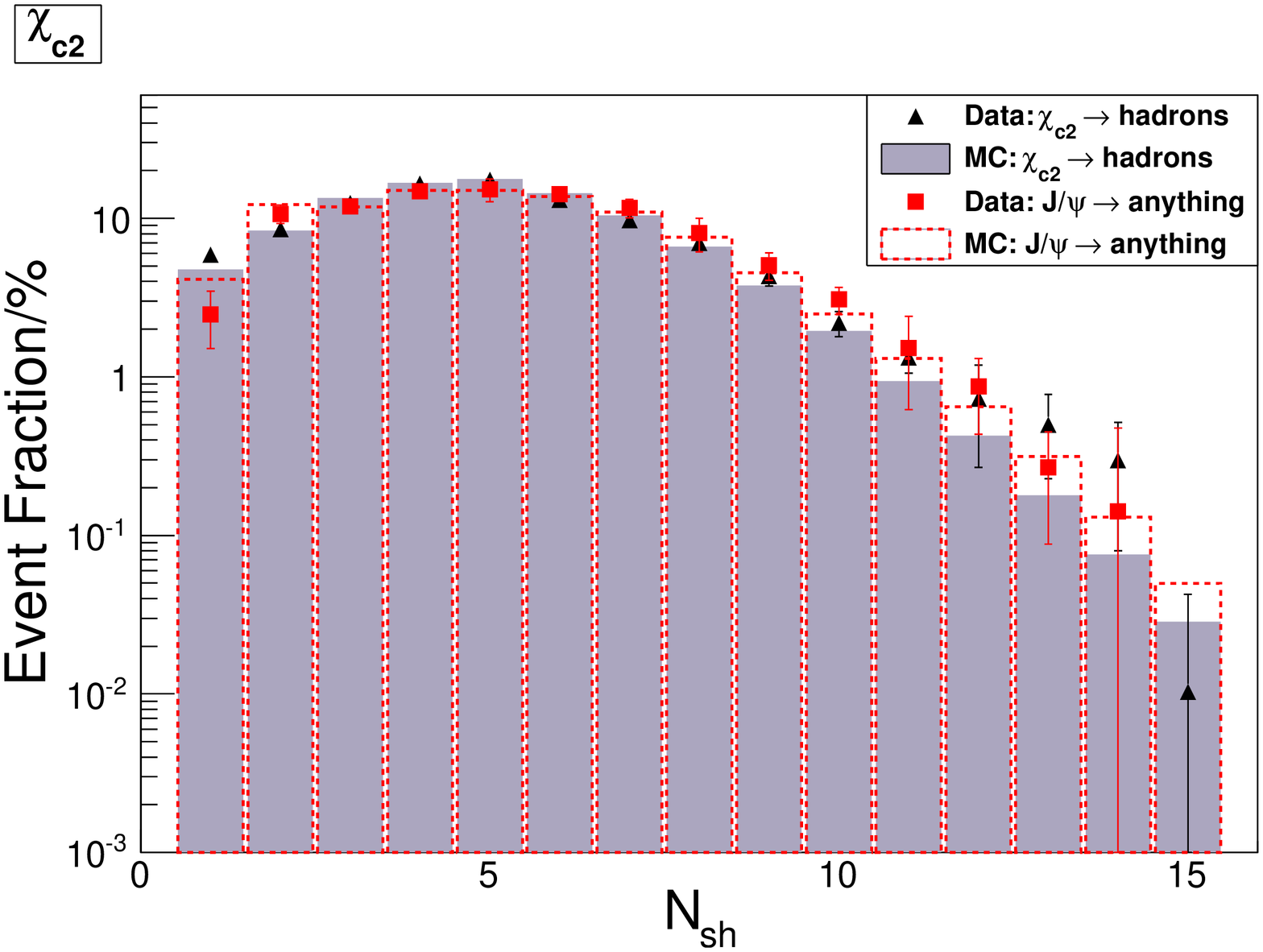}}
\put(-165,-26){\bf \large {(f)}}
\caption{\label{gam_results2} (Color online) Comparisons of the
  event fractions of data and those for scaled MC simulation events
  versus $N_{\rm sh}$ for (a) $\chi_{c0} \to$ hadrons, (c) $\chi_{c1}
  \to$ hadrons and $\chi_{c1} \to \gamma J/\psi,\:J/\psi \to$
  anything, and (e) $\chi_{c2} \to$ hadrons and $\chi_{c2} \to \gamma
  J/\psi,\:J/\psi \to$ anything, while (b)(d)(f) are the corresponding
  logarithmic plots.}
\end{center}
\end{figure*}

  In Figs.~\ref{gam_results2} (a), (c), and (e) the comparisons of the
  $N_{\rm sh}$ fractions between data and the scaled MC simulated sample are shown,
  and Figs.~\ref{gam_results2} (b), (d), and (f) are the corresponding
  plots in logarithmic scale.  For $\chi_{cJ} \to$ hadrons, the
  distributions in Fig.~\ref{gam_results2} are similar for the three
  $\chi_{cJ}$ decays, and data are above MC simulation for $N_{\rm sh} = 1$
  and $N_{\rm sh} > 7$ and below for $N_{\rm sh} = 3$ and 6.  For $J/\psi \to$
  anything ($\chi_{c1}$ and $\chi_{c2} \to \gamma J/\psi$), there is
  only minor disagreement between data and MC simulation for the
  $N_{\rm sh}$ distributions.

\section{\boldmath Multiplicity distribution of the number of $\pi^0$s}
\label{npi0_results}
\begin{figure}[htbp]
\begin{center}
  {\includegraphics*[width=3.0in]{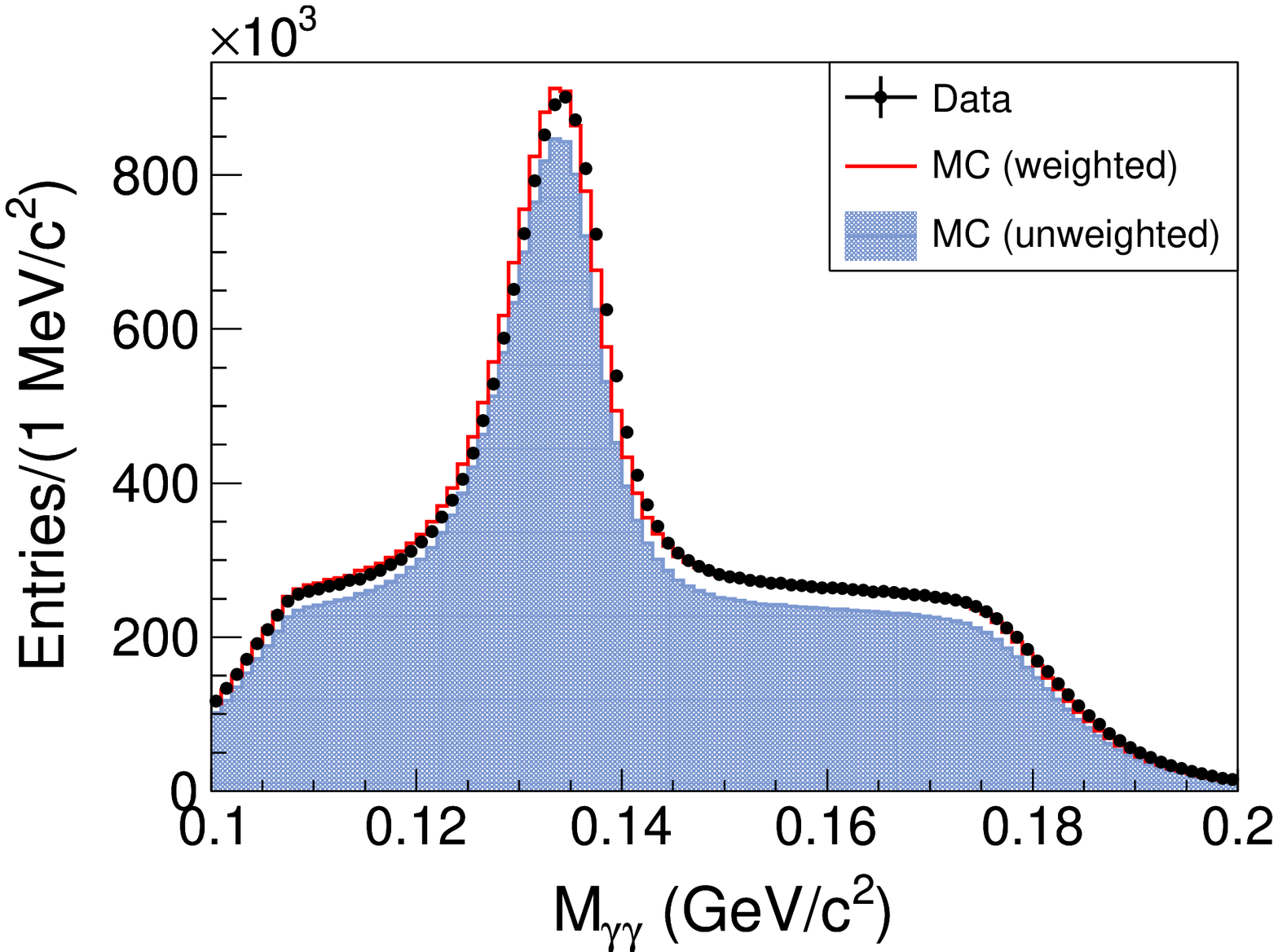}}
  \caption{\label{m_pi0} The $M_{\gamma\gamma}$
    distribution of $\pi^0$ candidates reconstructed without the tight
    $\pi^0$ mass selection requirement.  Data are represented by dots,
    and the MC sample by the red and shaded histograms for the
    MC events weighted by $wt_{\pi^0}$ and unweighted events, respectively.}
\end{center}
\end{figure}

An even more complicated case is the distribution of the number of
$\pi^0$s, $N_{\pi^0}$.  Here, as for the $N_{\rm ch} = 0$ case, the
$N_{\pi^0}$ distribution is considered in more detail. The $\gamma
\gamma$ invariant mass, $M_{\gamma \gamma}$, distribution of the
$\pi^0$ candidates is shown in Fig.~\ref{m_pi0}, where there are a
large number of $\gamma \gamma$ miscombinations in the plot.  A
somewhat better estimate of $N_{\pi^0}$ is made with the restrictive
requirement ${\red 0.120}  < M_{\gamma \gamma} < $ 0.145 GeV$/c^2$, which was
the requirement used when vetoing EMCSHs that might be part of a
$\pi^0$ combination from the $E_{\rm sh}$ distribution used in the fitting
for the number of $\psi(3686) \to \gamma \chi_{cJ}$ and $\chi_{cJ} \to
\gamma J/\psi$ events~\cite{bam248}.  However, even with this
requirement there are still many $\gamma \gamma$ miscombinations.

To determine the fraction, $R$, of the $\pi^0$ candidates that are
valid $\pi^0$s, we fit the $M_{\gamma\gamma}$ distributions for ${\red
  0.120}
 < M_{\gamma \gamma} < $ 0.145 GeV$/c^2$ for each $N_{\pi^0}$ for both
data and the MC simulated sample to a signal shape and first order
Chebychev polynomial background.  The basic signal shape was
determined using the MC truth information to identify correct $\gamma
\gamma$ combinations in simulated data. For data, the basic signal
shape is convolved with a bifurcated Gaussian function to account for
the difference in resolution between data and the MC simulated sample.  $R$ is the
fraction of signal events in the region ${\red 0.120}  < M_{\gamma \gamma} < $
0.145 GeV$/c^2$.  The values of $R$ versus $N_{\pi^0}$ are listed in
Table~\ref{fvalues}.

Note that $N_{\pi^0}$ may not fully determine the number of valid
$\pi^0$s.  For instance, $N_{\pi^0}=3$ may include the cases of three
valid $\pi^0$s, two valid $\pi^0$s and one miscombination, one valid
$\pi^0$ and two miscombinations, and three miscombinations.

\begin{table}[bth]
\begin{center}
\caption{Fraction $R$ of events that are valid $\pi^0$s versus $N_{\pi^0}$. For $N_{\pi^0} = 0$, $R = 1$ is assumed.
\label{fvalues}}
\begin{footnotesize}
\begin{tabular}{l|r|r} \hline
\T $N_{\pi^0}$ & $R(data)$ (\%) & $R(MC)$ (\%)    \\ \hline
all  &  $56.09\pm0.23$   & $56.71\pm0.04$ \B \\
0  &  100 (assumed) & 100 (assumed)  \\
1  &  $80.32\pm0.23$  & $78.36\pm0.09$  \\
2  &  $67.30\pm0.20$  & $65.49\pm0.08$ \\
3  &  $56.10\pm0.34$ & $56.14\pm0.09$  \\
4  &  $50.10\pm0.39$ & $50.04\pm0.11$  \\
5  &  $45.88\pm0.45$ & $45.69\pm0.13$  \\
6  &  $41.60\pm0.18$ & $42.21\pm0.15$  \\
7  &  $39.74\pm0.15$ & $39.54\pm0.18$  \\
8  &  $36.91\pm0.19$ & $37.53\pm0.22$  \\
9  &  $32.37\pm0.12$ & $33.02\pm0.15$  \\
\hline
\end{tabular}
\end{footnotesize}
\end{center}
\end{table}

The analysis for the detected $N_{\pi^0}$ distributions is similar to
those for $N_{\rm ch}$ and $N_{\rm sh}$.  Here 10 $E_{\rm sh}$ distributions of
data are constructed for $N_{\pi^0}$ ranging from 0 to $\ge9$.  For
more direct comparison of data with MC simulation, MC events are
weighted only by $wt_{\rm trans}$.

Using the number of detected data events, $D$, the MC determined
efficiencies, $\epsilon$, and $R(data)$ versus $N_{\pi^0}$, we
determine the efficiency corrected $N$ distributions of data for
$\chi_{cJ} \to$ anything and $\chi_{cJ} \to \gamma J/\psi,\:J/\psi \to$
anything, where $N = R\cdot D/\epsilon$, which gives a better
representation of the $N_{\pi^0}$ distribution.  Results are listed in the appendix in
Table~\ref{result_chic_to_all_pi0} for $\chi_{cJ} \to$ anything and
Table~\ref{result_jpsi_to_all_pi0} for $\chi_{c1/2} \to \gamma J/\psi,
J/\psi \to$ anything. The $N_{\pi^0}$ fractions, $F$, are also
determined and are listed in Table~\ref{fraction1_pi0} for $\chi_{cJ}
\to$ hadrons and Table~\ref{fraction2_pi0} for $\chi_{c1/2} \to \gamma
J/\psi,\:J/\psi \to$ anything.  For comparison, scaled MC simulation
numbers, $N^{\rm MC}$, multiplied by $R(MC)$ are listed in
Tables~\ref{result_chic_to_all_pi0} and \ref{result_jpsi_to_all_pi0}
and MC fractions, $F^{\rm MC}$, are listed in
Tables~\ref{fraction1_pi0} and \ref{fraction2_pi0}.

\begin{table*}[bth]
\begin{center}
\caption{Comparison of fraction of events in \% with $N_{\pi^0}$ for data and scaled MC simulated sample for $\chi_{cJ} \to$ hadrons. Both $F$ and $F^{\rm MC}$ are based on numbers of events multiplied by $R$, the fraction of valid $\pi^0$s.
  The first uncertainties are the
  uncertainties from the fits to the inclusive $E_{\rm sh}$
  distributions and $R$, and the second are the systematic
  uncertainties, described in Section~\ref{systematics}.
\label{fraction1_pi0}}
\begin{footnotesize}
\begin{tabular}{l|cc|cc|cc} \hline
\T $N_{\pi^0}$ & $F_{\chi_{c0}}$ & $F^{\rm MC}_{\chi_{c0}}$ & $F_{\chi_{c1}}$ & $F^{\rm MC}_{\chi_{c1}}$ &  $F_{\chi_{c2}}$ & $F^{\rm MC}_{\chi_{c2}}$ \B \\ \hline
0  &  $47.59\pm0.08\pm1.43$ & 45.53 & $42.75\pm0.12\pm1.47$ & 44.79 & $42.85\pm0.09\pm3.27$ & 44.27\\
1  &  $27.08\pm0.10\pm1.33$ & 31.27 & $28.00\pm0.12\pm0.93$ & 29.57 & $26.88\pm0.11\pm1.99$ & 29.37\\
2  &  $13.27\pm0.07\pm0.88$ & 14.08 & $14.79\pm0.09\pm0.47$ & 14.09 & $14.83\pm0.09\pm1.14$ & 14.32\\
3  &   $5.66\pm0.06\pm0.44$ &  5.29 &  $6.90\pm0.08\pm0.18$ &  6.18 &  $6.98\pm0.08\pm0.34$ &  6.36\\
4  &   $2.79\pm0.04\pm0.50$ &  2.16 &  $3.66\pm0.07\pm0.27$ &  2.80 &  $3.60\pm0.06\pm0.33$ &  2.91\\
5  &   $1.52\pm0.04\pm0.15$ &  0.92 &  $1.86\pm0.06\pm0.12$ &  1.30 &  $2.00\pm0.06\pm0.26$ &  1.38\\
6  &   $0.84\pm0.03\pm0.11$ &  0.40 &  $0.89\pm0.05\pm0.06$ &  0.62 &  $1.15\pm0.06\pm0.17$ &  0.68\\
7  &   $0.57\pm0.03\pm0.83$ &  0.18 &  $0.48\pm0.05\pm0.48$ &  0.31 &  $0.55\pm0.04\pm0.56$ & 0.34\\
8  &   $0.29\pm0.02\pm0.43$ &  0.08 &  $0.37\pm0.03\pm0.53$ &  0.16 &  $0.35\pm0.02\pm0.66$ & 0.17\\
$\ge9$ & $0.39\pm0.02\pm0.23$ &  0.07 &  $0.28\pm0.03\pm0.29$ &  0.18 &  $0.81\pm0.04\pm2.78$ & 0.20\\
\hline
\end{tabular}
\end{footnotesize}
\end{center}
\end{table*}

\begin{table}[bth]
\begin{center}
  \caption{Comparison of fraction of events in \% with $N_{\pi^0}$
    for data and scaled MC simulated sample for $\chi_{c1,2} \to
    \gamma J/\psi, J/\psi \to$ anything.  Both $F$ and $F^{\rm MC}$ are
    based on numbers of events multiplied by $R$.
\label{fraction2_pi0}}
\begin{footnotesize}
\begin{tabular}{l|cc|cc} \hline
\T $N_{\pi^0}$ & $F_{J/\psi_1}$ &  $F^{\rm MC}_{J/\psi_1}$ &  $F_{J/\psi_2}$ &   $F^{\rm MC}_{J/\psi_2}$  \B  \\ \hline
0  & $52.68\pm0.14\pm3.27$ & 54.72 & $46.94\pm0.21\pm6.85$ & 53.29\\
1  & $24.84\pm0.15\pm1.10$ & 25.16 & $27.88\pm0.21\pm3.34$ & 25.23\\
2  & $11.51\pm0.11\pm1.70$ & 10.79 & $13.71\pm0.17\pm1.11$ & 11.25\\
3  &  $5.14\pm0.09\pm0.50$ &  4.80 &  $5.53\pm0.12\pm0.80$ &  5.14\\
4  &  $2.50\pm0.07\pm0.42$ &  2.24 &  $2.84\pm0.10\pm0.65$ &  2.46\\
5  &  $1.35\pm0.07\pm0.19$ &  1.09 &  $1.37\pm0.09\pm0.27$ &  1.23\\
6  &  $0.85\pm0.06\pm0.38$ &  0.56 &  $0.76\pm0.08\pm0.11$ &  0.63\\
7  &  $0.51\pm0.06\pm0.74$ &  0.29 &  $0.53\pm0.08\pm1.06$ &  0.34\\
8  &  $0.23\pm0.04\pm0.41$ &  0.16 &  $0.28\pm0.06\pm1.14$ &  0.19\\
$\ge9$ & $0.40\pm0.04\pm0.58$ &  0.20 &  $0.17\pm0.05\pm0.45$ &  0.25\\
\hline
\end{tabular}
\end{footnotesize}
\end{center}
\end{table}

\begin{figure*}[htbp]
\begin{center}
\rotatebox{-90}
{\includegraphics*[width=2.0in]{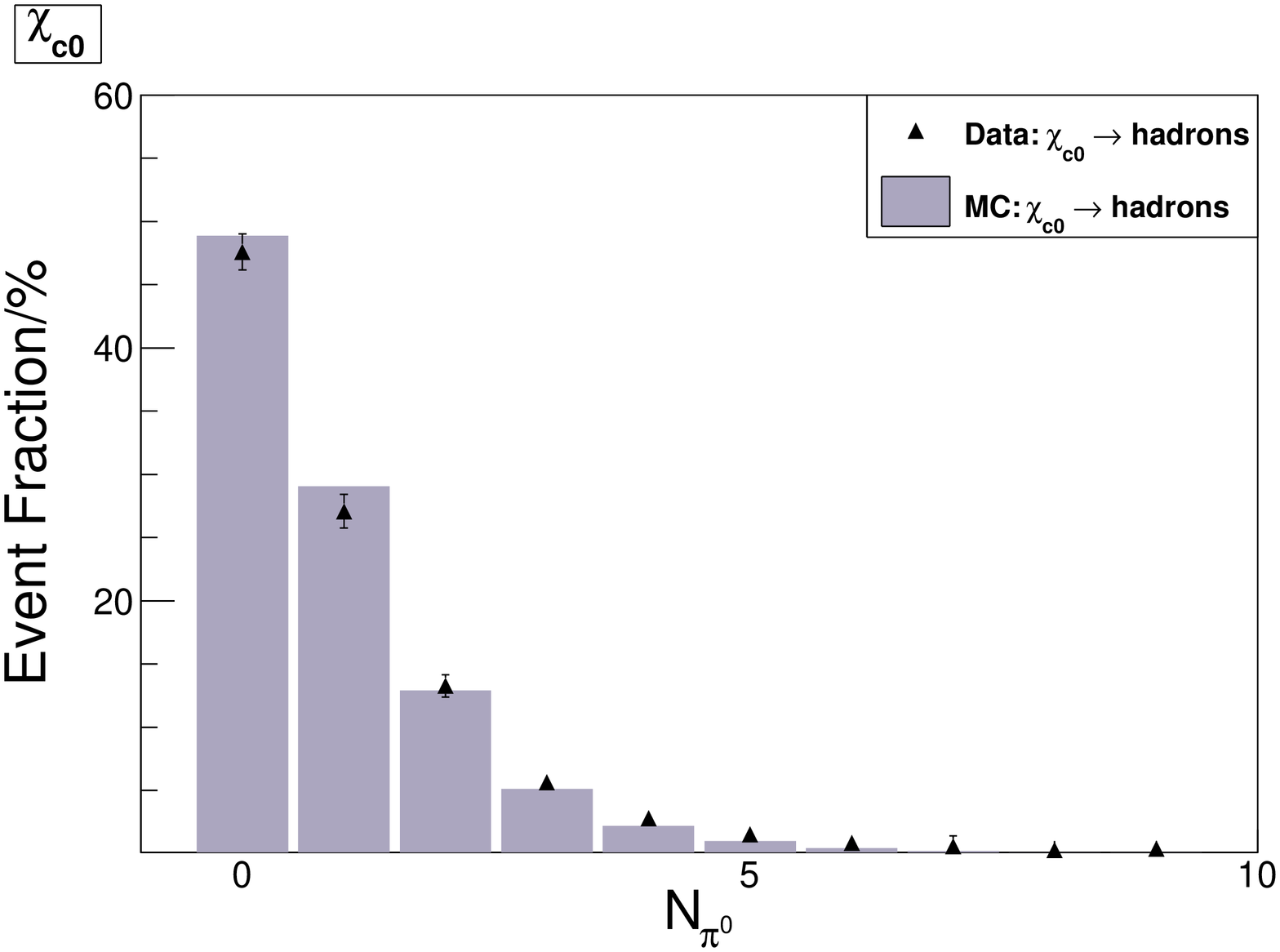}}
\put(-115,-35){\bf \large  {(a)}}
\rotatebox{-90}
{\includegraphics*[width=2.0in]{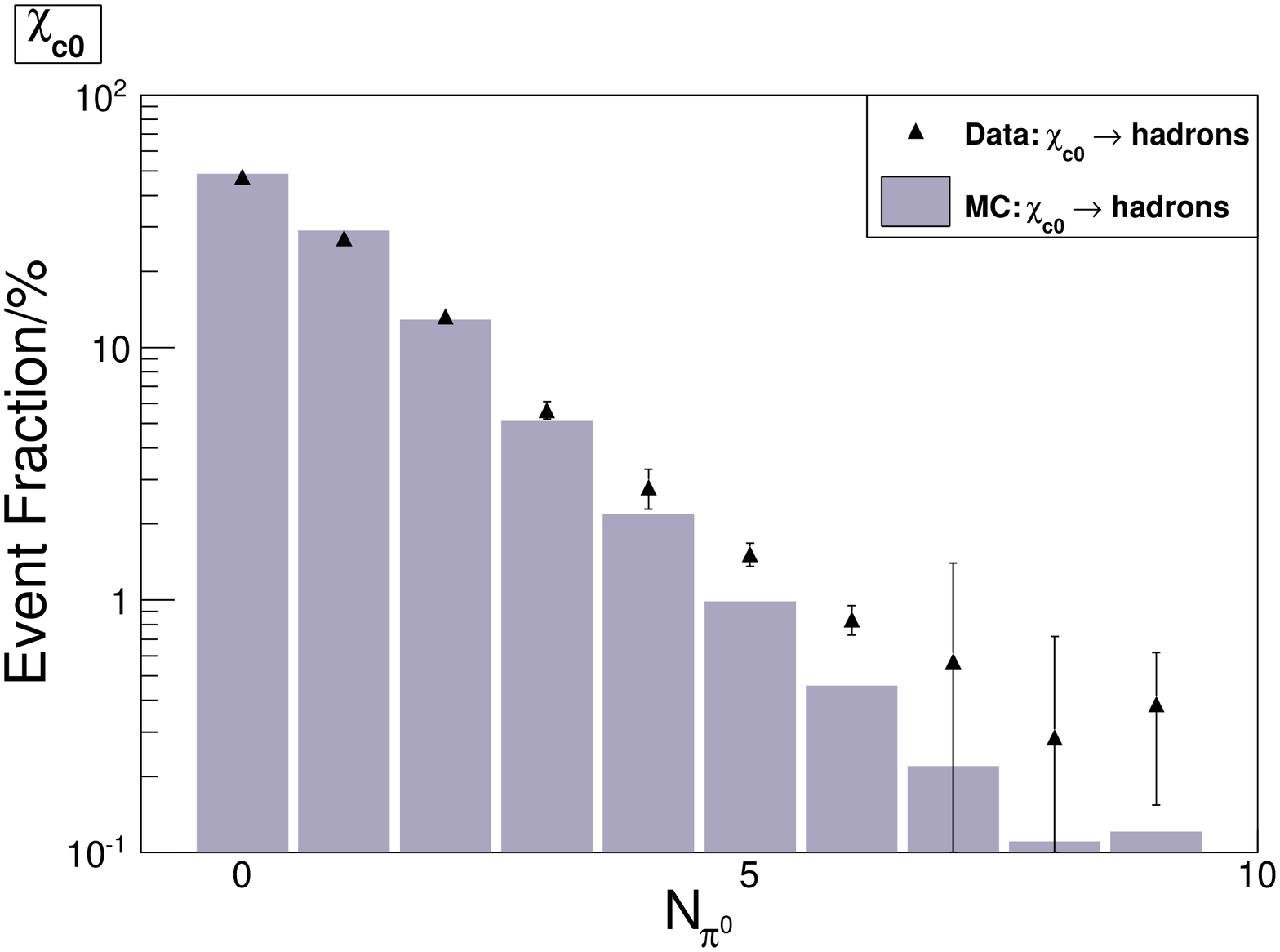}}
\put(-115,-35){\bf \large {(b)}}

\rotatebox{-90}
{\includegraphics*[width=2.0in]{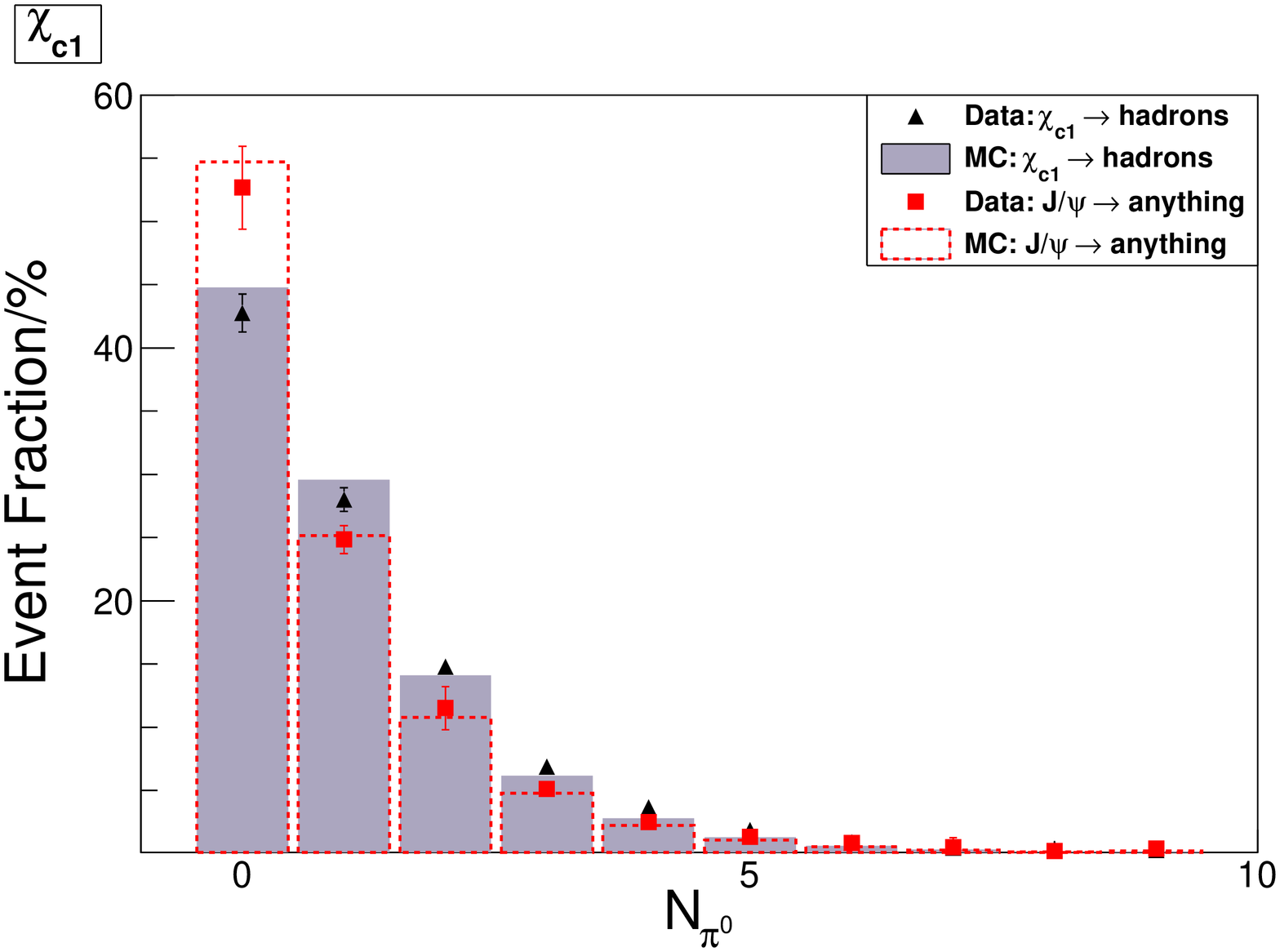}}
\put(-115,-35){\bf \large  {(c)}}
\rotatebox{-90}
{\includegraphics*[width=2.0in]{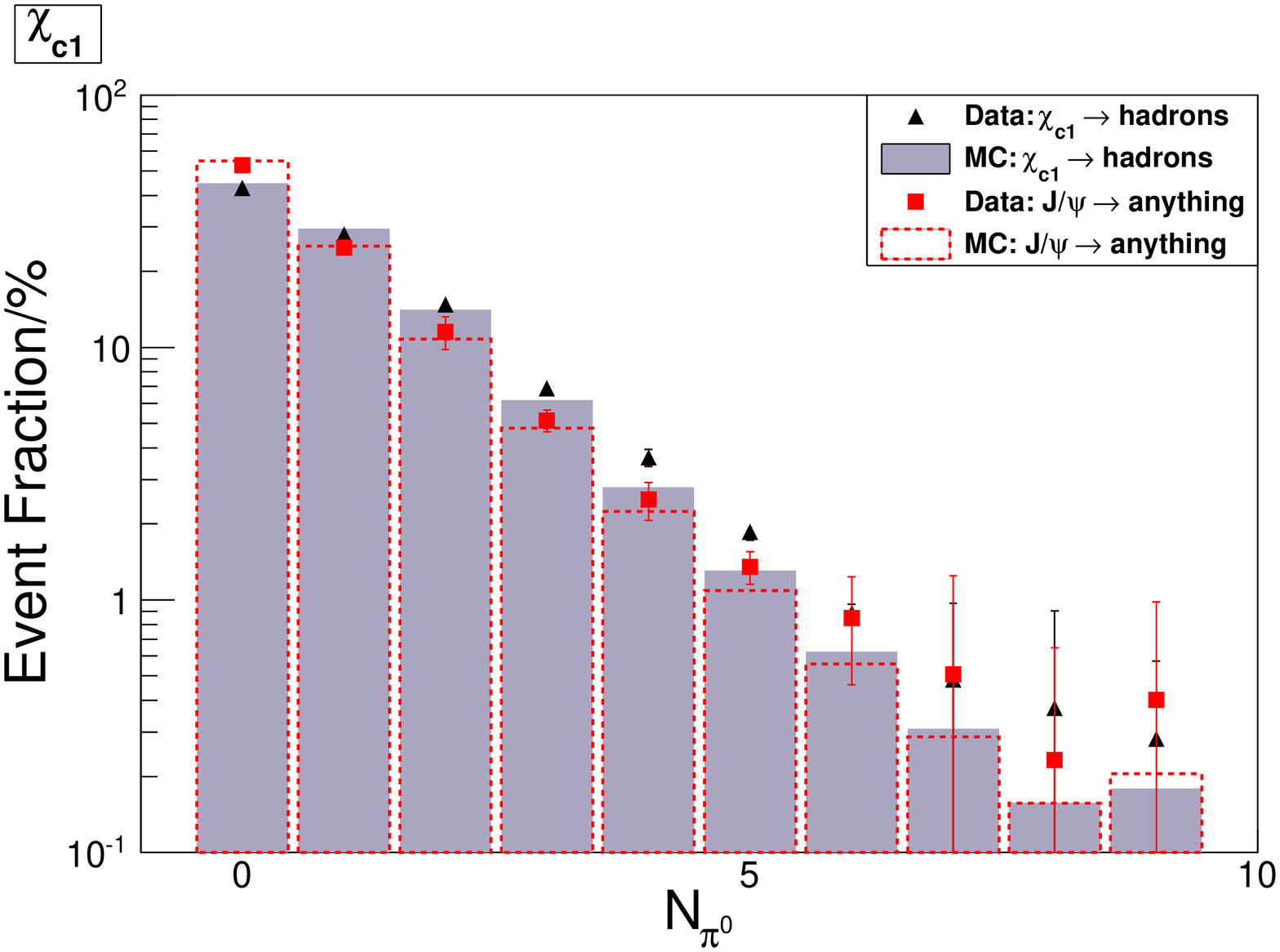}}
\put(-115,-35){\bf \large  {(d)}}

\rotatebox{-90}
{\includegraphics*[width=2.0in]{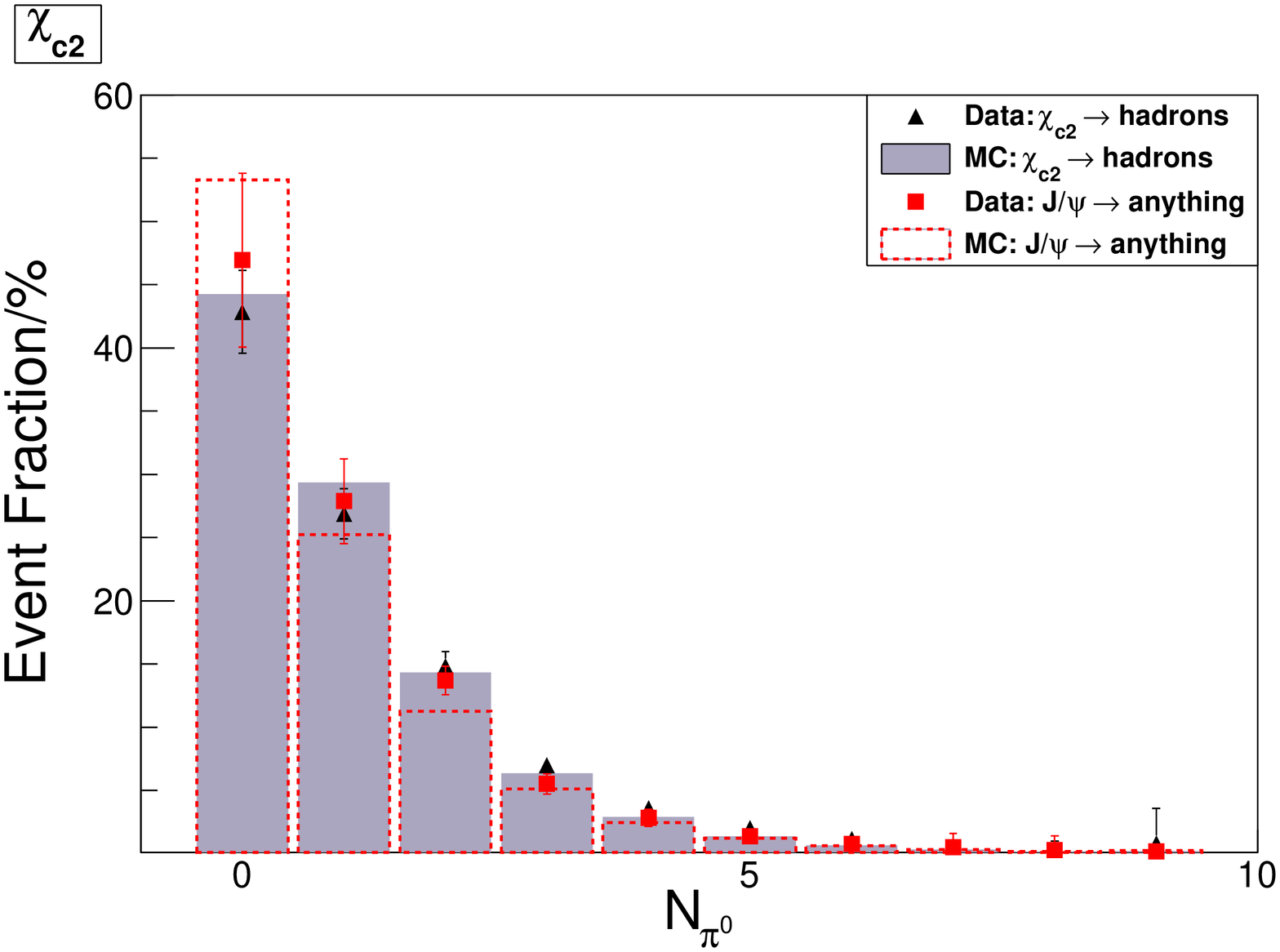}}
\put(-115,-35){\bf \large {(e)}} \rotatebox{-90}
{\includegraphics*[width=2.0in]{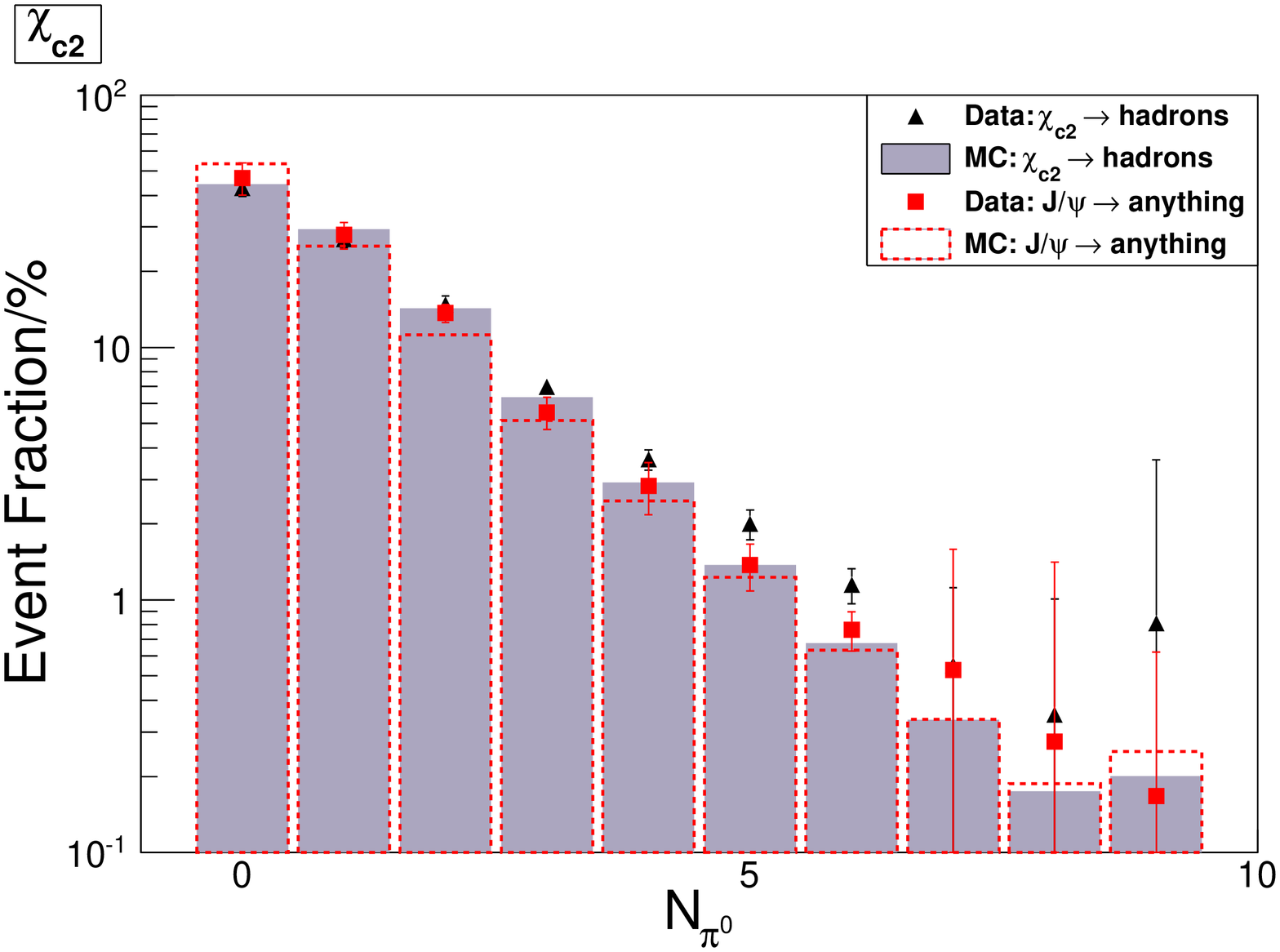}}
\put(-115,-35){\bf \large {(f)}}
\caption{\label{pi0_results2} (Color online) Comparisons of the
  event fractions of data and those for scaled MC simulation events
  versus $N_{\pi^0}$ for (a) $\chi_{c0} \to$ hadrons, (c) $\chi_{c1}
  \to$ hadrons and $\chi_{c1} \to \gamma J/\psi,\:J/\psi \to$
  anything, and (e) $\chi_{c2} \to$ hadrons and $\chi_{c2} \to \gamma
  J/\psi,\:J/\psi \to$ anything, while (b)(d)(f) are the corresponding
  logarithmic plots.  The uncertainties are the uncertainties from the
  fits to the inclusive $E_{\rm sh}$ distribution combined with the
  uncertainty in $R(data)$ added in quadrature with the systematic
  uncertainties, described in Section~\ref{systematics}.}
\end{center}
\end{figure*}


In Figs.~\ref{pi0_results2} (a), (c), and (e) comparisons of the
$N_{\pi^0}$ fractions between data and scaled MC simulated samples are shown, and
Figs.~\ref{pi0_results2} (b), (d), and (f) provide logarithmic
versions.  For $\chi_{cJ} \to$ hadrons, the $N_{\pi^0}$ distribution,
data are above MC simulation for $N_{\pi^0} >2$.  For $J/\psi \to$
anything ($\chi_{c1}$ and $\chi_{c2} \to \gamma J/\psi$), data are
above MC simulation for $N_{\pi^0} > 5$, but the uncertainties are
bigger for these decays.

\section{Produced distributions}
\label{produced_section}
So far, we have only dealt with the distributions of the
efficiency-corrected number of detected charged tracks, EMCSHs{\red ,} or pions. These
depend on the geometry and performance of the BESIII detector.  Of
more interest are the actual physics distributions in the decays of
the $\chi_{cJ}$ and $J/\psi$.

To determine these distributions from data, we construct detection
matrices using the $\chi_{cJ} \to $ hadrons and $\chi_{cJ} \to \gamma
J/\psi,\:J/\psi \to $ anything events in the inclusive $\psi(3686)$ MC
events.  The matrix ($M$) times the produced vector ($P$) determines
the detected vector ($D$), where ($P_i$) is the number of events
with {\red $i$} charged tracks, photons, or $\pi^0$s, etc.

\begin{equation} \left(\begin{array}{c}D_0\\D_1\\ \vdots
    \\D_Q \end{array} \right) = \left(\begin{array}{c}M_{00} \; M_{01}
    \cdots M_{0N} \\ M_{10} \; M_{11} \cdots M_{1N} \\ \vdots
    \\ M_{Q0} \; M_{Q1} \cdots M_{QN} \end{array} \right)
\left( \begin{array}{c}P_0\\P_1\\ \vdots \\P_N \end{array} \right) \label{eq:matrix} \end{equation}

The elements of $M$ are determined using the MC ``truth'' information
by tallying the detected versus the produced track information for
each event.  The detection matrix $M$ is then assumed to apply to
data, as well as to MC simulation.  Detected histograms are
constructed corresponding to each element in the $P$ vector using the
matrix equation {\red (\ref{eq:matrix})}.  These are used to give a
set of probability density functions (PDFs) {\red with
which to perform a $\chi^2$ fit of the detected} distributions of data to
determine the values for $P_0, \ldots, P_N$.

\subsection{\boldmath $P_{N^{\rm P}_{\rm ch}}$ distributions}
\label{nch_prod}
\begin{figure*}[htbp]
\begin{center}

\rotatebox{-90}{
\includegraphics*[width=1.7in]{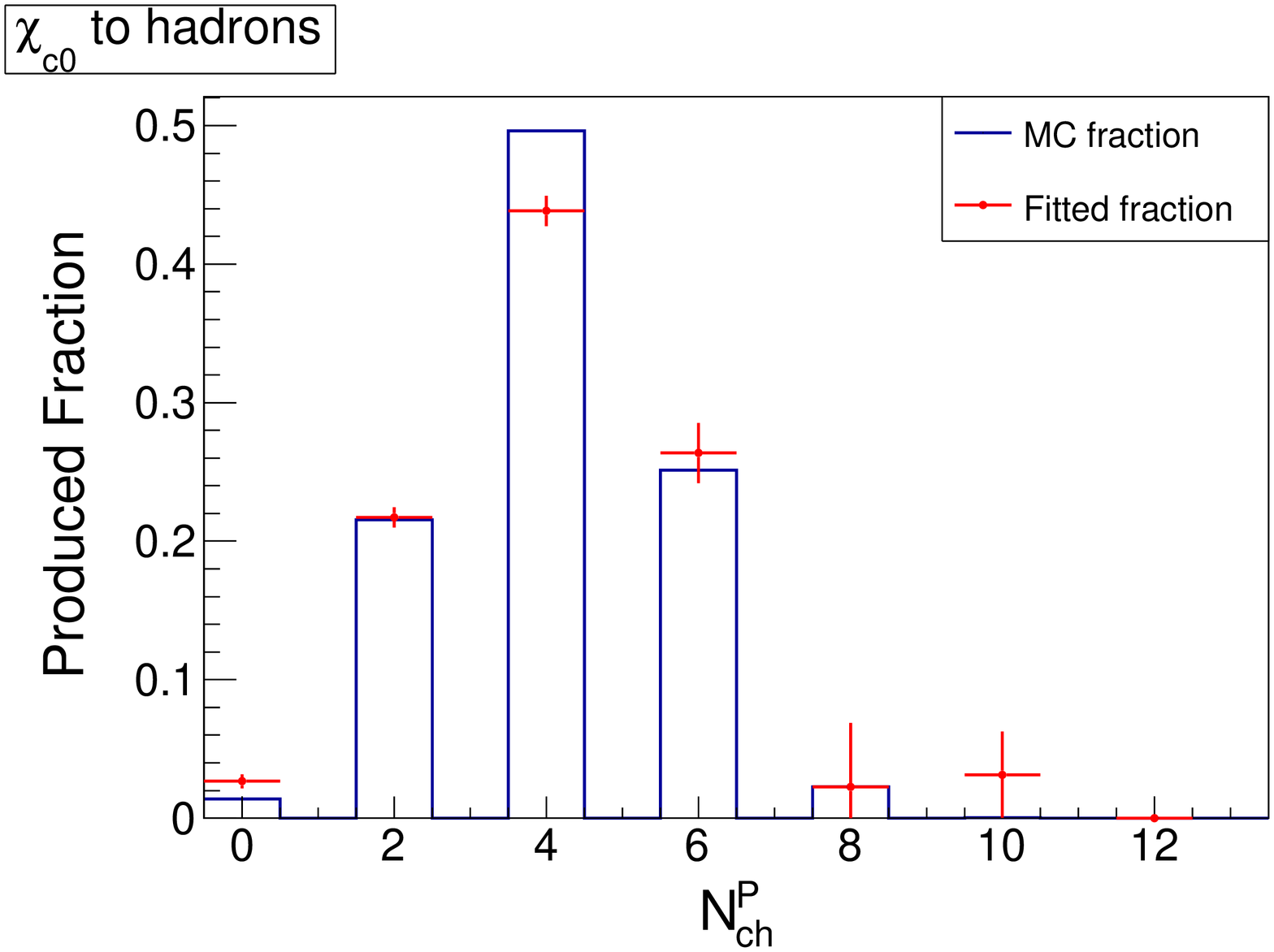}}
\rotatebox{-90}{
\includegraphics*[width=1.7in]{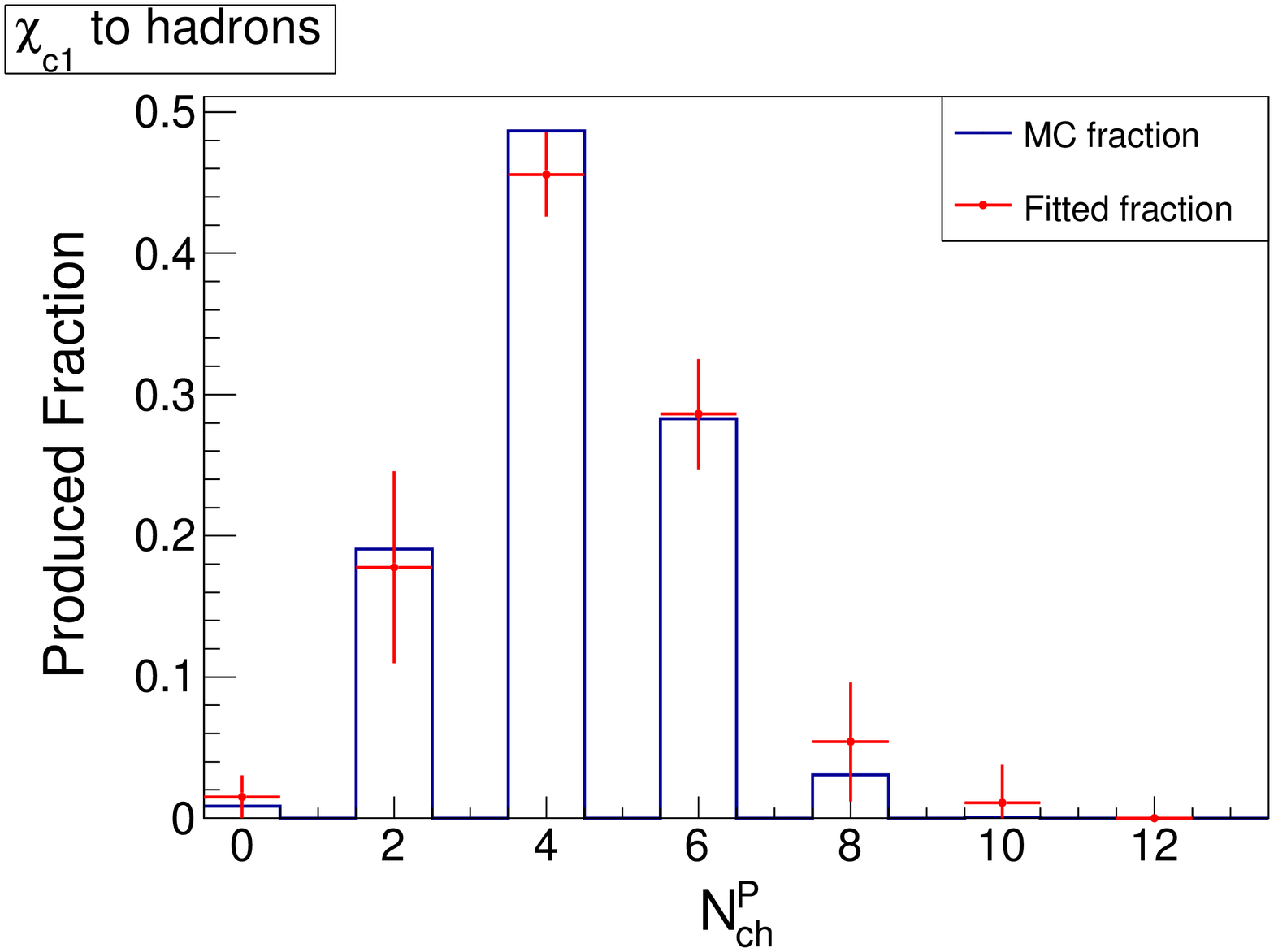}}
\rotatebox{-90}{
\includegraphics*[width=1.7in]{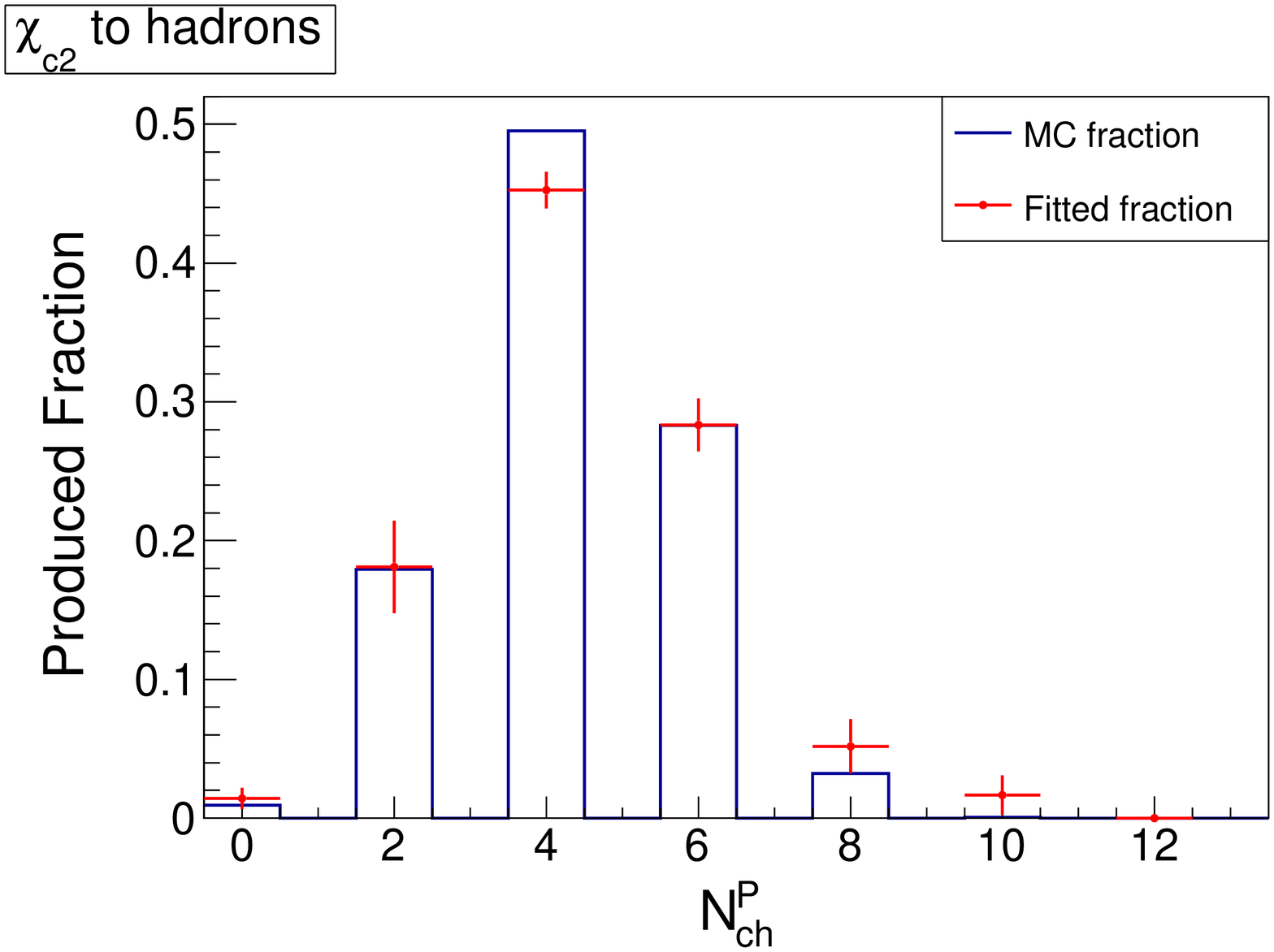}}
\put(-465,-35){\bf \large {(a)}}
\put(-295,-35){\bf \large {(b)}}
\put(-125,-35){\bf \large {(c)}}

\rotatebox{-90}{
\includegraphics*[width=1.7in]{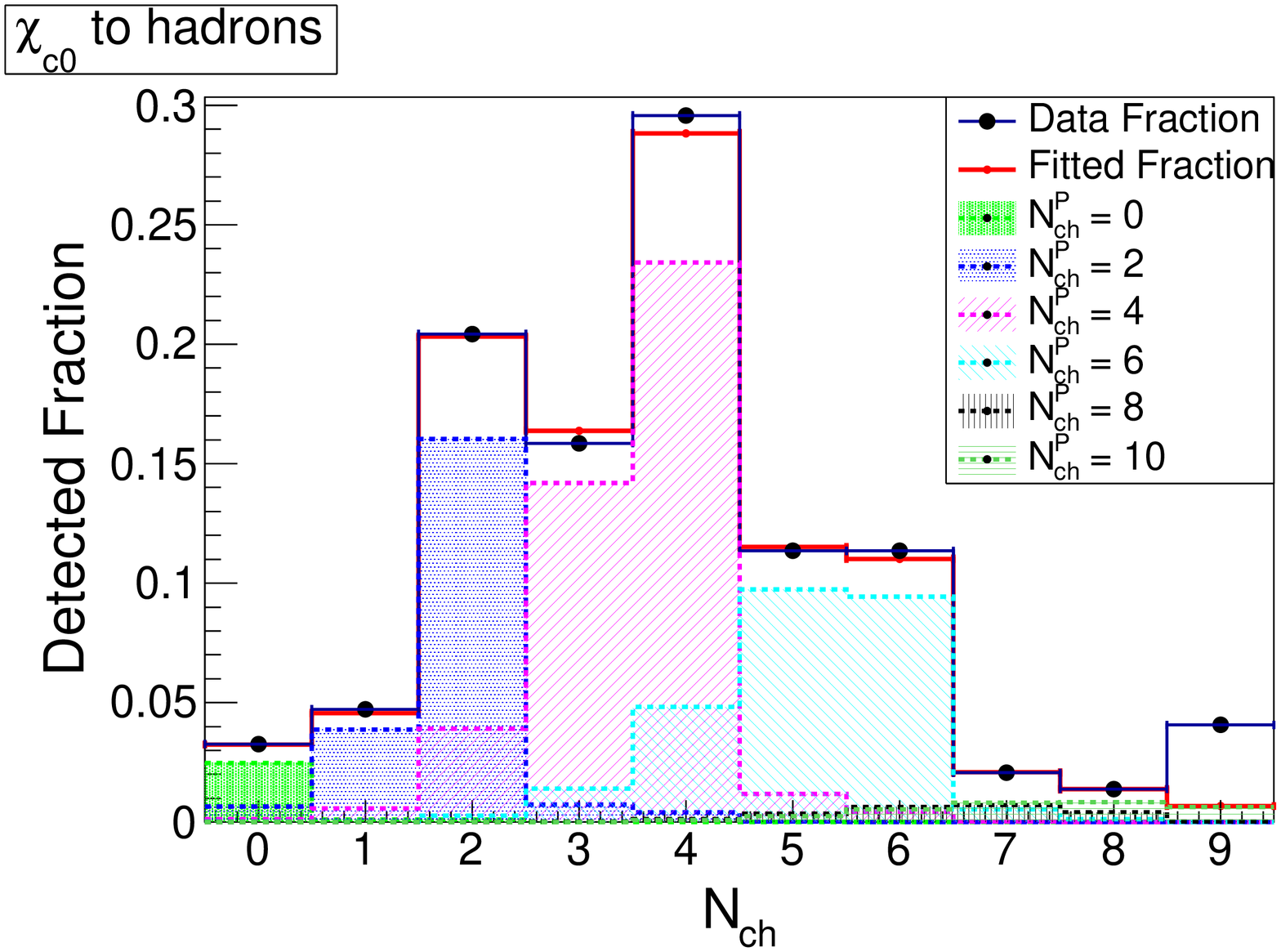}}
\rotatebox{-90}{
\includegraphics*[width=1.7in]{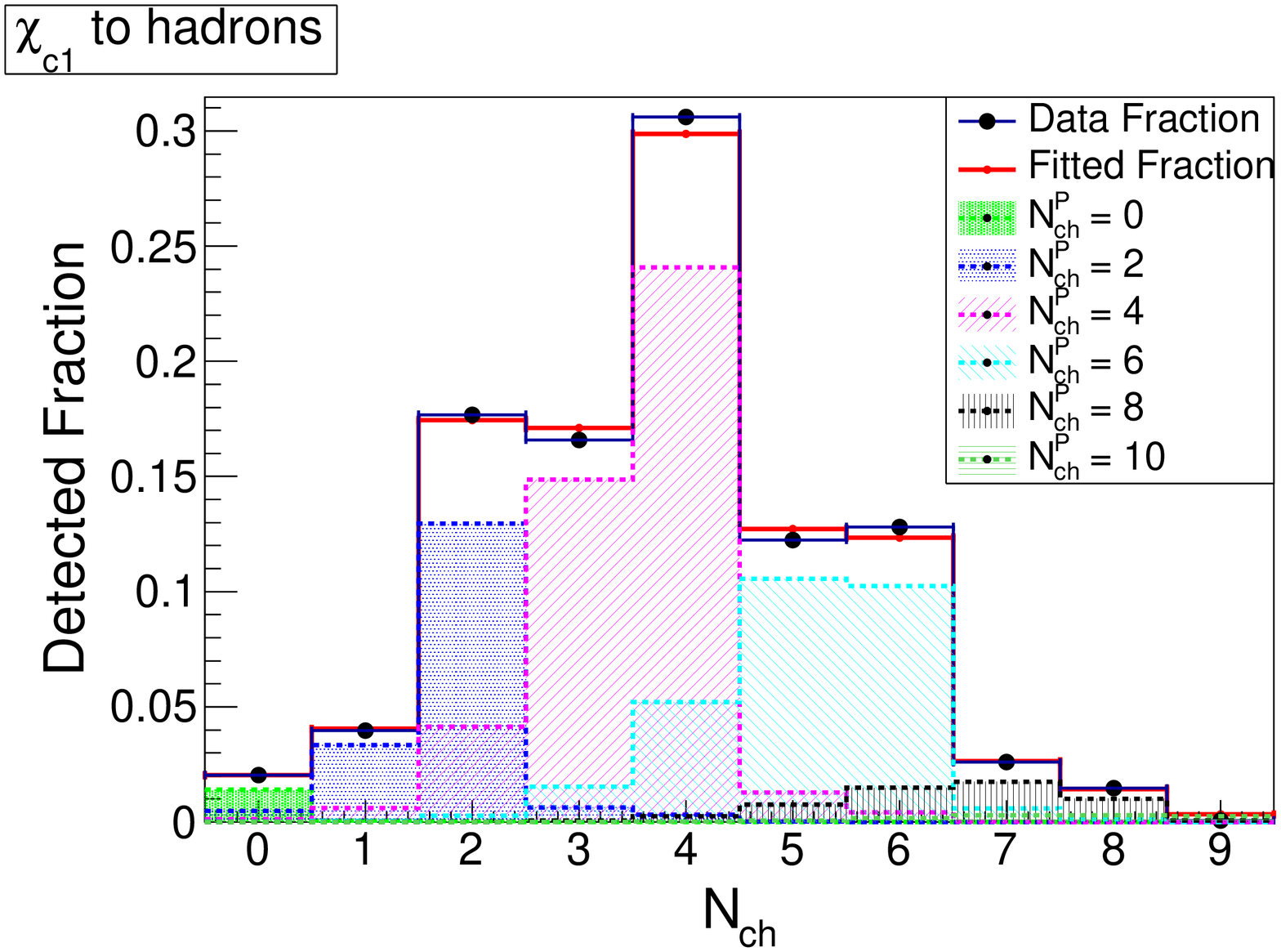}}
\rotatebox{-90}{
  \includegraphics*[width=1.7in]{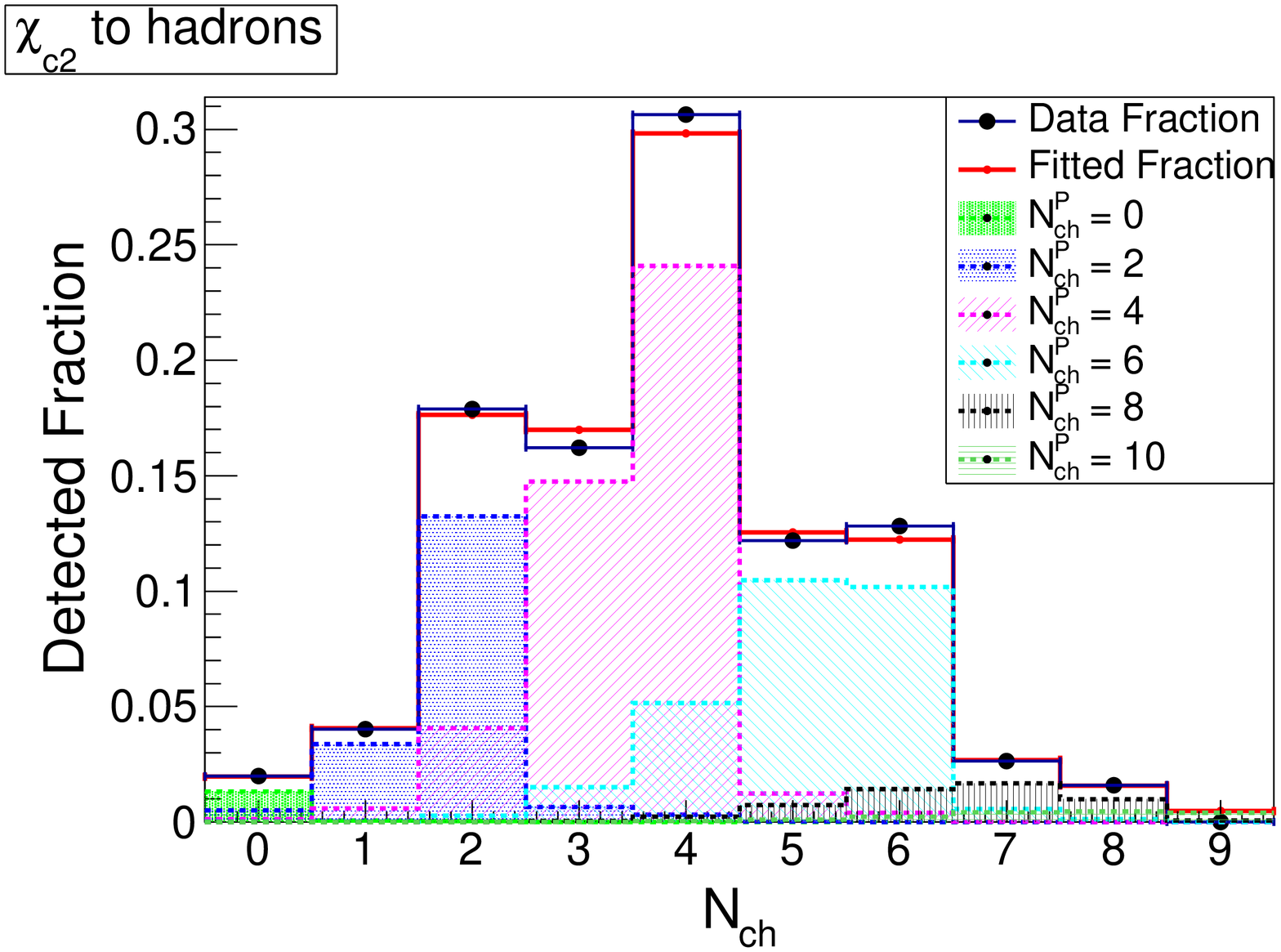}}
\put(-465,-35){\bf \large {(d)}} \put(-295,-35){\bf \large {(e)}}
\put(-125,-35){\bf \large {(f)}}
\caption{\label{prod_ch_1} The distributions are the MC and fitted
  fractions versus $N_{\rm ch}^{\rm P}$ for (a) $\chi_{c0}$, (b) $\chi_{c1}$, and (c) $\chi_{c2}
  \to$ hadrons.  For $N_{\rm ch}^{\rm P} = 12$, the value is fixed to the MC
  result in the fitting. The distributions in (d) - (f) are the
  corresponding detected fractions. Here and in Figs.~\ref{prod_ch_2}
  through \ref{prod_gam_2} below, the produced uncertainties are the
  uncertainties from the fits for $P_{N^{\rm P}_{\rm ch}}$ combined in
  quadrature with the systematic errors, described in
  Section~{\ref{systematics}}. The data uncertainties and the fitted
  fraction uncertainties in (d) - (f) are the uncertainties from the
  fits to the inclusive $E_{\rm sh}$ distributions and the uncertainties
  from the fits for $P_{N^{\rm P}_{\rm ch}}$, respectively.  Also shown in these
  plots are the PDFs used in the fits.  The distribution is fitted
  over bins $N_{\rm ch} = 0 - 8$.}
\end{center}
\end{figure*}

The results of the fits to the detected charged track distributions of
data to determine the produced charged track distributions
$P_{N^{\rm P}_{\rm ch}}$ are shown in Fig.~\ref{prod_ch_1} for $\chi_{cJ} \to$
hadrons. Here $N^{\rm P}_{\rm ch}$ refers to the number of produced tracks.
Shown in Figs.~\ref{prod_ch_1} (a) - (c) are the MC fractions and the
results from the fits to the detected distributions of data.  Charge
conservation requires that $N^{\rm P}_{\rm ch}$ be even. Shown in
Figs.~\ref{prod_ch_1} (d) - (f) are the detected data fractions and
the fractions determined from the fit results, as well as the PDFs
used in the fits.  The distributions in Figs.~\ref{prod_ch_1} (a) -
(c) are similar, and the fit results are below the MC fractions for
$N^{\rm P}_{\rm ch} = 4$ and somewhat above for $N^{\rm P}_{\rm ch} = 0, 8,$ and $10$.

\begin{figure*}[htbp]
\begin{center}
\rotatebox{-90}{
\includegraphics*[width=1.7in]{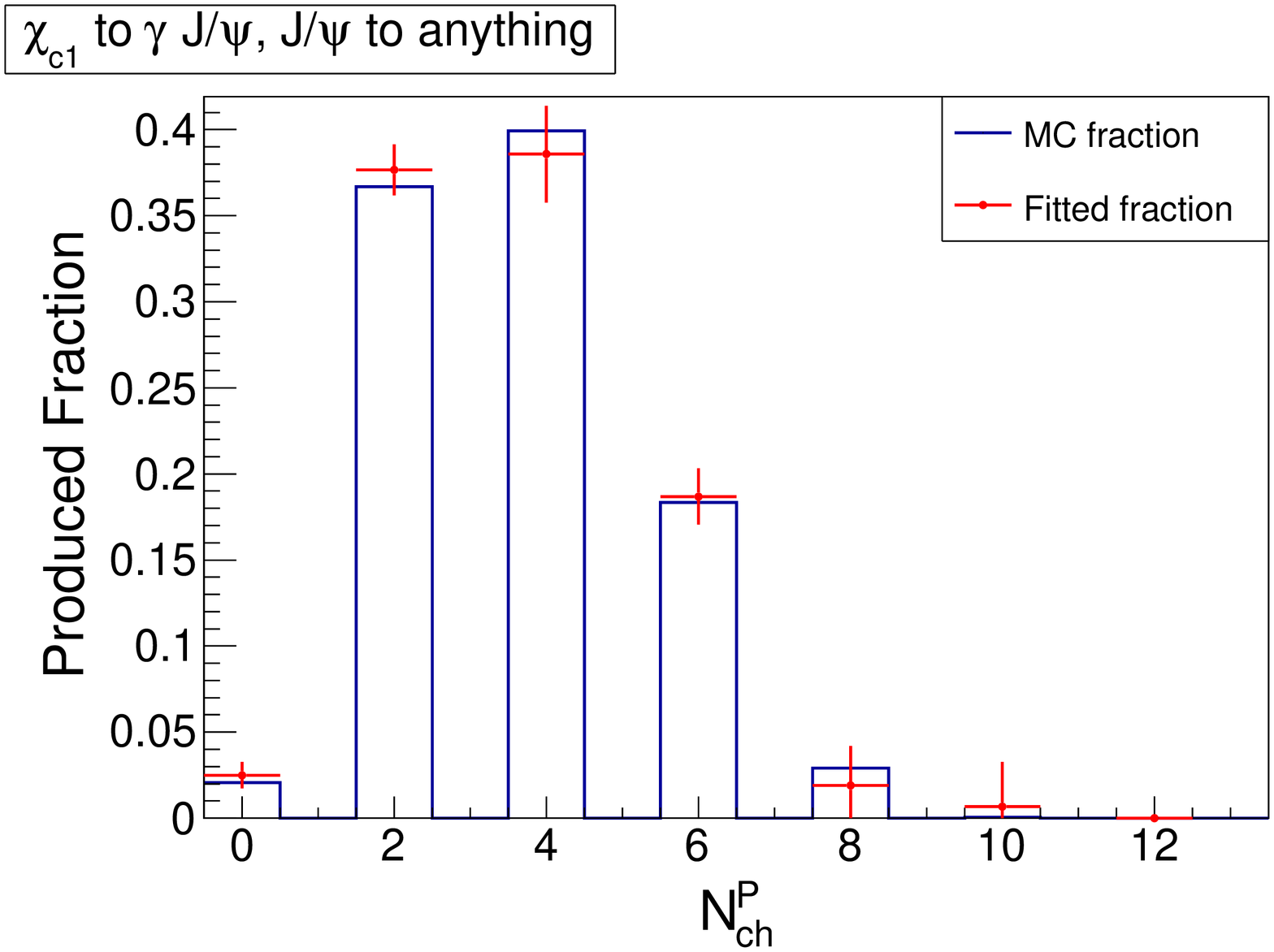}}
\rotatebox{-90}{
\includegraphics*[width=1.7in]{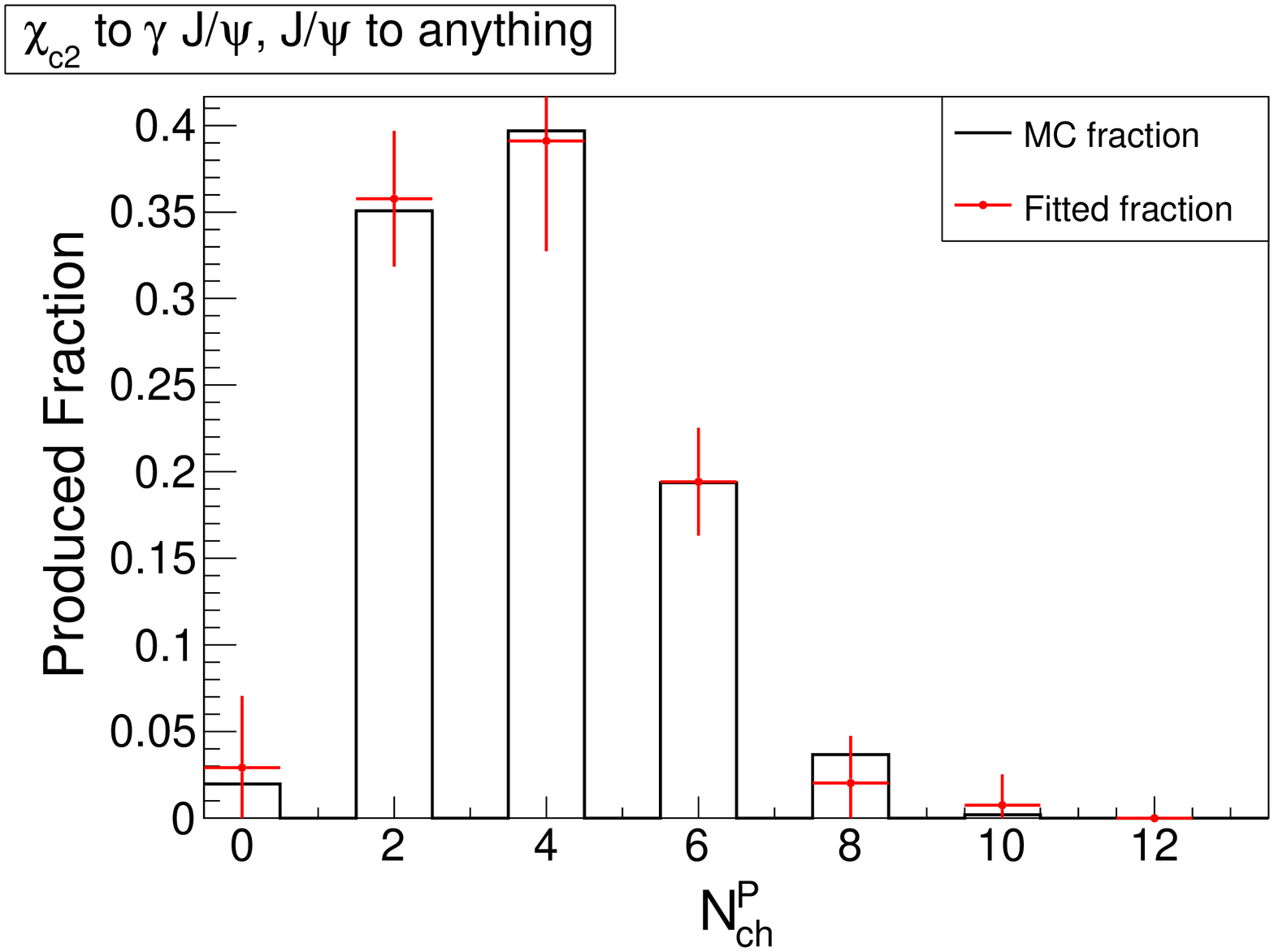}}
\put(-235,-35){\bf \large {(a)}}
\put(-70,-35){\bf \large {(b)}}

\rotatebox{-90}{
\includegraphics*[width=1.7in]{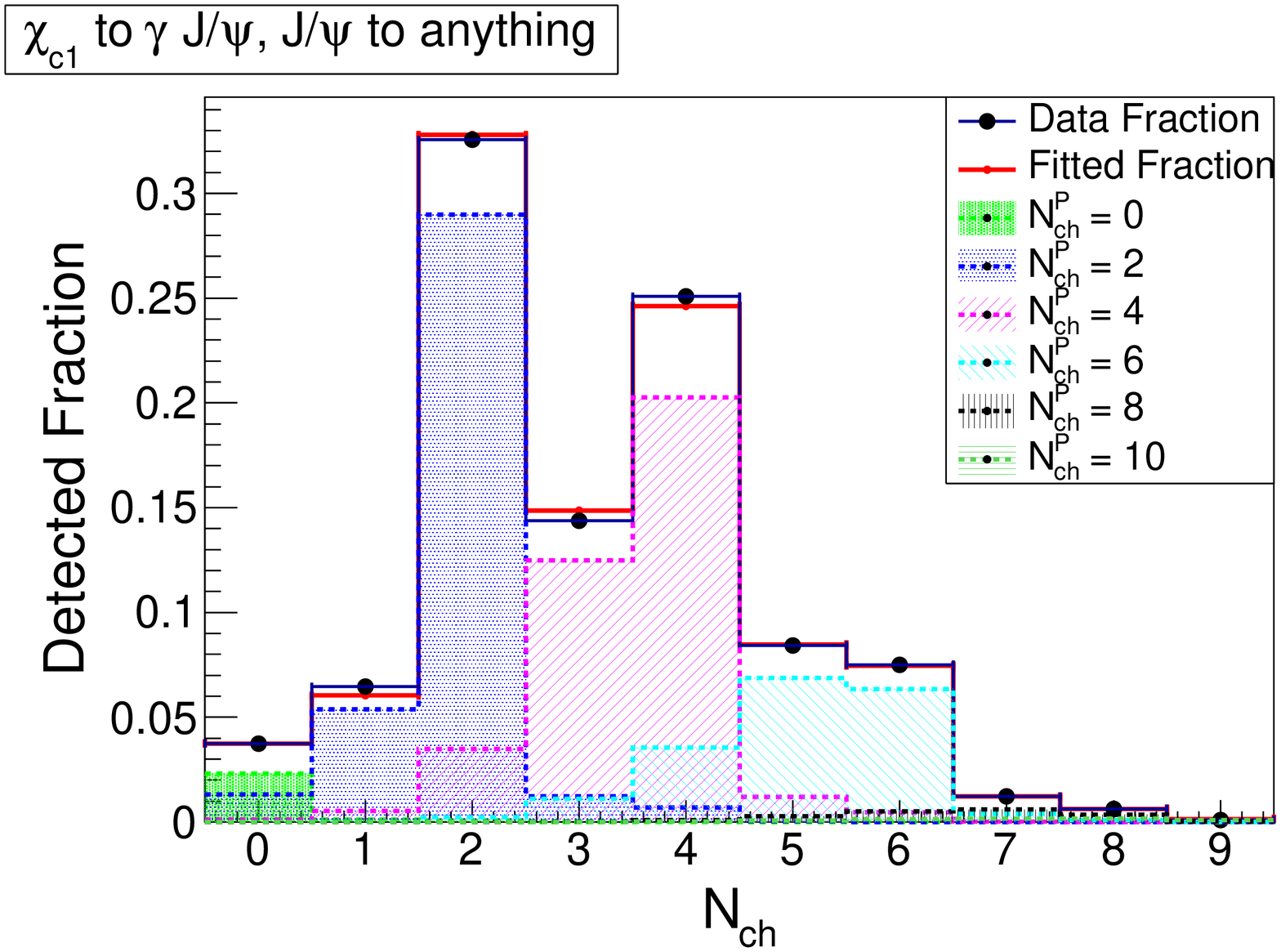}}
\rotatebox{-90}{
  \includegraphics*[width=1.7in]{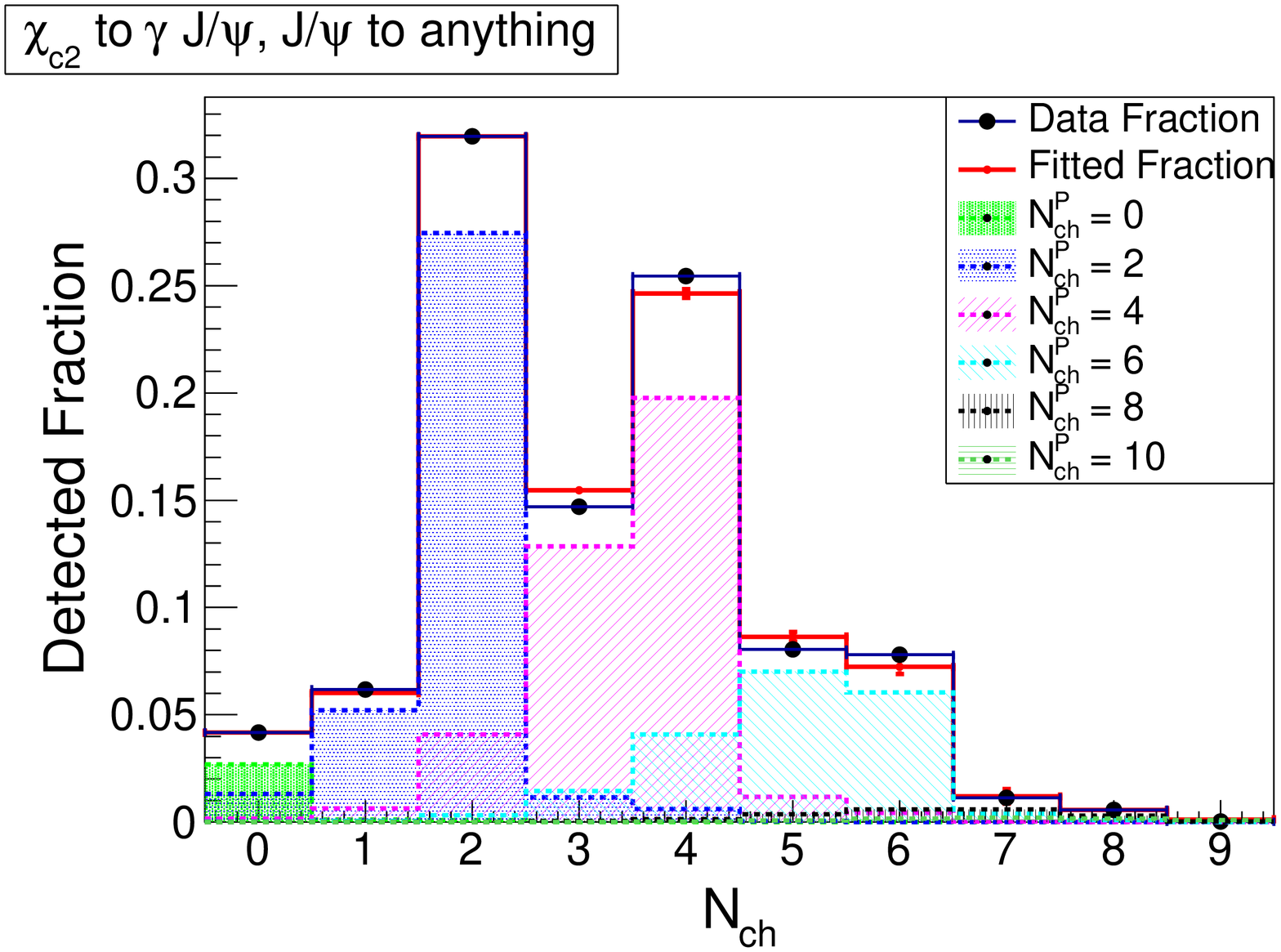}}
\put(-235,-35){\bf \large {(c)}} \put(-70,-35){\bf \large {(d)}}
\caption{\label{prod_ch_2} MC and fitted fraction
  distributions  versus $N_{\rm ch}^{\rm P}$ for (a) $\chi_{c1} \to \gamma J/\psi,\:J/\psi \to$
  anything and (b) $\chi_{c2} \to \gamma J/\psi,\:J/\psi \to$ anything.
  In the fit, the value of the fraction for $N_{\rm ch}^{\rm P} = 12$
  is fixed to the MC result.  The distributions in (c) and (d) are the
  corresponding detected fractions.  Also shown in these plots are the
  PDFs used in the fits.  The distribution is fitted over bins $N_{\rm ch}
  = 0 - 8$.}
\end{center}
\end{figure*}

Results for $\chi_{c1,2} \to \gamma J/\psi,\:J/\psi \to$ anything are
shown in Fig.~\ref{prod_ch_2}.  Shown in Figs.~\ref{prod_ch_2} (a) -
(b) are the MC fractions and the results from the fits to the detected
distributions of data.  Shown in Figs.~\ref{prod_ch_2} (c) - (d) are
the detected data fractions and the fit results, as well as the PDFs
used in the fits.  The distributions in Figs.~\ref{prod_ch_2} (a) -
(b) are similar, and the fitted fractions are in reasonable agreement
with the MC fractions.

In Figs.~\ref{prod_ch_1} (d) - (f) and Figs.~\ref{prod_ch_2} (c) -
(d), 9 bins of detected data are fitted with 6 parameters ($P_0$
through $P_{10}$) and with $P_{12}$ fixed to the MC values.  Fractions
$F^{\rm P}$ of $\chi_{cJ} \to$ hadrons and $\chi_{c1/2} \to
J/\psi,\:J/\psi \to$ anything are listed in Tables~\ref{prod-ch} and
~\ref{prod-cha}, respectively.  The $\chi^2/ndf$ values for the five
cases are 65, 52, 85, 18, and 28. Alternative fits with $P_{12}$ free
give the same results as shown in Tables~\ref{prod-ch} and
\ref{prod-cha}.  Comparing the fits and the PDFs in
Figs.~\ref{prod_ch_1} and \ref{prod_ch_2} suggests that the MC PDFs do
not describe data well, which contributes to the large $\chi^2/ndf$.
{\blue However, corrections to the PDFs to improve the fits to the
  detected charged track distributions, as described in
  Section~{\ref{systematics}}, result in small changes to the
  $P_{N^{\rm P}_{\rm ch}}$ values compared with the systematic
  uncertainties shown in Table~\ref{systematics-prod} and are
  neglected.}


\begin{table*}[htb]
\begin{center}
  \caption{$P_{N^{\rm P}_{\rm ch}}$ event fractions in \% for data
    $F_{\chi_{cJ}}^{\rm P}$ and MC simulated sample $F^{\rm MC}_{\chi_{cJ}}$ for
    $\chi_{cJ} \to$ hadrons. In the fit, the value of the fraction for
    $N_{\rm ch}^{\rm P} = 12$ is fixed to the MC result.  Here and below, the
    first uncertainties shown are the uncertainties from the fits for
    $P_{N^{\rm P}_{\rm ch}}$ and the second are the systematic uncertainties
    described in Section~{\ref{systematics}}.
\label{prod-ch}}
\begin{footnotesize}
\begin{tabular}{r|cr|cr|cr} \hline
\T $N^{\rm P}_{\rm ch}$ & $F^{\rm P}_{\chi_{c0}}$ & $F^{\rm MC}_{\chi_{c0}}$ & $F^{\rm P}_{\chi_{c1}}$ & $F^{\rm MC}_{\chi_{c1}}$ &  $F^{\rm P}_{\chi_{c2}}$ & $F^{\rm MC}_{\chi_{c2}}$  \B \\ \hline
0  & $2.67\pm0.04\pm0.49$  & 1.41  & $1.51\pm0.06\pm1.50$  &  0.86 & $1.43\pm0.06\pm0.76$  & 0.94  \\
2  & $21.72\pm0.08\pm0.78$ & 21.55 & $17.77\pm0.17\pm6.80$ & 19.04 & $18.11\pm0.11\pm3.57$ & 17.92 \\
4  & $43.84\pm0.11\pm1.11$ & 49.61 & $45.57\pm0.31\pm2.99$ & 48.67 & $45.26\pm0.14\pm1.37$ & 49.53 \\
6  & $26.36\pm0.13\pm2.17$ & 25.11 & $28.61\pm0.32\pm3.92$ & 28.30 & $28.34\pm0.16\pm2.27$ & 28.31 \\
8  & $2.26\pm0.27\pm4.66$  & 2.27  & $5.41\pm0.34\pm4.19$  &  3.07 & $ 5.19\pm0.29\pm1.94$ & 3.23  \\
10  & $3.14\pm0.24\pm3.11$ & 0.05  & $1.11\pm0.35\pm2.65$  &  0.07 & $ 1.67\pm0.26\pm1.43$ & 0.08 \\
12  & $ 0.00\pm0.00\pm0.00 $   & 0  & $0.00\pm0.0\pm0.00$  &  0    & $ 0.00\pm0.00\pm0.00$ &  0   \\
\hline
\end{tabular}
\end{footnotesize}
\end{center}
\end{table*}


\begin{table}[htb]
\begin{center}
  \caption{$P_{N^{\rm P}_{\rm ch}}$ event fractions in \% for data
    $F^{\rm P}_{J/\psi_1}$$(F^{\rm P}_{J/\psi_2})$ and MC simulated sample
    $F^{\rm MC}_{J/\psi_1}$$(F^{\rm MC}_{J/\psi_2})$ for $\chi_{c1}\to \gamma
    J/\psi,\:J/\psi \to$ anything ($\chi_{c2}\to \gamma
    J/\psi,\:J/\psi \to$ anything). In the fit, the value for $N_{\rm ch}^{\rm P}
    = 12$ is fixed to the MC result.\label{prod-cha}}
\begin{footnotesize}
\begin{tabular}{r|cr|cr} \hline
\T $N^{\rm P}_{\rm ch}$ & $F^{\rm P}_{J/\psi_1}$ & $F^{\rm MC}_{J/\psi_1}$ &  $F^{\rm P}_{J/\psi_2}$ & $F^{\rm MC}_{J/\psi_2}$ \B \\ \hline
0  & $2.50\pm0.09\pm0.81$  & 2.07  & $2.91\pm0.14\pm4.14$ & 1.99 \\
2  & $37.65\pm0.18\pm1.57$ & 36.68 & $35.78\pm0.25\pm3.91$ & 35.08\\
4  & $38.58\pm0.20\pm2.81$ & 39.92 & $39.10\pm0.31\pm6.34$ & 39.68\\
6  & $18.69\pm0.18\pm1.64$ & 18.35 & $19.43\pm0.37\pm3.09$ & 19.37\\
8  & $1.90\pm0.41\pm2.35$  & 2.91  & $ 2.04\pm0.89\pm2.57$ & 3.67\\
10  & $0.69\pm0.41\pm2.64$  & 0.06  & $0.74\pm0.76\pm1.62$ & 0.21\\
12  & $0.00\pm0.00\pm0.00$  & 0     & $0.00\pm0.00\pm0.00$ & 0 \\
\hline
\end{tabular}
\end{footnotesize}
\end{center}
\end{table}

\vspace*{0.1cm}
\paragraph*{Mean charged multiplicity and dispersion \\}

\begin{figure}[hbt]
\begin{center}
\rotatebox{-90}{
  \includegraphics*[width=2.35in]{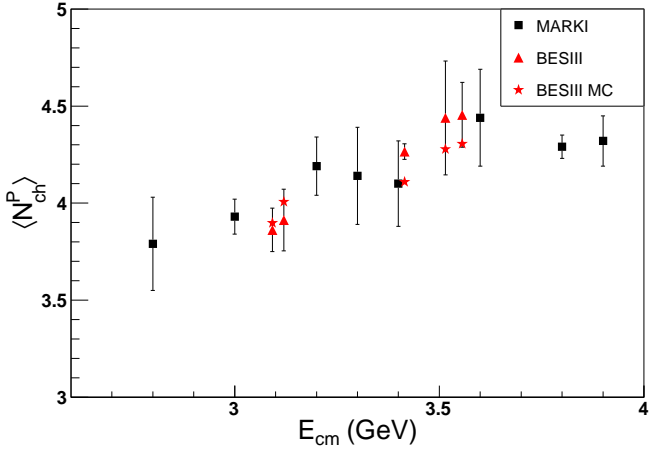}}
\caption{\label{means} Plot of $\langle
  N^{\rm P}_{\rm ch}\rangle$ versus center-of-mass energy for MARK~I $e^+ e^-
  \to$ hadrons and BESIII $J/\psi$ and $\chi_{cJ}$ to hadrons. While
  BESIII results include systematic uncertainties, MARK~I results do not.
  The two results for $J/\psi \to$ hadrons have been offset in $E_\mathrm{cm}$ for
  visualization purposes.  Also shown are the values for the BESIII MC.}
\end{center}
\end{figure}

\begin{table*}[hbt]
\begin{center}
 \caption{Mean charged multiplicity $\langle N^{\rm P}_{\rm ch}\rangle$,
    dispersion $D=\sqrt{ \langle [N^{\rm P}_{\rm ch}]^2\rangle - \langle
      N^{\rm P}_{\rm ch}\rangle^2}$ , and $\langle N^{\rm P}_{\rm ch}\rangle/D$ for
    $\chi_{cJ}$ and $J/\psi$ to hadrons.\label{mean}}
\begin{footnotesize}
\begin{tabular}{l|c|c|c|c} \hline
   \T                    & $E_{\rm cm}$ (GeV)     & $\langle N^{\rm P}_{\rm ch}\rangle$   & $D$ & $\langle N^{\rm P}_{\rm ch}\rangle/D$ \B  \\ \hline
$\chi_{c0} \to$ hadrons & $3.415$  & $4.265\pm0.007\pm0.043$ & $1.942\pm0.012\pm0.133$ & $2.196\pm0.026\pm0.049$ \\
$\chi_{c1} \to$ hadrons & $3.511$  & $4.439\pm0.031\pm0.293$ & $1.781\pm0.038\pm0.179$ & $2.493\pm0.096\pm0.335$ \\
$\chi_{c2} \to$ hadrons & $3.556$  & $4.455\pm0.008\pm0.170$ & $1.820\pm0.013\pm0.085$ & $2.449\pm0.032\pm0.194$ \\
$\chi_{c1} \to \gamma J/\psi,\:J/\psi \to$ hadrons & $3.097$  & $3.862\pm0.014\pm0.113$ & $1.754\pm0.030\pm0.199$ & $2.201\pm0.067\pm0.128$ \\
$\chi_{c2} \to \gamma J/\psi,\:J/\psi \to$ hadrons & $3.097$  & $3.913\pm0.022\pm0.160$ & $1.779\pm0.050\pm0.223$ & $2.200\pm0.110\pm0.186$ \\
\hline
\end{tabular}
\end{footnotesize}
\end{center}
\end{table*}

We determine values of the mean multiplicity $\langle N^{\rm P}_{\rm ch}\rangle$, 
dispersion  $D=\sqrt{\langle [N^{\rm P}_{\rm ch}]^2\rangle - \langle
  N^{\rm P}_{\rm ch}\rangle^2}$, and $\langle N^{\rm P}_{\rm ch}\rangle/D$.
Such measurements have been performed for $e^+e^-
\to$ hadrons at LEP~\cite{opal}, and also at lower energies with the MARK I experiment~\cite{marki}.
  The results of these measurements from our data  are
listed in Table~\ref{mean}.  Although we measure $J/\psi \to$ anything
via $\chi_{cJ} \to \gamma J/\psi,\:J/\psi \to$ anything, we can
calculate the $J/\psi \to$ hadron distribution using the branching
fractions of $J/\psi \to e^+ e^-$ and $\mu^+ \mu^-$~\cite{PDG} and
assuming that these events populate $N^{\rm P}_{\rm ch} = 2$ only.  The
calculated values are also listed in Table~\ref{mean}.

Our values for $\langle N^{\rm P}_{\rm ch}\rangle$ can be compared with those of
MARK I for $e^+e^- \to$ hadrons~\cite{marki}.  The MARK I values from
{\red 2.8} to 4.0 GeV are plotted in Fig.~\ref{means} along with our
values for both $J/\psi$ and $\chi_{cJ}$ to hadrons.  While our
results include statistical and systematic uncertainties, those of
MARK I do not include systematic uncertainties, which range from 25\%
at 2.6 GeV to 15\% at 6 GeV and above.  Still, the agreement between
the results is very good.

\subsection{\boldmath $P_{N^{\rm P}_{\gamma}}$ distributions}
\label{ngam_prod}
\begin{figure*}[htbp]
\begin{center}
\centering
\rotatebox{-90}{
\includegraphics*[width=1.67in]{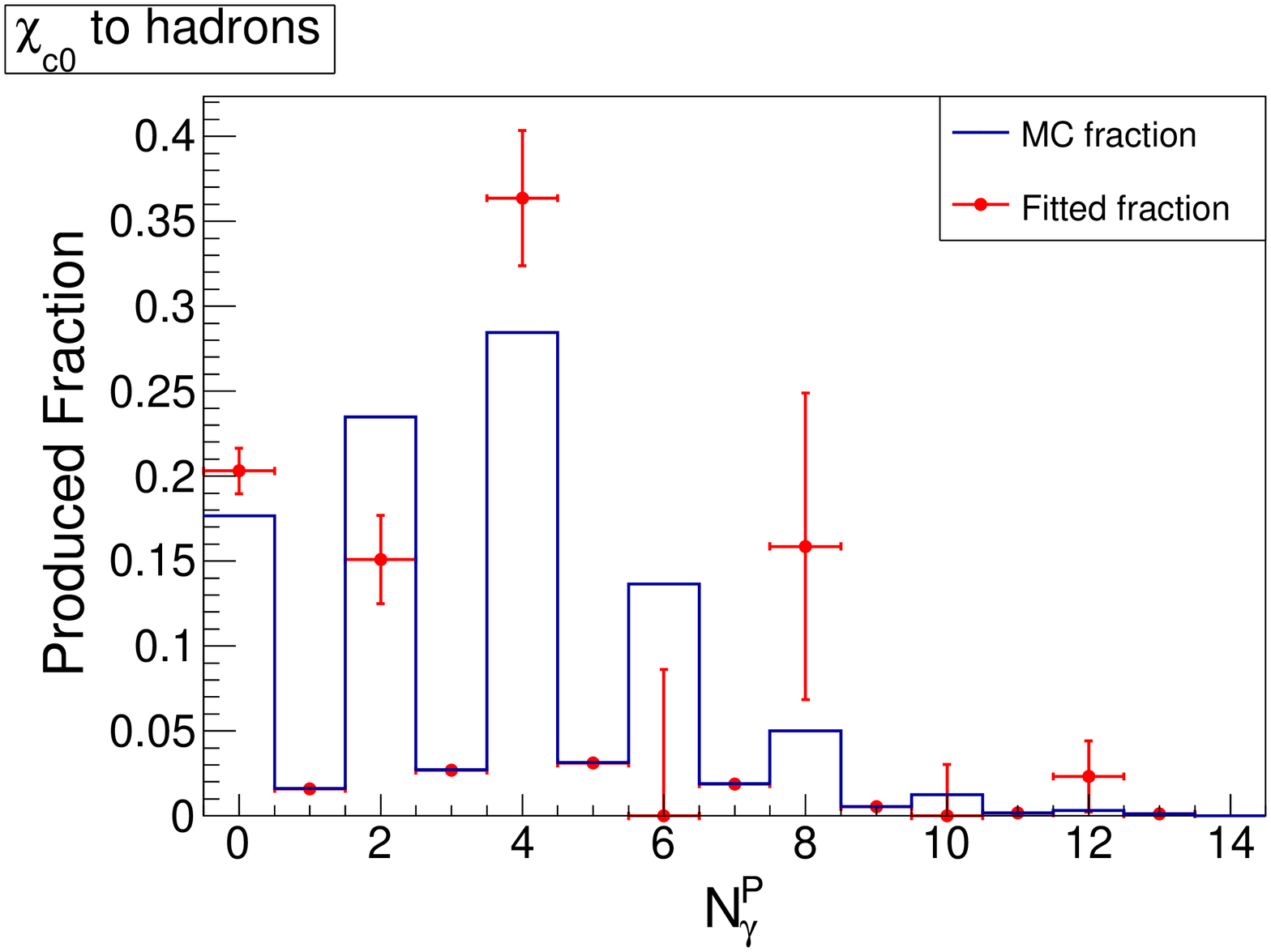}}
\rotatebox{-90}{
\includegraphics*[width=1.67in]{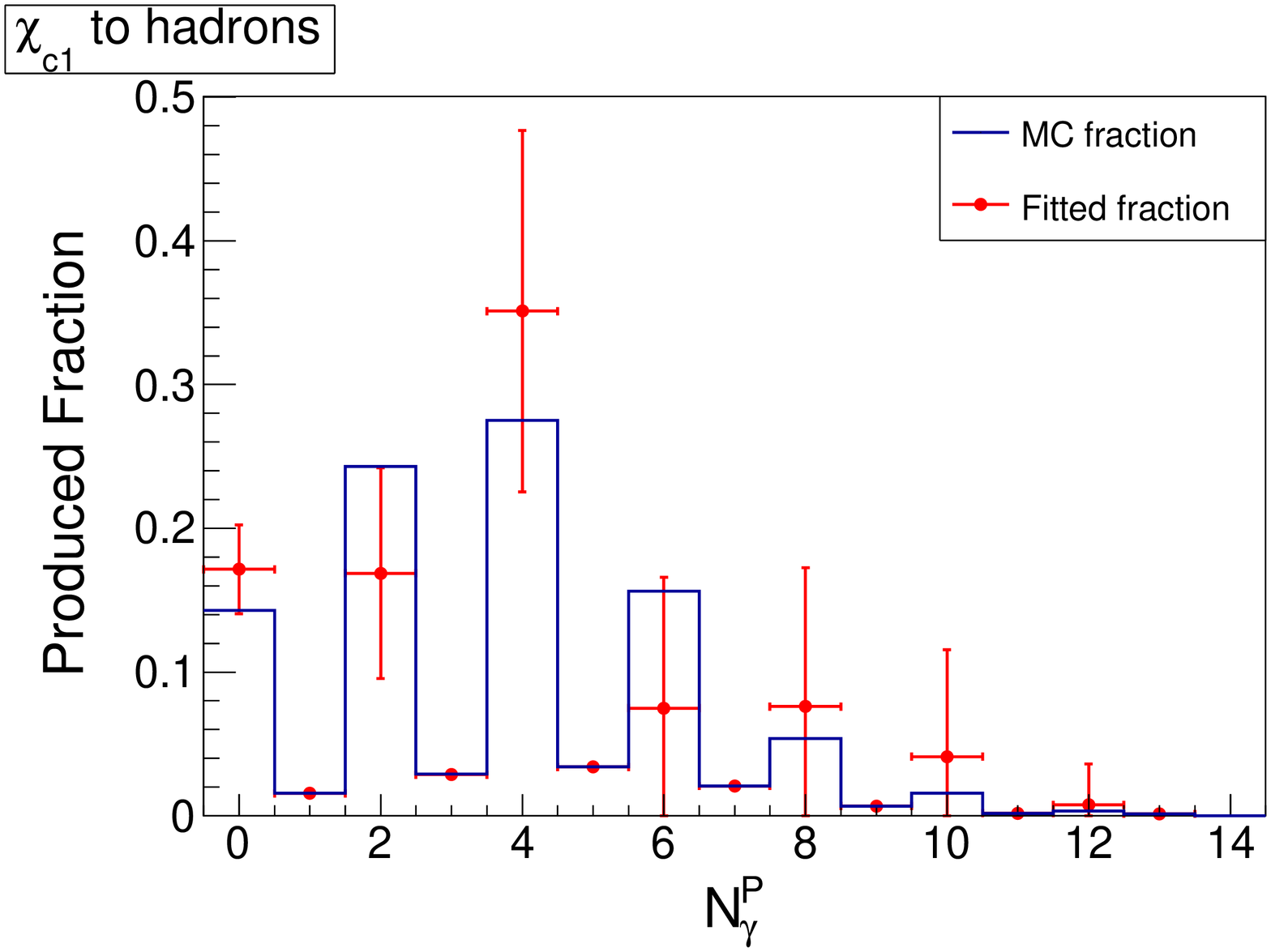}}
\rotatebox{-90}{
\includegraphics*[width=1.67in]{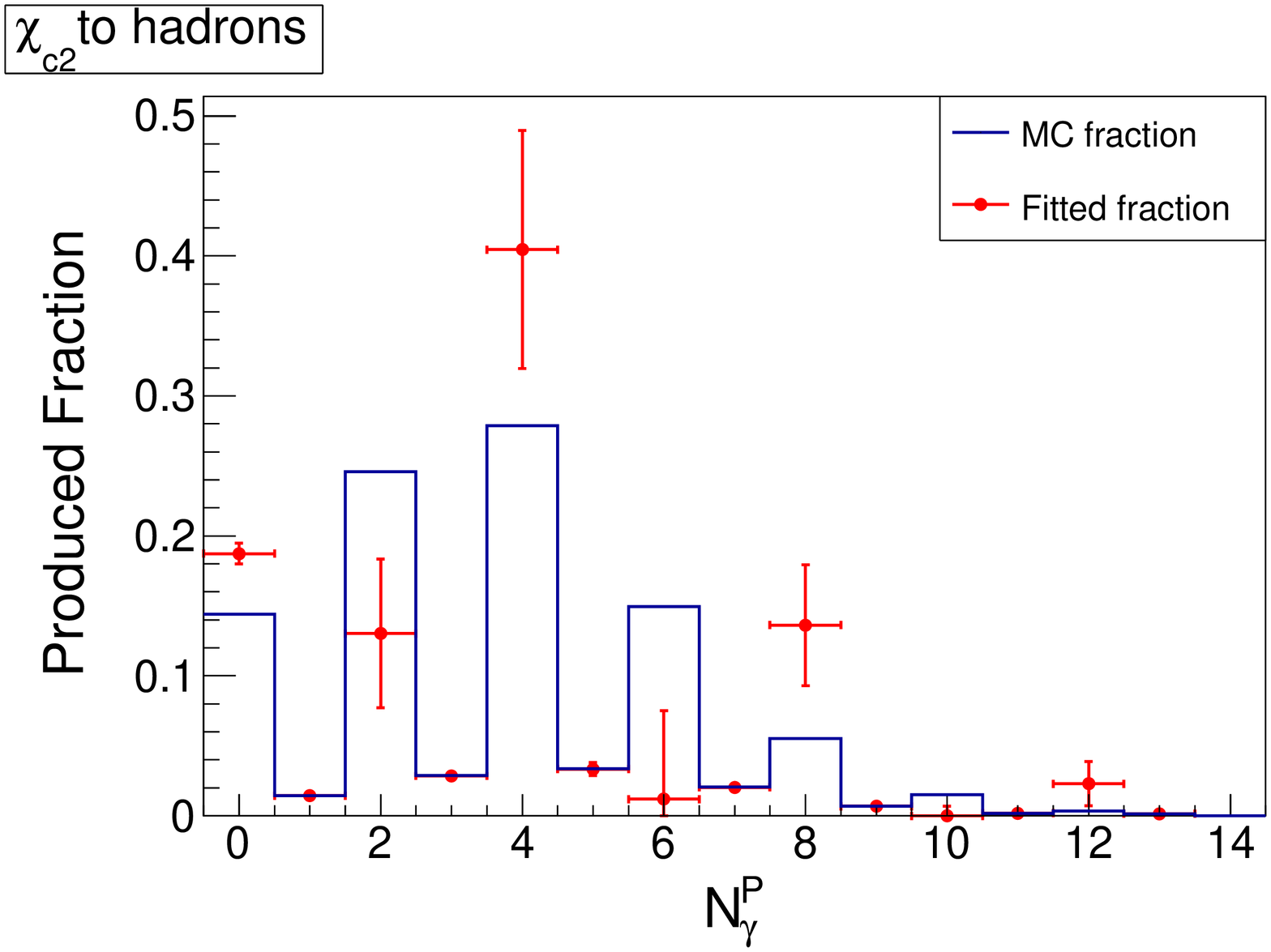}}
\put(-455,-35){\bf \large {(a)}}
\put(-290,-35){\bf \large {(b)}}
\put(-130,-35){\bf \large {(c)}}

\rotatebox{-90}{
\includegraphics*[width=1.67in]{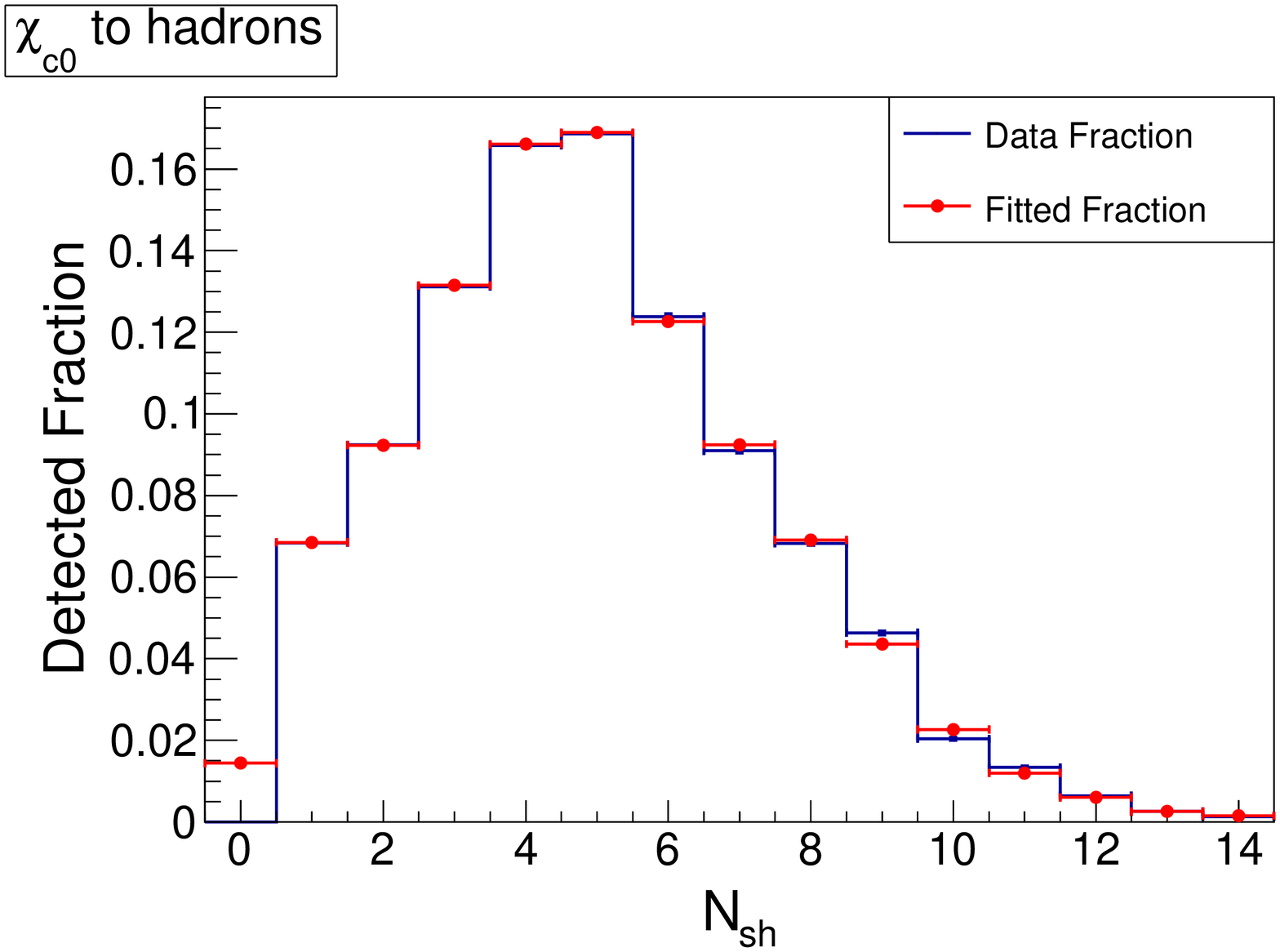}}
\rotatebox{-90}{
\includegraphics*[width=1.67in]{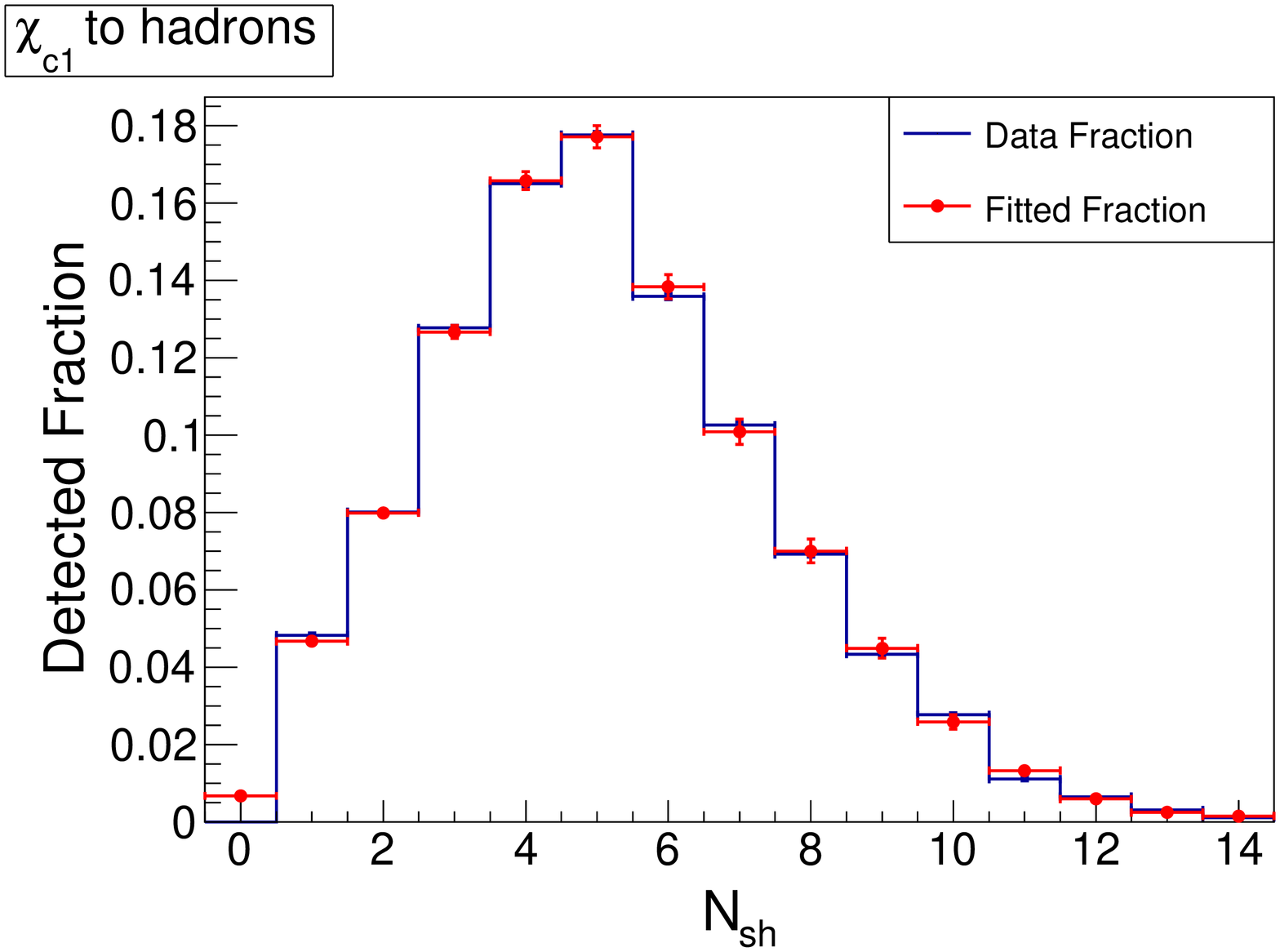}}
\rotatebox{-90}{
  \includegraphics*[width=1.67in]{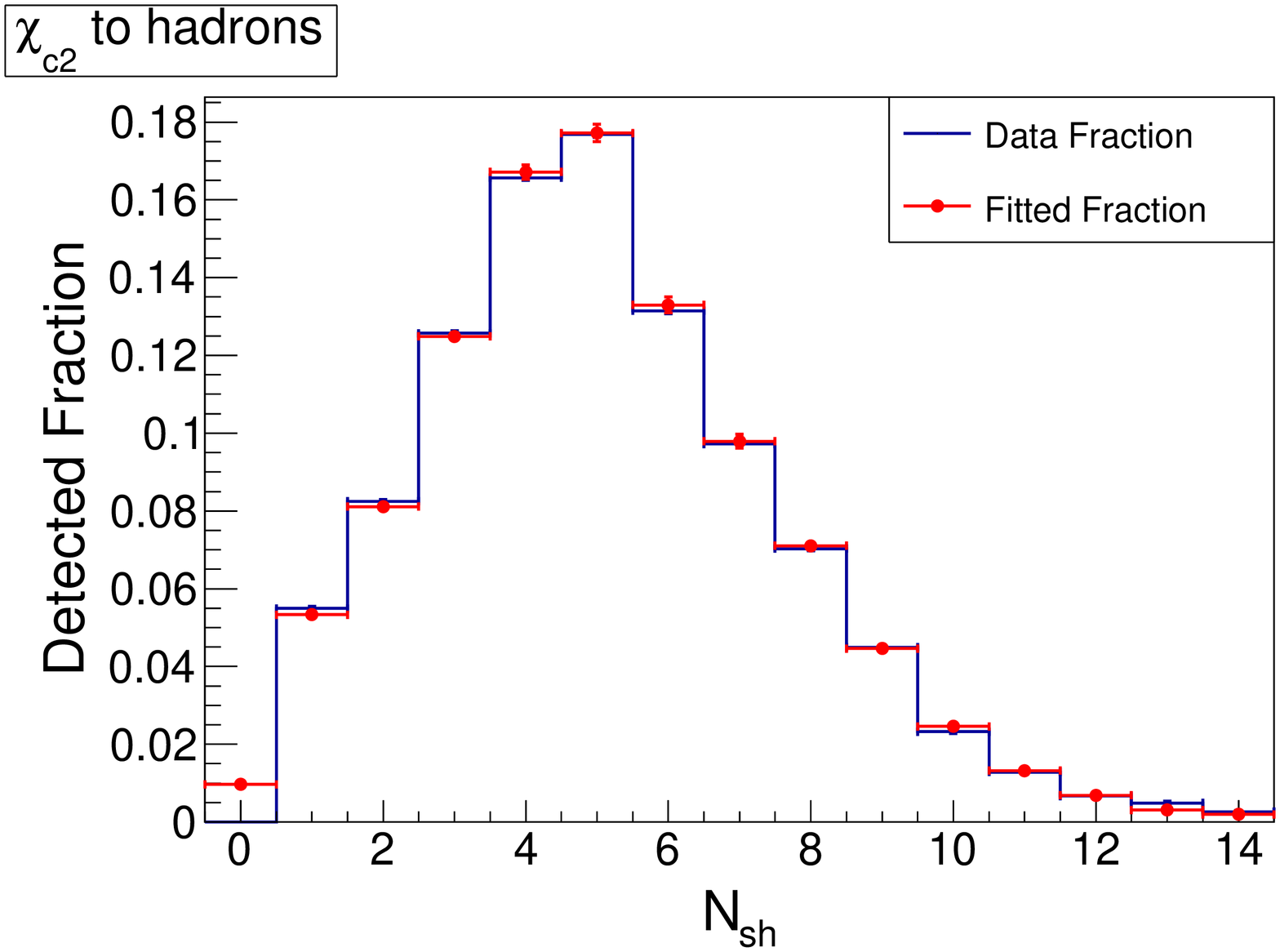}}
\put(-455,-35){\bf \large {(d)}} \put(-290,-35){\bf \large {(e)}}
\put(-130,-35){\bf \large {(f)}}
\caption{\label{prod_gam_1} MC and fitted fractions versus $N^{\rm P}_{\gamma}$ for (a)
  $\chi_{c0}$, (b) $\chi_{c1}$, and (c) $\chi_{c2} \to$ hadrons.  Odd
  bins are fixed to MC result values.  Radiative photons from
  $\psi(3686) \to \gamma \chi_{cJ}$ are not counted so the lowest bin
  is $N^{\rm P}_{\gamma} = 0$.  The distributions in (d) - (f) are the
  corresponding detected fractions versus $N_{\rm sh}$.  Since at least
  one EMCSH must be detected, the data fraction for $N_{\rm sh} = 0$ is
  empty.}
\end{center}
\end{figure*}

$P_{N^{\rm P}_{\gamma}}$ distributions are studied in an analogous way.
Here the $P_{N^{\rm P}_{\gamma}}$ distributions correspond to the MC-tagged
photons, described in Section~{\ref{ngam_results}}, and the detected
distributions are the EMC shower distributions, which include both
good and bad shower matches.  The results of the fits for the
$P_{N^{\rm P}_{\gamma}}$ distributions are shown in Fig.~\ref{prod_gam_1}
for $\chi_{cJ} \to$ hadrons.  Shown in Figs.~\ref{prod_gam_1} (a) -
(c) are the MC fractions and the results from the fits to the detected
distributions of data.  The radiative photons from $\psi(3686) \to
\gamma \chi_{cJ}$ are not counted, so the lowest bin is $N^{\rm P}_{\gamma}
= 0$.  Even bins are much larger than odd ones since most photons are
from $\pi^0\to\gamma\gamma$ decays.  Photons from final-state radiation (FSR) and
radiative photons from intermediate-state decays are counted and
contribute to odd bins.  While fit results for bins $N^{\rm P}_{\gamma} =$
2, 6, and 10 are smaller than MC, those for $N^{\rm P}_{\gamma} =$ 0, 4, 8,
and 12, which correspond to an even number of $\pi^0$s, are much
larger than MC.  The detected data fractions as a function of $N_{\rm sh}$
and the fractions determined from the fit results are shown in
Figs.~\ref{prod_gam_1} (d) - (f).

\begin{figure*}[htbp]
\begin{center}
\rotatebox{-90}{
\includegraphics*[width=1.67in]{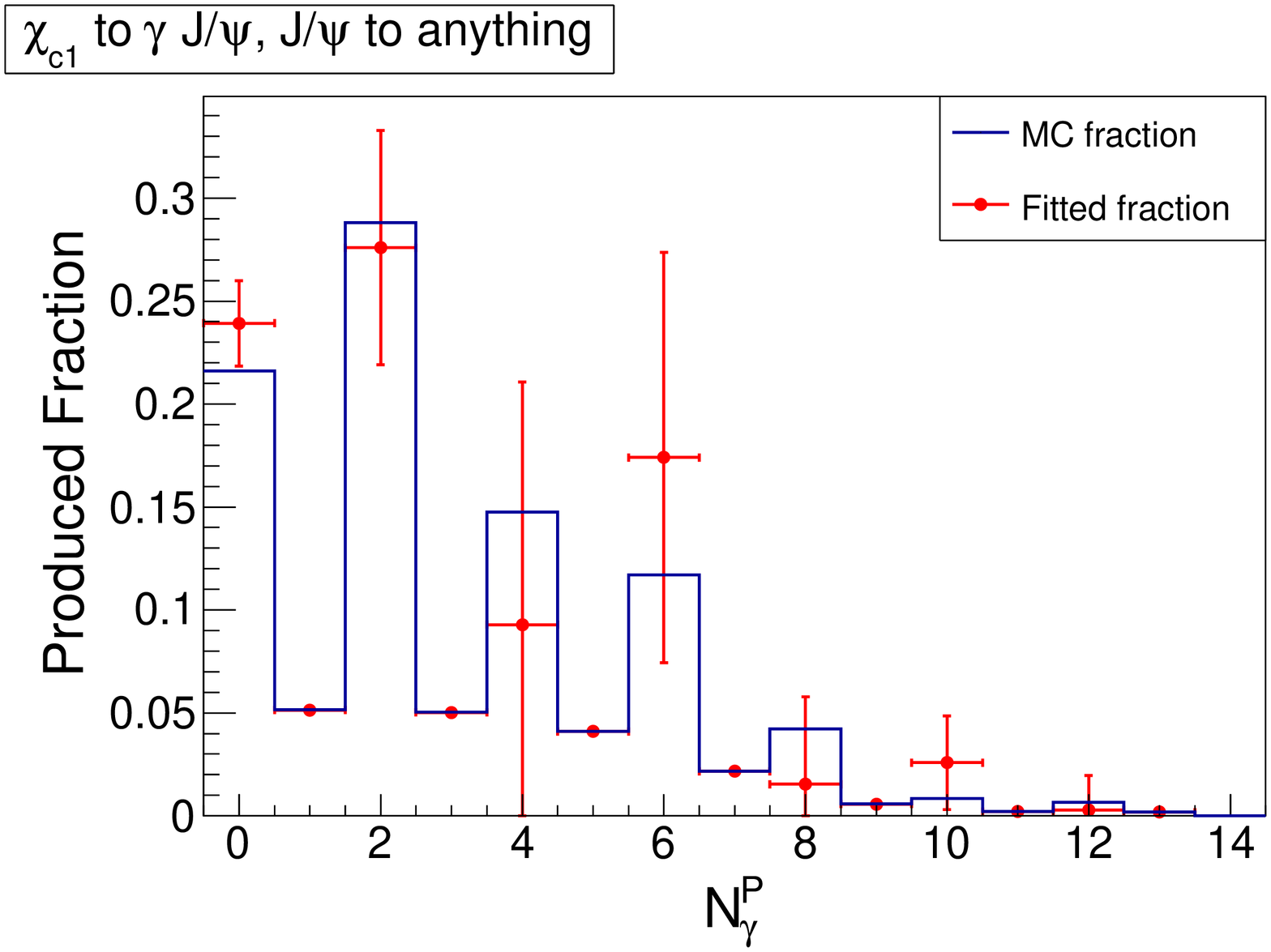}}
\rotatebox{-90}{
\includegraphics*[width=1.67in]{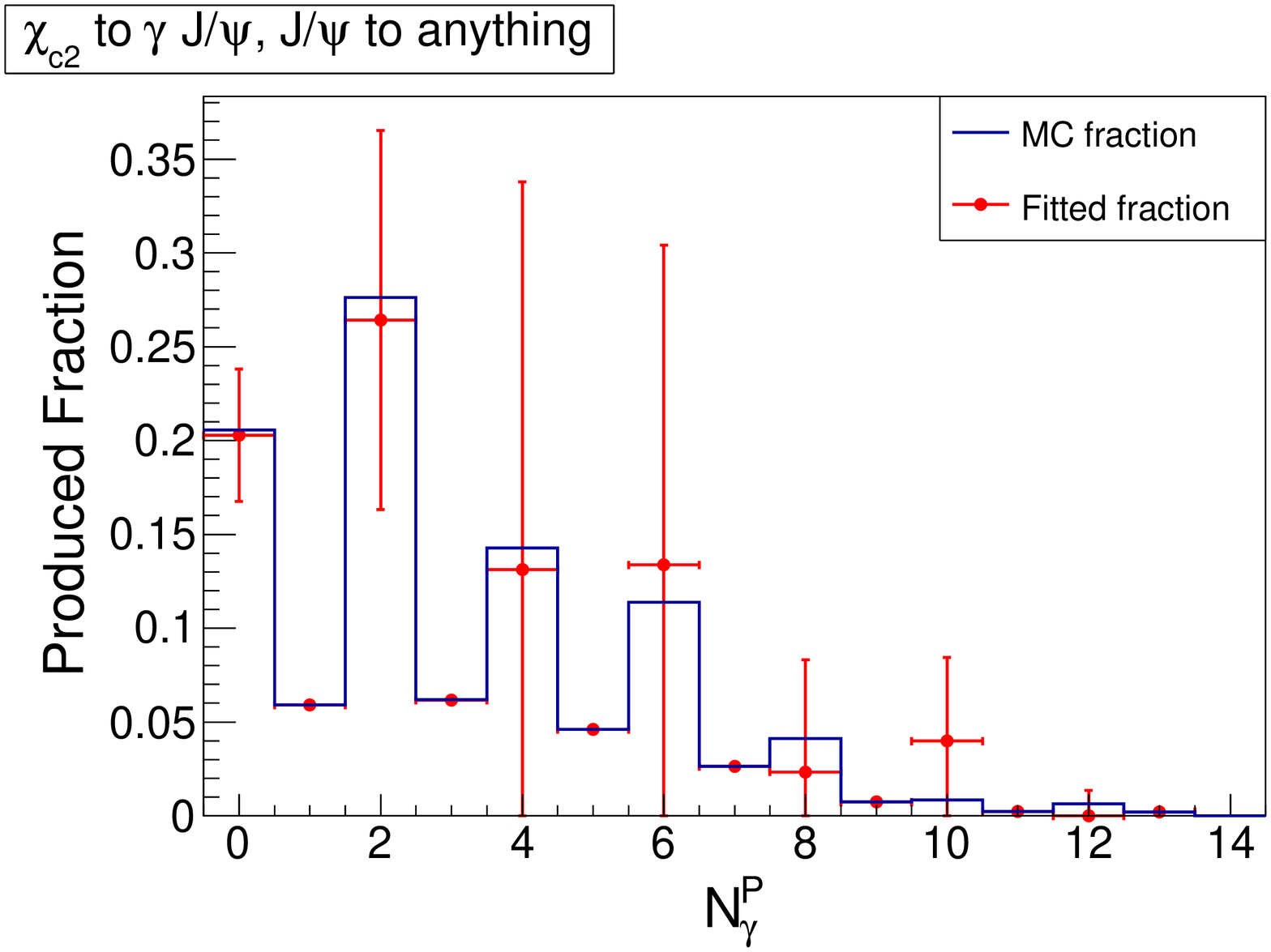}}
\put(-235,-30){\bf \large {(a)}}
\put(-70,-30){\bf \large {(b)}}

\rotatebox{-90}{
\includegraphics*[width=1.67in]{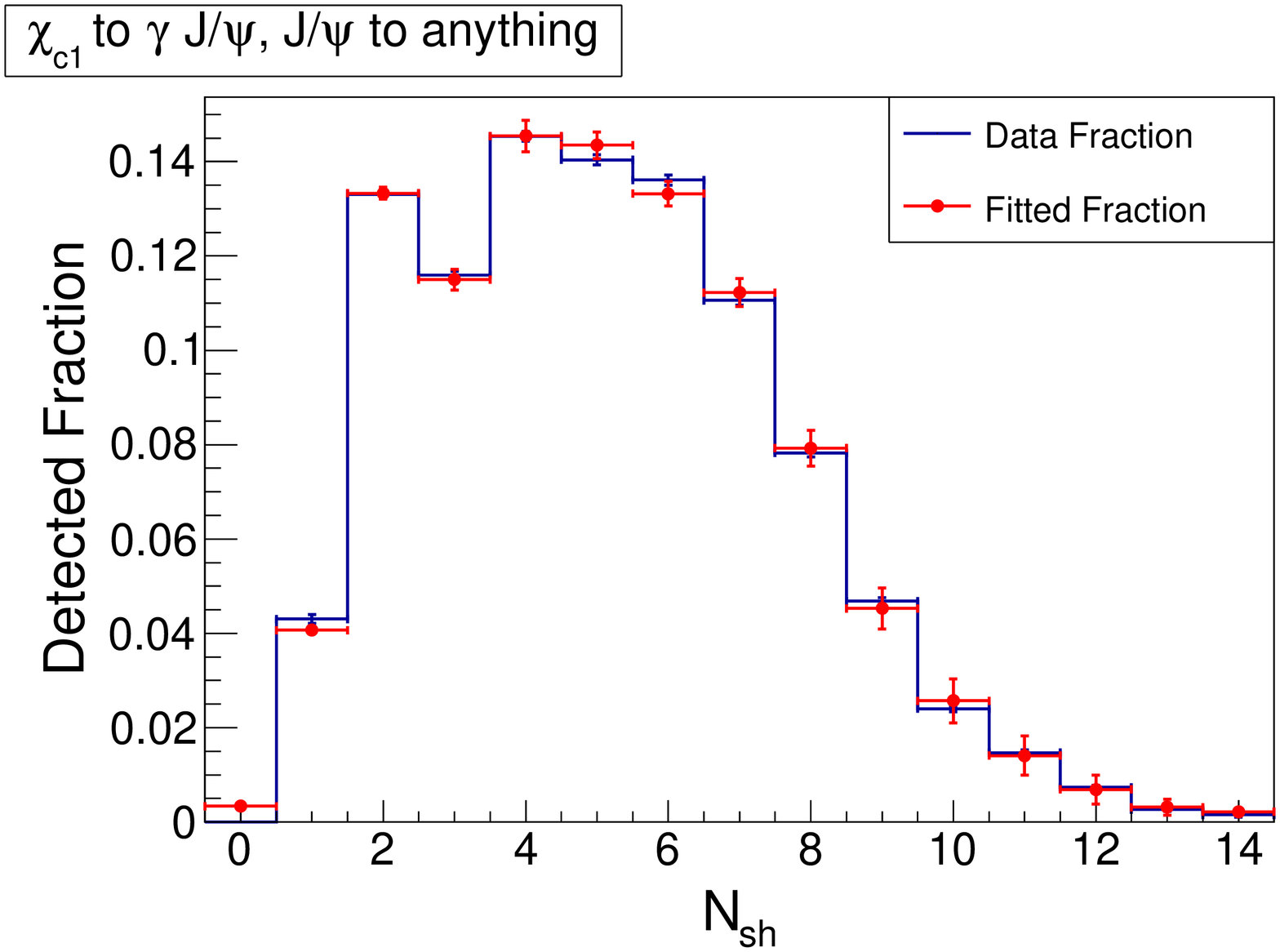}}
\rotatebox{-90}{
  \includegraphics*[width=1.67in]{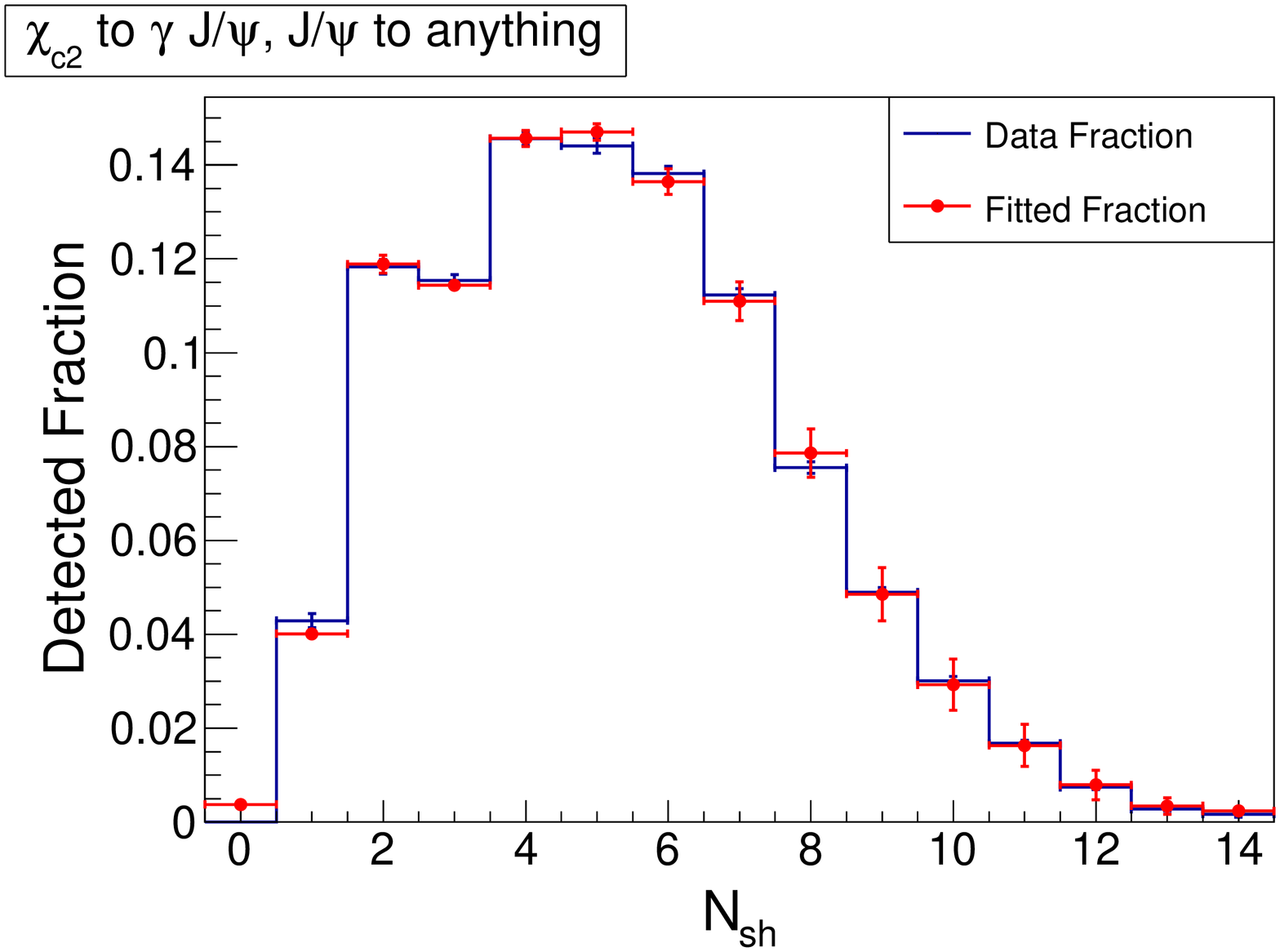}}
\put(-235,-30){\bf \large {(c)}}
\put(-70,-30){\bf \large {(d)}}
\caption{\label{prod_gam_2} MC and fitted 
  fractions versus $N^{\rm P}_{\gamma}$ for (a) $\chi_{c1} \to \gamma J/\psi,\:J/\psi \to$ anything
  and (b) $\chi_{c2} \to \gamma J/\psi,\:J/\psi \to$ anything.  Odd
  bins are fixed to MC result values.  Radiative photons from
  $\psi(3686) \to \gamma \chi_{cJ}$ and $\chi_{cJ} \to \gamma J/\psi$
  are not counted so the lowest bin is $N^{\rm P}_{\gamma} = 0$. The
  distributions in (c) and (d) are the corresponding detected
  fractions versus $N_{\rm sh}$.}
\end{center}
\end{figure*}

  
  Results for $\chi_{c1/2} \to \gamma J/\psi,\:J/\psi \to$ anything
  are shown in Fig.~\ref{prod_gam_2}.  The MC
  fractions and the results from the fits to the detected
  distributions of data are shown in
  Figs.~\ref{prod_gam_2} (a) - (b).  Since radiative photons from
  $\chi(3686) \to \gamma \chi_{cJ}$ and $\chi_{cJ} \to \gamma J/\psi$
  are not counted, the lowest bin is also $N^{\rm P}_{\gamma} = 0$.  Here,
  for bins with $N^{\rm P}_{\gamma} =$ 2, 6, and 10, which correspond to a
  preference for an odd number of $\pi^0$s, fit results are slightly
  larger than MC, but uncertainties are large.  The detected data
  fractions versus $N_{\rm sh}$ and the fit results are shown in
  Figs.~\ref{prod_gam_2} (c) - (d).  The $G$-parity for $\chi_{cJ}$s
  is positive, suggesting that decays to pions should favor an even
  number of $\pi$s, while $G$-parity for the $J/\psi$ is negative,
  implying that decays to pions favor an odd number of $\pi$s.  These
  preferences in the distributions of the number of photons are
  observed above for data, but MC simulation does not reflect this.

In Figs.~\ref{prod_gam_1} (d) - (f) and Figs.~\ref{prod_gam_2} (c) -
(d), 14 bins of detected data are being fit with 7 parameters ($P_0$,
$P_2$, $P_4$, $P_6$, $P_8$, $P_{10}$, and $P_{12}$) with in-between
bins fixed to MC values. The number of degrees of freedom is $ndf$ = 7.
Fractions $F^{ndf=7}$ of $\chi_{cJ} \to$ hadrons are listed in
Table~\ref{prod-gam1} and $\chi_{c1/2} \to \gamma J/\psi,\:J/\psi \to$
anything are listed in Table~\ref{prod-gam2}.  The $\chi^2/ndf$ values
for the five cases are 17, 7.8, 4.3, 5.7, and 2.9.


\begin{table*}[bth]
\begin{center}
\caption{$P_{N^{\rm P}_{\gamma}}$ event fractions in \% for data, $F_{\chi_{cJ}}^{ndf=7}$, and MC simulated
  sample, $F^{\rm MC}_{\chi_{cJ}}$,  for $\chi_{cJ} \to$ hadrons. Odd
  bins are fixed to MC result values, so only the systematic
  uncertainties are shown. {\red Here and in Table XV, it is the number of
  events that is fixed, so the fractions may differ slightly.}
\label{prod-gam1}}
\begin{footnotesize}
\begin{tabular}{r|cr|cr|cr} \hline
\T $N^{\rm P}_{\gamma}$ & $F_{\chi_{c0}}^{ndf=7}$ & $F^{\rm MC}_{\chi_{c0}}$ & $F_{\chi_{c1}}^{ndf=7}$  & $F^{\rm MC}_{\chi_{c1}}$ & $F_{\chi_{c2}}^{ndf=7}$ & $F^{\rm MC}_{\chi_{c2}}$ \B  \\ \hline
0  & $20.31\pm0.16\pm1.33$ & 17.66 & $17.15\pm0.28\pm3.09$ & 14.29 & $18.74\pm0.22\pm0.69$ & 14.40\\
1  & $1.59\pm0.02$     &  1.61 & $1.56\pm0.01$     &  1.57 & $1.43\pm0.20$     & 1.44\\
2  & $15.09\pm0.16\pm2.59$ & 23.48 & $16.87\pm0.85\pm7.27$ & 24.31 & $13.03\pm0.18\pm5.30$ & 24.59 \\
3  & $2.68\pm0.03$     & 2.72  & $2.87\pm0.01$     & 2.89  & $2.85\pm0.39$     & 2.87\\
4  & $36.36\pm0.13\pm3.98$ & 28.45 & $35.11\pm0.36\pm12.56$ & 27.51& $40.46\pm0.17\pm8.50$   & 27.90\\
5  & $3.10\pm0.03$     &  3.15 & $3.40\pm0.01$     & 3.42  & $3.35\pm0.46$     & 3.37\\
6  & $0.00\pm0.32\pm8.61$  & 13.64 & $7.48\pm0.61\pm9.08$  & 15.61 & $1.19\pm0.52\pm6.30$  & 14.95 \\
7  & $1.88\pm0.02$     &  1.90 & $2.07\pm0.01$     & 2.09  & $2.04\pm0.28$ & 2.06\\
8  & $15.86\pm0.36\pm9.02$ & 5.00  & $7.61\pm0.83\pm9.63$  & 5.39  & $13.62\pm0.70\pm4.26$  & 5.54 \\
9  & $0.54\pm0.01$     & 0.55  & $0.67\pm0.00$     & 0.68  & $0.68\pm0.09$      & 0.69 \\
10 & $0.00\pm0.45\pm2.98$  & 1.25  & $4.12\pm0.99\pm7.38$  & 1.56  & $0.00\pm0.69\pm0.00$   & 1.50 \\
11 & $0.16\pm 0.00$    & 0.17  & $0.18\pm0.00$     & 0.18  & $0.18\pm0.03$      & 0.19 \\
12 & $2.32\pm0.18\pm2.09$  & 0.32  & $0.76\pm1.08\pm2.62$  & 0.34  & $2.29\pm0.45\pm1.51$   & 0.36 \\
13 &$0.11\pm0.00$      & 0.11  & $0.14\pm0.00$     & 0.14  & $0.14\pm0.02$      & 0.15 \\
\hline
\end{tabular}
\end{footnotesize}
\end{center}
\end{table*}

\begin{table}[bth]
\begin{center}
  \caption{$P_{N^{\rm P}_{\gamma}}$ event fractions in \% for data
    $F_{J/\psi_1}^{ndf=7}(F_{J/\psi_2}^{ndf=7})$ and MC simulated
    sample $F^{\rm MC}_{J/\psi_1}(F^{\rm MC}_{J/\psi_2})$ for 
    $J/\psi \to \chi_{c1}(\chi_{c2}),\: J/\psi \to$ anything.  Odd
    bins are fixed to MC result values, so only the systematic
    uncertainty is shown.
  \label{prod-gam2}}
\begin{footnotesize}
\begin{tabular}{r|cr|cr} \hline
\T $N^{\rm P}_{\gamma}$ & $F_{J/\psi_1}^{ndf=7}$ & $F^{\rm MC}_{J/\psi_1}$ & $F_{J/\psi_2}^{ndf=7}$  & $F^{\rm MC}_{J/\psi_2}$ \B \\ \hline
0  & $23.92\pm0.26\pm2.05$ & 21.60 & $20.27\pm0.46\pm3.50$  & 20.57 \\
1  & $5.14\pm0.03$     & 5.15  & $5.91\pm0.08$      & 5.92 \\
2  & $27.60\pm1.15\pm5.57$ & 28.82 & $26.42\pm0.58\pm10.08$ & 27.61 \\
3  & $5.02\pm0.03$     & 5.03  & $6.16\pm0.08$      & 6.18 \\
4  & $9.29\pm0.55\pm11.76$ & 14.76 & $13.12\pm0.19\pm20.66$ & 14.28 \\
5  & $4.10\pm0.03$     & 4.11  & $4.59\pm0.06$      & 4.61 \\
6  & $17.41\pm0.62\pm9.95$ & 11.69 & $13.38\pm0.97\pm17.00$ & 11.39 \\
7  & $2.16\pm0.01$     & 2.17  & $2.63\pm0.04$      & 2.64 \\
8  & $1.53\pm1.12\pm4.11$  & 4.22  & $2.34\pm1.63\pm5.74$   & 4.11\\
9  & $0.57\pm0.00$     & 0.57  & $0.74\pm0.01$      & 0.74 \\
10  & $2.58\pm1.44\pm1.76$ & 0.84  & $3.99\pm1.82\pm4.07$   & 0.84 \\
11  & $0.21\pm0.00$    & 0.21  & $0.23\pm0.00$      & 0.23 \\
12  & $0.29\pm1.40\pm0.92$ & 0.65  & $0.00\pm1.37\pm0.08$   & 0.64  \\
13  & $0.18\pm0.00$    & 0.18  & $0.22\pm0.00$      & 0.22 \\
\hline
\end{tabular}
\end{footnotesize}
\end{center}
\end{table}

\subsection{\boldmath $P_{N^{\rm P}_{\pi^0}}$ distributions}
\label{npi0_prod}
$P_{N^{\rm P}_{\pi^0}}$ distributions are studied in a similar fashion.  Here, the
situation is complicated because events for a given $N_{\pi^0}$
contain miscombinations as well as real $\pi^0$s.  We assume that the
MC simulation correctly describes the miscombinations in data and do
not multiply by $R$.  Unlike the cases above, the alternate bins are
not suppressed, so adjacent PDFs are similar in shape, 
which results in larger fit uncertainties for  the  values of $P_{N^{\rm P}_{\pi^0}}$.

The low sensitivity that the fit has to $P_{N^{\rm P}_{\pi^0}}$ has other consequences.
For most of the variations used
in Section~{\ref{systematics}} to determine $N^{\rm P}_{\pi^0}$ systematic
uncertainties, the fits of the detected distributions of data fail, with only 
three successful fits out of a total of nine.   See Section~{\ref{systematics}} for
details.  In conclusion, we are not able to determine the systematic
uncertainties for the $P_{N^{\rm P}_{\pi^0}}$ distributions corresponding to the
detected $\pi^0$ distributions and therefore the event
fractions themselves.

\subsection{Input-Output Check}

The procedures above have been repeated using MC detected
distributions as input. The output produced distributions determined
by the analyses should then agree closely with the MC truth
distributions.  We divide the MC data in half and use the first half
to construct the detection matrices and use them in fitting the
detected distributions of the second half.  We compare the fitting
results with the MC truth fractions of the second half.  For this
check, the uncertainties on the detected distributions are taken as
the statistical uncertainties on the number of detected events
combined in quadrature with the statistical uncertainties on the
number of MC events.
The output fitted fractions $F^{\rm P}$ and input MC fractions
$F^{\rm MC}$ versus $N^{\rm P}_{\rm ch}$ are given in Table~\ref{input_output_t1},
where the agreement is very good.  The $\chi^2/ndf$ values for the
fits are 1.2, 0.9, 0.8, 0.5, and 0.7.

The output fitted fractions $F^{\rm P}$ and input MC fractions
$F^{\rm MC}$ versus $N^{\rm P}_{\gamma}$ are given in
Table~\ref{input_output_t2}.  The agreement for these cases is not as
good as for the $N^{\rm P}_{\rm ch}$ cases.  The $\chi^2/ndf$ values for
$N^{\rm P}_{\gamma}$ are 1.2, 0.5, 0.2, 1.5, and 2.2.  In all cases, the
differences between input and output are small compared to the
systematic uncertainties detailed in Table~\ref{systematics-prod} and
are neglected.

\begin{table*}[bth]
\begin{center}
\caption{Results from the MC input-output test.  The output
   ($F^{\rm P}$) and input MC ($F^{\rm MC}$) fractions of events in \% with
  $N^{\rm P}_{\rm ch}$ for $\chi_{cJ} \to$ hadrons ($F_{\chi_{cJ}}$) and
  $\chi_{c1/2} \to \gamma J/\psi$ ($F_{J/\psi_{1/2}})$. The
  $P_{N^{\rm P}_{\rm ch}}$ values for $N^{\rm P}_{\rm ch} \ge 12$ are fixed to those of
  the MC in the fitting and are not listed.
\label{input_output_t1}}
\begin{footnotesize}
\begin{tabular}{lcc|cc|cc|cc|cc} \hline
 \T $N_{\rm ch}^{\rm P}$ & $F^{\rm P}_{\chi_{c0}}$ & $F^{\rm MC}_{\chi_{c0}}$ &
  $F^{\rm P}_{\chi_{c1}}$ & $F^{\rm MC}_{\chi_{c1}}$ &
  $F^{\rm P}_{\chi_{c2}}$ & $F^{\rm MC}_{\chi_{c2}}$ &
  $F^{\rm P}_{J/\psi_1}$ & $F^{\rm MC}_{J/\psi_1}$ &
  $F^{\rm P}_{J/\psi_2}$ & $F^{\rm MC}_{J/\psi_2}$ \B \\ \hline
0 & $1.401\pm0.010$ & 1.413 & $0.851\pm0.011$ & 0.854 &
$0.914\pm0.010$ & 0.930& $2.037\pm0.022$ & 2.068 & $1.952\pm0.029$ &
1.985 \\
2 & $21.55\pm0.03$  & 21.56 & $19.08\pm0.04$  & 19.07 & $17.95\pm0.04$
& 17.94 & $36.75\pm0.06$  & 36.71 & $35.14\pm0.09$ & 35.14\\
4 & $49.60\pm0.04$  & 49.61 & $48.60\pm0.06$  & 48.66 & $49.49\pm0.05$
& 49.49 & $39.96\pm0.07$ & 39.94 & $39.71\pm0.10$ & 39.68\\
6 & $25.14\pm0.04$  & 25.10 & $28.36\pm0.05$  & 28.28 & $28.34\pm0.05$
& 28.33 & $18.25\pm0.06$ & 18.31 & $19.26\pm0.10$ & 19.32\\
8 & $2.241\pm0.020$ & 2.271 & $3.051\pm0.028$ & 3.062 &
$3.217\pm0.026$ & 3.229 & $2.98\pm0.040$ & 2.911 & $3.752\pm0.074$ & 3.655\\
10 & $0.066\pm0.007$ & 0.048 &  $0.058\pm0.009$ & 0.069 &
$0.084\pm0.008$ & 0.081 &  $0.016\pm0.019$ & 0.063 &   $0.177\pm0.039$
& 0.205 \\ \hline
\end{tabular}
\end{footnotesize}
\end{center}
\end{table*}

\begin{table*}[bth]
\begin{center}
\caption{Results from the input-output test.  The output ($F^{\rm P}$) and
  input MC ($F^{\rm MC}$) fractions of events in \% with $N^{\rm P}_{\gamma}$ for
  $\chi_{cJ} \to$ hadrons ($F_{\chi_{cJ}}$) and $\chi_{c1/2} \to
  \gamma J/\psi$ ($F_{J/\psi_{1/2}})$. The $P_{N^{\rm P}_{\gamma}}$ values
  for $N^{\rm P}_{\gamma}$ odd are fixed to those of the MC in the fitting.
\label{input_output_t2}}
\begin{footnotesize}
\begin{tabular}{lcc|cc|cc|cc|cc} \hline
  \T $N_{\gamma}^{\rm P}$ & $F^{\rm P}_{\chi_{c0}}$ & $F^{\rm MC}_{\chi_{c0}}$ &
  $F^{\rm P}_{\chi_{c1}}$ & $F^{\rm MC}_{\chi_{c1}}$ &
  $F^{\rm P}_{\chi_{c2}}$ & $F^{\rm MC}_{\chi_{c2}}$ &
  $F^{\rm P}_{J/\psi_1}$ & $F^{\rm MC}_{J/\psi_1}$ &
  $F^{\rm P}_{J/\psi_2}$ & $F^{\rm MC}_{J/\psi_2}$ \B \\ \hline
0 & $17.79\pm0.07$ & 17.67 & $13.97\pm0.09$ & 14.28 &
$14.02\pm0.08$ & 14.41& $21.58\pm0.10$ & 21.62 & $20.39\pm0.14$ &
20.55 \\
1 & $1.60$  & 1.60 & $1.57$  & 1.57 & $1.44$
& 1.44 & $5.15$  & 5.15 & $5.92$ & 5.92\\
2 & $23.34\pm0.12$  & 23.49 & $24.20\pm0.17$  & 24.34 & $24.35\pm0.16$
& 24.61 & $28.85\pm0.27$  & 28.82 & $28.03\pm0.41$ & 27.61\\
3 & $2.72$  & 2.72 & $2.89$  & 2.89 & $2.87$
& 2.87 & $5.04$ & 5.04 & $6.18$ & 6.18\\
4 & $28.74\pm0.05$  & 28.42 & $28.14\pm0.06$  & 27.49 & $28.58\pm0.06$
& 27.87 & $14.81\pm0.11$ & 14.75 & $14.17\pm0.17$ & 14.28\\
5 & $3.15$  & 3.15 & $3.43$  & 3.43 & $3.39$
& 3.39 & $4.11$ & 4.11 & $4.61$ & 4.61\\
6 & $13.51\pm0.18$  & 13.64 & $15.25\pm0.24$  & 15.62 & $14.60\pm0.23$
& 14.96 & $11.73\pm0.26$ & 11.69 & $11.03\pm0.36$ & 11.40\\
7 & $1.91$ & 1.91 & $2.08$ & 2.08 &
$2.06$ & 2.06 & $2.16$ & 2.16 & $2.65$ & 2.65\\
8 & $4.86\pm0.23$ & 5.00 & $5.57\pm0.32$ & 5.40 &
$6.03\pm0.30$ & 5.52 & $4.09\pm0.43$ & 4.21 & $4.41\pm0.60$ & 4.13\\
9 & $0.55$ & 0.55 &  $0.68$ & 0.68 &
$0.69$ & 0.69 &  $0.57$ & 0.57 &   $0.74$ & 0.74 \\
10 & $1.28\pm0.23$ & 1.25 &  $1.48\pm0.32$ & 1.56 &
$1.11\pm0.30$ & 1.49 &  $0.89\pm0.48$ & 0.84 &   $0.66\pm0.69$
& 0.83 \\
11 & $0.17$ & 0.17 &  $0.18$ & 0.18 &
$0.19$ & 0.19 &  $0.21$ & 0.21 &   $0.23$
& 0.23 \\
12 & $0.29\pm0.20$ & 0.32 &  $0.35\pm0.28$ & 0.34 &
$0.44\pm0.26$ & 0.36 &  $0.59\pm0.40$ & 0.66 &   $0.68\pm0.61$
& 0.63 \\
13 & $0.11$ & 0.11 &  $0.14$ & 0.14 &
$0.15$ & 0.15 &  $0.18$ & 0.18 &   $0.22$
& 0.22 \\ \hline
\end{tabular}
\end{footnotesize}
\end{center}
\end{table*}


\section{Systematic Uncertainties}
\label{systematics}
Extensive studies of systematic uncertainties were carried out in
Ref.~\cite{bam248}.  For the $\psi(3686) \to \gamma \chi_{cJ}$
branching fraction, they are under 4\,\% with the largest contribution
coming from fitting the $E_{\rm sh}$ distribution.  Many of the
uncertainties do not apply here.  For the distribution of the number
of charged tracks, the uncertainty from $N_{\rm ch} > 0$ does not apply,
since we include events with no charged tracks. The requirement
$N_{\rm sh} <17$ essentially includes all events, as does $E_{\rm vis} >
0.22E_{\rm cm}$, which has a small systematic uncertainty.  The
uncertainty from $N_{\psi(3686)}$ does not apply since we calculate
event fractions.  Those for MC signal shape, multipole correction, and
$|\cos \theta| < 0.80$ affect the selection of the radiative photon
candidate and the overall number of events, but should not affect the
various distributions.

Systematic uncertainties are determined here for detected event fractions in
Sections~{\ref{nch_results}}~-~{\ref{npi0_results}} and for produced
event fractions in Sections~{\ref{nch_prod}}~-~{\ref{ngam_prod}}
using samples selected with alternate selection criteria and with
modified fitting procedures.  Systematic uncertainties are the
differences from the standard procedure added in quadrature.

For all distributions, the fitting uncertainties are determined by
changing the background polynomial, changing the range, and fixing
small signals.  Background polynomials are changed from second order
to first, and the fit ranges are changed from 0.08-0.5 GeV to
0.08-0.35 GeV and 0.2-0.54 GeV. 

For the detected charged track event fraction systematic uncertainties
in Section~{\ref{nch_results}} and the $P_{N^{\rm P}_{\rm ch}}$ fraction
uncertainties in Section~{\ref{nch_prod}}, (a) the fitting uncertainties are
considered, and in addition, uncertainties from (b) the $\psi(3686)$
background veto, (c) the $\pi^+ \pi^- J/\psi$ veto, (d) the $\delta >
14^\circ$ requirement, and (e) the continuum energy difference are
determined.  The uncertainties from (b) - (d) are determined by
removing those requirements and comparing with the analyses making
them.  The uncertainty from (e) is determined by scaling the EMCSH
energies of the continuum events by the ratio of the center-of-mass
energies of $\psi(3686)$ data and the continuum data.

For the detected photon event fraction systematic uncertainties in
Section~{\ref{ngam_results}}, the  $P_{N^{\rm P}_{\gamma}}$ event fraction systematic
uncertainties in Section~{\ref{ngam_prod}}, and the detected pion event fraction
uncertainties in Section~{\ref{npi0_results}}, the fitting errors are
considered. In addition, uncertainties from the $\psi(3686)$
background veto, the $\pi^0 \pi^0 J/\psi$ veto, the $\delta > 14^\circ$
requirement, and continuum energy difference are determined.

An important question is what are the systematic uncertainties
associated with the determination of the produced distributions by
fitting the detected distributions in
Section~{\ref{produced_section}}. {\blue We study this in two ways. In
  the first, we modify the PDFs used in determining the $P_{N^{\rm
      P}_{\rm ch}}$ values.}  In Section~{\ref{nch_prod}} the fits
have large $\chi^2/ndf$, and the PDFs in Figs.~\ref{prod_ch_1} (d) -
(f) and Figs.~\ref{prod_ch_2} (c) - (d) do not appear to describe the
detected distributions of data well.  Data are above the fit in the
highest bin and below in the preceding bin for the $N^{\rm P}_{\rm ch}
= 4$ and $N^{\rm P}_{\rm ch} = 6$ PDFs in Figs.~\ref{prod_ch_1} (d) -
(f) and Figs.~\ref{prod_ch_2} (c) - (d). The PDFs determined from the
inclusive MC seem to be too broad.

{\blue The PDFs have been modified to see the effect on the $\chi^2/ndf$s and
further to determine the differences in the $P_{N^{\rm P}_{\rm ch}}$ results.}
We assume that part of the PDFs may be described approximately by a
binomial distribution.  For instance, we assume that the PDF for 
$N^{\rm P}_{\rm ch} = 4$ for $N_{\rm ch} = 0$ through $N_{\rm ch} = 4$ can be described
by a binomial distribution in terms of an efficiency $\epsilon_4$,
which includes the geometric, tracking, and vertexing efficiencies,
and the fraction of $N_{\rm ch} = 4$ in $N_{\rm ch} = 0$ through $N_{\rm ch} = 4$
is given according to a binomial distribution by $\epsilon^4_4$. The
PDFs of the MC being too wide is due to the efficiences being too
small.  We estimate the corrected efficiency approximately by
comparing the data fractions and the fitted fractions for the $N_{\rm ch}
= 4$ bins in Figs.~\ref{prod_ch_1} (d) - (f) and Figs.~\ref{prod_ch_2}
(c) - (d), $\epsilon_{4_{\rm corr}} = \,^4\!\!\!\surd
\overline{D\,\epsilon^4_4/F}$, where $D$ is the data fraction bin
content of $N_{\rm ch}=4$ and $F$ is the fitted fraction bin content.  For
Fig.~\ref{prod_ch_1} (d), $\epsilon_4 = 0.8630$ and
$\epsilon_{4_{\rm corr}} = 0.8685$; the difference is only 0.64\%.  We
then use the ratio of the binomial distributions in terms of the two
efficiencies in each bin to correct the MC PDFs and use the corrected
PDFs to fit the detected distributions.  The part of the PDF for
$N_{\rm ch} > 4$ is left unchanged.  We do analogous calculations for
$N^{\rm P}_{\rm ch} = 2$ and $N^{\rm P}_{\rm ch} = 6$.

The corrected PDFs fit the detected distributions much better now than
they did in Figs.~\ref{prod_ch_1} and \ref{prod_ch_2}, and the
$\chi^2/ndf$s become 15.5, 12.3, 17.7, 5.4, and 9.0, which are much
reduced compared to those in Section~{\ref{nch_prod}} (65, 52, 85, 18,
and 28). The differences with the uncorrected fractions are small
compared with the systematic uncertainties shown in
Table~\ref{systematics-prod} and are neglected.

{\blue In the second, more quantitative study, we modify the selection
  criteria for both charged tracks and EMCSHs by requiring that they
  satisfy $|\cos \theta| < 0.8$, corresponding to the barrel shower
  counter.  The detected distributions are greatly altered by such a
  requirement.  This is easy to understand: the probability of
  removing a charged track goes way up when there are a large number
  of charged tracks.  The detected distributions are pushed to lower
  values, while the produced truth distributions of MC are not
  affected.  For example, the means of the detected $N_{\rm {ch}}$
  distributions are 3.82, 3.58, and 3.68 (3.20, 3.14, and 3.25) for
  the standard ($|\cos \theta| < 0.8$) selection for $\chi_{c0}$,
  $\chi_{c1}$, and $\chi_{c2} \to$ hadrons, while the means of the
  produced distributions are 4.27, 4.44, and 4.46 (4.28, 4.44, and
  4.48), respectively, for the standard ($|\cos \theta| < 0.8$)
  selection. The differences with the standard selection for both
  $P_{N^P_{\rm ch}}$ and $P_{N^P_{\gamma}}$ values are taken as the
  systematic uncertainties associated with the determination of the
  produced distributions by fitting the detected distributions.}

The detected event fraction uncertainties and the
$P_{N^{\rm P}_{\rm ch/\gamma}}$ event fraction uncertainties are listed in
Tables~\ref{systematics-nch}~ and \ref{systematics-prod},
respectively.  The uncertainties are the individual uncertainties for
all cases added in quadrature.

\begin{table*}[bth]
\begin{center}
  \caption{$N_{\rm ch}$, $N_{\rm sh}$ and $N_{\pi^0}$ detected event
    fraction systematic uncertainties in \%. In the table, $\chi_{cJ}$
    represents $\chi_{cJ} \to$ hadrons and $J/\psi_{1/2}$ represents
    $\chi_{c1/2} \to \gamma J/\psi,\:J/\psi \to
    $anything.\label{systematics-nch}}
\begin{footnotesize}
\begin{tabular}{lccccc| lccccc | lccccc} \hline
 \T $N_{\rm ch}$ & $\chi_{c0}$ & $\chi_{c1}$ & $\chi_{c2}$ & $J/\psi_1$ &  $J/\psi_2$ & $N_{\rm sh}$ & $\chi_{c0}$ & $\chi_{c1}$ & $\chi_{c2}$ & $J/\psi_1$ &  $J/\psi_2$ & $N_{\pi^0}$ & $\chi_{c0}$ & $\chi_{c1}$ & $\chi_{c2}$ & $J/\psi_1$ &  $J/\psi_2$\B  \\  \hline
\T 0 &  9.63 & 20.1 &  7.57  &  16.4  &  28.1 & & & & & & & 0  & 3.01 & 3.50 & 7.64 & 6.21 & 14.6 \\
 1 & 7.70 & 5.97 & 4.59 & 21.9 & 16.3 & 1 & 4.82 & 5.04 & 3.34 & 12.5 & 22.4 & 1 & 4.91 & 3.33 & 7.40 & 4.41 & 12.0 \\
 2 & 1.98 & 1.10 & 1.48 & 3.94 & 8.23 & 2 & 6.49 & 3.37 & 4.76 & 6.43 & 10.5& 2 & 6.64 & 3.24 & 7.68 & 14.7 & 8.09 \\
 3 & 2.69 & 1.47 & 1.19 & 6.44 & 5.77 & 3 & 2.21 & 2.53 & 3.62 & 5.28 & 9.28& 3 & 7.84 & 2.70 & 4.90 & 9.75 & 14.4 \\
 4 & 1.80 & 1.07 & 1.92 & 3.96 & 6.79 & 4 & 2.36 & 1.54 & 3.06 & 8.75 & 9.13& 4 & 17.9 & 7.49 & 9.11 & 16.9 & 23.0 \\
 5 & 5.74 & 2.61 & 2.70 & 10.2 & 10.3 & 5 & 3.21 & 2.05 & 3.71 & 7.86 & 17.4& 5 & 10.2 & 6.61 & 13.2 & 13.8 & 20.1 \\
 6 & 2.91 & 2.21 & 1.50 & 6.66 & 7.90 & 6 & 4.61 & 1.71 & 3.02 & 6.40 & 7.02 & 6 & 12.8 & 6.37 & 15.0 & 44.9 & 14.4 \\
 7 & 30.4 & 14.0 & 11.0 & 16.9 & 27.8 &7 & 5.79 & 3.30 & 3.40 & 6.59 & 14.0 & 7 & 145 & 100 & 103 & 146 & 210 \\
 8 & 5.60 & 8.01 & 16.3 & 40.4 & 22.2 & 8 & 9.04 & 4.60 & 4.48 & 12.1 & 24.7& 8 & 151 & 143 & 188 & 179 & 414 \\
 9 & 126 & 43.3 & 75.3 & 457 & 374 & 9 & 24.4 & 8.28 & 10.1 & 11.7 & 21.1 & 9 & 60.6 & 102 & 344 & 144 & 270 \\
& & & & & & 10 & 41.9 & 13.0 & 12.0 & 27.5 & 24.3 \\
& & & & & & 11 & 28.7 & 18.5 & 10.4 & 29.7 & 59.6 \\
& & & & & & 12 & 39.3 & 26.3 & 50.5 & 52.0 & 73.5 \\
& & & & & & 13 & 57.5 & 35.6 & 43.5 & 57.0 & 47.4 \\
& & & & & & 14 & 96.5 & 33.1 & 55.4 & 114 & 188 \\ \hline
\end{tabular}

\end{footnotesize}
\end{center}
\end{table*}

\begin{table}[bth]
\begin{center}
  \caption{$P_{N^{\rm P}_{\rm ch}}$ and $P_{N^{\rm P}_{\gamma}}$ event fraction
    systematic uncertainties in \%. 
    Bins are
    fixed to MC result values for $N^{\rm P}_{\rm ch} = 12$ and odd
    $N^{\rm P}_{\gamma}$ bins. \label{systematics-prod}}
\begin{footnotesize}
\begin{tabular}{lccccc| lccccc } \hline
 \T $N^{\rm P}_{\rm ch}$ & $\chi_{c0}$ & $\chi_{c1}$ & $\chi_{c2}$ & $J/\psi_1$ &  $J/\psi_2$ & $N^{\rm P}_{\gamma}$ & $\chi_{c0}$ & $\chi_{c1}$ & $\chi_{c2}$ & $J/\psi_1$ &  $J/\psi_2$ \B  \\  \hline
\T 0 & 0.49 & 1.50 & 0.76 & 0.81 & 4.14 & 0 & 1.33 & 3.09 & 0.69 & 2.05 & 3.50 \\
   2 & 0.78 & 6.80 & 3.57 & 1.57 & 3.91 & 1 & 0.02 & 0.01 & 0.20 & 0.03 & 0.08 \\
   4 & 1.11 & 2.99 & 1.37 & 2.81 & 6.34 & 2 & 2.59 & 7.27 & 5.30 & 5.57 & 10.1 \\
   6 & 2.17 & 3.92 & 2.27 & 1.64 & 3.09 & 3 & 0.03 & 0.01 & 0.39 & 0.03 & 0.08 \\
   8 & 4.66 & 4.19 & 1.94 & 2.35 & 2.57 & 4 & 3.98 & 12.56 & 8.50 & 11.8 & 20.7 \\
  10 & 3.11 & 2.65 & 1.43 & 2.64 & 1.62 & 5 & 0.03 & 0.01 & 0.46 & 0.03 & 0.06 \\
  12 &   0 &   0 &   0 &   0 &   0 & 6 & 8.61 & 9.08 & 6.30 & 9.95 & 17.0 \\
     &     &     &     &     &     & 7 & 0.02 & 0.01 & 0.28 & 0.01 & 0.04 \\
     &     &     &     &     &     & 8 & 9.02 & 9.63 & 4.26 & 4.11 & 5.74 \\
     &     &     &     &     &     & 9 & 0.01 & 0.00 & 0.09 & 0.00 & 0.01 \\
& & & & & & 10 & 2.98 & 7.38 & 0.00 & 1.76 & 4.07 \\
& & & & & & 11 & 0.00 & 0.00 & 0.03 & 0.00 & 0.00 \\
& & & & & & 12 & 2.09 & 2.62 & 1.51 & 0.92 & 0.08 \\
& & & & & & 13 & 0.00 & 0.00 & 0.02 & 0.00 & 0.00 \\ \hline

\end{tabular}

\end{footnotesize}
\end{center}
\end{table}

\section{Summary}
\label{summary}
The study of $\chi_{cJ}$ decays is important since they are expected
to be a source for glueballs, and their simulation is a necessary part
of their understanding.  Since a large fraction of their hadronic
decay modes are unmeasured, the close modeling of their inclusive
decays is very important.

Using 106 million $\psi(3686)$ decays, we study $\chi_{cJ} \to$ anything,
$\chi_{cJ} \to$ hadrons, and $J/\psi \to$ anything distributions.
Distributions of event fractions for data are compared with MC
simulation versus the number of detected charged tracks, EMCSHs and
$\pi^0$s in Figs.~\ref{ch_results2}, \ref{gam_results2} and
\ref{pi0_results2}, respectively.  For all comparisons, the agreement
is reasonable. However, there are differences.

To start with $\chi_{cJ} \to$ anything, for the $N_{\rm ch}$
distributions, data are above MC simulation for $N_{\rm ch} = 0$ and
$N_{\rm ch} >5$ and below for $N_{\rm ch} = 3$ and 6. For the $N_{\rm sh}$
distributions, data are above MC simulation for $N_{\rm sh} = 1$ and
$N_{\rm sh} > 7$ and below for $N_{\rm sh} = 3$, and for the
$N_{\pi^0}$ distributions, data are above MC simulation for $N_{\pi^0}
>2$.

For $J/\psi \to$ anything ($\chi_{c1}$ and $\chi_{c2} \to \gamma
J/\psi$), the agreement between data and MC simulation is good for the
$N_{\rm ch}$ distributions. There is some disagreement for the $N_{\rm sh}$
distributions, and for the $N_{\pi^0}$ distributions data are above MC
simulation for $N_{\pi^0} >5$, but the uncertainties are bigger.
Better agreement is expected for $J/\psi \to$ anything
distributions, since MC tuning was performed on the $J/\psi \to$ anything
events.


For $\chi_{cJ} \to$ hadron charged track distributions, fit results
shown in Fig.~\ref{prod_ch_1} for $P_{N^{\rm P}_{\rm ch}}$ are below the MC
fractions for $N^{\rm P}_{\rm ch} = 4$ and above for $N^{\rm P}_{\rm ch} = 0, 8,$ and
$10$.  $P_{N^{\rm P}_{\rm ch}}$ results for $\chi_{c1/2} \to \gamma
J/\psi,\:J/\psi \to$ anything charged track distributions are shown in
Fig.~\ref{prod_ch_2}. The distributions are similar, and the fit
fractions are in reasonable agreement with the MC fractions. The means
of the above $N^{\rm P}_{\rm ch}$ distributions in Figs.~\ref{prod_ch_1} and
\ref{prod_ch_2} are determined and plotted along with results from
MARK I for $e^+ e^- \to$ hadrons in the same energy range in
Fig.~\ref{means}.  The charmonium decays to hadrons and $e^+ e^- \to$
hadrons results are consistent.

The results for the $P_{N^{\rm P}_{\gamma}}$ distributions are shown in
Fig.~\ref{prod_gam_1} (a) - (c) for $\chi_{cJ} \to$ hadrons.  The
content of even bins are much larger than those of odd ones since most
photons are from the decay of $\pi^0$s.  While fit results for bins
$N^{\rm P}_{\gamma} =$ 2, 6, and 10 are smaller than MC, those for
$N^{\rm P}_{\gamma} =$ 0, 4, 8, and 12, which correspond to an even number
of $\pi^0$s, are much larger than MC.  Results for $\chi_{c1/2} \to
\gamma J/\psi,\:J/\psi \to$ anything for photons are shown in
Fig.~\ref{prod_gam_2}.  Here, bins with $N^{\rm P}_{\gamma} =$ 2, 6, and 10,
which correspond to a preference for an odd number of $\pi^0$s, appear
to have fit results slightly larger than MC.

The $G$-parity for $\chi_{cJ}$s is positive, suggesting that decays
should favor an even number of $\pi$s, while $G$-parity for the
$J/\psi$ is negative, implying that decays favor an odd number of
$\pi$s.  These preferences in the distributions of the number of
produced photons are observed for data, but MC simulation
does not adequately reflect this.

While the agreement between data and MC simulation is reasonable
at present, it should be improved for future studies of $\chi_{cJ}$
decays and measurements of the $\psi(3686) \to \gamma \chi_{cJ}$
branching fractions with even larger data sets.  This can be
accomplished with further MC tuning or by weighting the present or
future MC simulation to give better agreement with data.


\begin{acknowledgments}
The BESIII collaboration thanks the staff of BEPCII and the IHEP
computing center for their strong support. This work is supported in
part by National Key Basic Research Program of China under Contract
No. 2015CB856700; National Natural Science Foundation of China (NSFC)
under Contracts Nos. 11625523, 11635010, 11735014, 11822506, 11835012;
the Chinese Academy of Sciences (CAS) Large-Scale Scientific Facility
Program; Joint Large-Scale Scientific Facility Funds of the NSFC and
CAS under Contracts Nos. U1532257, U1532258, U1732263, U1832207; CAS
Key Research Program of Frontier Sciences under Contracts
Nos. QYZDJ-SSW-SLH003, QYZDJ-SSW-SLH040; 100 Talents Program of CAS;
INPAC and Shanghai Key Laboratory for Particle Physics and Cosmology;
ERC under Contract No. 758462; German Research Foundation DFG under
Contracts Nos. Collaborative Research Center CRC 1044, FOR
2359, GRK 214; Istituto Nazionale di Fisica Nucleare, Italy;
Koninklijke Nederlandse Akademie van Wetenschappen (KNAW) under
Contract No. 530-4CDP03; Ministry of Development of Turkey under
Contract No. DPT2006K-120470; National Science and Technology fund;
STFC (United Kingdom); The Knut and Alice Wallenberg Foundation
(Sweden) under Contract No. 2016.0157; The Royal Society, UK under
Contracts Nos. DH140054, DH160214; The Swedish Research Council;
U. S. Department of Energy under Contracts Nos. DE-FG02-05ER41374,
DE-SC-0010118, DE-SC-0012069; University of Groningen (RuG) and the
Helmholtzzentrum fuer Schwerionenforschung GmbH (GSI), Darmstadt
\end{acknowledgments}

\clearpage
\onecolumngrid
\appendix*
\section{Additional Material}
\vspace{-10mm}
\begin{table}[!h]
\begin{center}
\caption{Detected data events, $D$, efficiencies, $\epsilon$, efficiency corrected events, $N$, and number of scaled simulated events $N^{\rm MC}$ for $\chi_{cJ} \to$ anything. }
\begin{footnotesize}
\begin{tabular}{l|rrrr|rrrr|rrrr} \hline
 \T $N_{\rm sh}$ & $D{\chi_{c0}}$ & $\epsilon_{\chi_{c0}}$ &
  $N_{\chi_{c0}}$  &  $N^{\rm MC}_{\chi_{c0}}$ & $D_{\chi_{c1}}$ &
  $\epsilon_{\chi_{c1}}$ &  $N_{\chi_{c1}}$  &  $N^{\rm MC}_{\chi_{c1}}$  & $D_{\chi_{c2}}$ & $\epsilon_{\chi_{c2}}$ &  $N_{\chi_{c2}}$ &  $N^{\rm MC}_{\chi_{c2}}$\\
  & & (\%) & & & & (\%) & & & & (\%) & &  \B \\ \hline
  1  &  330758  & 49.4 &  669238  &  615384 & 229324  & 47.4 &  483430
  & 427139 &  237868  & 44.7 &  532591 & 461904\\
  2  &  439022  & 47.6 &  921785  &  926793 & 531929  & 51.8 & 1027261
  & 996329 &  416190  & 46.2 &  901025 & 913571 \\
  3  &  638938  & 49.7 & 1286718  & 1375251 & 700129  & 54.0 &
      1295902 & 1316389 & 596027  & 47.7 & 1250544 & 1314188 \\
  4  &  803512  & 49.9 & 1609754  & 1674111 & 883514  & 53.4 & 1655182
  & 1671070 &  761699  & 46.5 & 1638571 & 1645770\\
  5  &  846497  & 51.6 & 1640005  & 1712936 & 888728  & 51.8 & 1717062
  & 1745654 &  776282  & 45.1 & 1719400 & 1716841\\
  6  &  589146  & 49.1 & 1198819  & 1322729 & 683444  & 48.4 & 1411174
  & 1483861 &  556411  & 41.5 & 1340519 & 1428027\\
  7  &  412950  & 46.1 &  895086  &  921471 & 489771  & 44.8 & 1093266
  & 1113650 &  383015  & 37.7 & 1016905 & 1053385\\
  8  &  272287  & 42.2 &  645279  &  564767 & 308651  & 40.3 &  765307
  &  725934 &  241913  & 33.4 &  724569 &  681520\\
  9  &  148286  & 37.2 &  398285  &  311757 & 168098  & 34.9 &  482143
  &  416583 &  127628  & 28.4 &  449811 & 390813\\
 10  &    68326  & 32.8 &  208052  & 154672 &  802275 & 29.4 &  273469
  &  219793 &   56201  & 23.7 &  237515 & 205847\\
 11  &    30494  & 24.4 &  125022 &   72763 &  33641  & 26.2 &  128294
  &  108885 &   26640   & 19.4    &  137025 & 100920 \\
 12  &    13037  & 23.7 &   55089 &   31841 &  14393  & 19.4 &   74292
  &   50682 &   11927   & 15.7    &   76107 & 46826 \\
 13  &     6089  & 23.0 &   26495 &   13265 &   5171  & 17.0 &   30430
  &   22417 &    4627   & 10.0    &   46364 & 20569 \\
 14  &     3549  & 25.8 &   13768 &    5385 &   1945  & 14.2 &   13659
  &    9332 &    1810   &  6.6    &   27262 & 8646 \\
 $\ge15$  &     1810  & 31.1 &    5824 &    2103 &    769  & 13.8 &    5564
  &    3664 &      98   & 11.4    &     857 & 3263 \\
\hline
\end{tabular}
\label{result_chic_to_all_gam}
\end{footnotesize}
\end{center}
\end{table}

\begin{table}[bth]
\begin{center}
\caption{Detected data events, $D$, efficiencies, $\epsilon$, efficiency corrected events, $N$, and  number of scaled simulated events $N^{\rm MC}$ for $\chi_{c1/2} \to \gamma J/\psi,\:J/\psi \to$ anything. }
\begin{footnotesize}
\begin{tabular}{l|rrrr|rrrr} \hline
$N_{\rm sh}$ & $D_{J/\psi_1}$ & $\epsilon_{J/\psi_1}$
& $N_{J/\psi_1}$  & $N^{\rm MC}_{J/\psi_1}$ &  $D_{J/\psi_2}$ & $\epsilon_{J/\psi_2}$ &  $N_{J/\psi_2}$  &  $N^{\rm MC}_{J/\psi_2}$\\
  & & (\%) & & & & (\%) & &   \\ \hline
1  &   37240  & 24.0 &  155156  & 135859 &  20743  & 25.7 &   44857 &
80865\\
2  &  235504  & 48.8 &  483059  & 451104 & 104954  & 46.2 &  192872
& 239863 \\
3  &  228811  & 54.3 &  421096  & 415890 & 117186  & 54.4 &  215350 &
231364 \\
4  &  298314  & 58.9 &  506781  & 544528 & 157503  & 58.9 &  267321 &
295275 \\
5  &  294228  & 57.7 &  510057  & 545477 & 159820  & 58.2 &  274596 &
295923 \\
6  &  271070  & 56.8 &  477675  & 493600 & 147040  & 57.1  &  257544 &
270282 \\
7  &  217292  & 54.5 &  399037  & 394206 & 116281  & 55.3 &  210384 &
215396 \\
8  &  141923  & 51.2 &  276988  & 274819 &  75507  & 51.8 &  145684 &
149717 \\
9  &   76562  & 45.1 &  169632  & 161328 &  41794  & 45.8 &   91345 &
 89043 \\
10  &  33602  & 38.6 &   87068  &  88704 &  21989  & 39.5 &   55648 & 49176\\
11 &   15673  & 29.2 &   53626  &  45834 &  10894  & 39.7 &   27438 &
25702 \\
12 &    5588  & 26.9 &   20780  &  22788 &   4294  & 27.4 &   15701 & 12718\\
13 &    1657  & 13.1 &   12771  &  10753 &   1328  & 27.3 &    4858 & 6165\\
14 &     611  &  9.9 &    6145  &   4545 &    324  & 12.6 &    2566 & 2558\\
$\ge15$ &     374  & 22.1 &    1693  &   1809 &      0  &  8.9 &       0 &  976\\
\hline
\end{tabular}
\end{footnotesize}
\label{result_jpsi_to_all_gam}
\end{center}
\end{table}

\begin{table}[bth]
\begin{center}
\caption{Detected data events, $D$, efficiencies, $\epsilon$,
  efficiency corrected events, $N$, and number of scaled simulated
  events $N^{\rm MC}$ for $\chi_{cJ} \to$
  anything. Here $N$ has been multiplied by $R(data)$ and $N^{\rm MC}$
  by $R(MC)$, the fractions of valid $\pi^0$s.}
\begin{footnotesize}
\begin{tabular}{l|rrrr|rrrr|rrrr} \hline
  $N_{\pi^0}$ & $D_{\chi_{c0}}$ & $\epsilon_{\chi_{c0}}$ &
  $N_{\chi_{c0}}$  &  $N^{\rm MC}_{\chi_{c0}}$ & $D_{\chi_{c1}}$ &
  $\epsilon_{\chi_{c1}}$ &  $N_{\chi_{c1}}$  &  $N^{\rm MC}_{\chi_{c1}}$  & $D_{\chi_{c2}}$ & $\epsilon_{\chi_{c2}}$ &  $N_{\chi_{c2}}$ &  $N^{\rm MC}_{\chi_{c2}}$\\
  &  & (\%) &  &  &  & (\%) & & & & (\%) & & \\ \hline
  0  &  2120810  & 54.7 &  3877450  &  3895611 & 2361620  & 59.9 &  3944259
  & 4008941 &  1960130  & 54.6 &  3592178 & 3669625\\
  1  &  1394550  & 50.9 &  2199410  &  2307965 & 1502400  & 52.6 &  2295310
  & 2322469 &  1288210  & 46.4 &  2228712 & 2271990\\
  2  &  674368  & 42.2 &  1075983  &  1023269 & 679189  & 39.3 & 1164391
  & 1071370 &  584461  & 32.7 &  1202462 & 1091068 \\
  3  &  261673  & 32.0 &   458967  &   406560 & 274281  & 28.7 &  536108
  & 471751 & 219636  & 22.4 & 551204 & 486484 \\
  4  &  111118  & 24.6 & 226317  & 174196 & 115964  & 20.9 & 277841
  & 215496 &  88399  & 15.6 & 284253 & 224716\\
  5  &  47410  & 17.7 & 122876  & 78627 & 47504  & 15.2 & 143682
  & 101846 &  36508  & 10.8 & 154817 & 107356\\
  6  &  20810  & 12.8 & 67830  & 36456 & 19718  & 11.0 & 74533
  & 49819 &  14077  & 6.62 & 88404 & 53125\\
  7  &  10644  & 9.17 &  46153  &  17555 & 7856  & 7.45 & 41914
  & 25008 &  5895  & 5.22 & 44884 & 26762\\
  8  &  5628  & 9.00 &  23078  &  8819 & 3690  & 4.92 &  27678
  &  13039 &  2736  & 3.64 &  27714 &  14105\\
  $\ge9$  &  8981  & 9.33 &  31171  &  9703 & 3144  & 3.68 &  27662
  &  15608 &  2943  & 1.69 &  56302 & 16785\\
\hline
\end{tabular}
\label{result_chic_to_all_pi0}
\end{footnotesize}
\end{center}
\end{table}

\begin{table}[bth]
\begin{center}
\caption{Detected data events, $D$, efficiencies, $\epsilon$, efficiency corrected events, $N$, and  number of scaled simulated events $N^{\rm MC}$ for $\chi_{c1/2} \to \gamma J/\psi,\:J/\psi \to$ anything.  Here $N$ has been multiplied by $R(data)$ and $N^{\rm MC}$ by $R(MC)$, the fractions of valid $\pi^0$s.}
\begin{footnotesize}
\begin{tabular}{l|rrrr|rrrr} \hline
$N_{\pi^0}$ & $D_{J/\psi_1}$ & $\epsilon_{J/\psi_1}$
& $N_{J/\psi_1}$  & $N^{\rm MC}_{J/\psi_1}$ &  $D_{J/\psi_2}$ & $\epsilon_{J/\psi_2}$ &  $N_{J/\psi_2}$  &  $N^{\rm MC}_{J/\psi_2}$\\
  & & (\%) & & & & (\%) & &   \\ \hline
0  &   891038  & 56.7 &  1571227  & 1618965 &  433812  & 55.8 &  750852 &
854740\\
1  &   529136  & 57.4 &  740963  & 744487 &  280425  & 57.8 &   445982 &
404714\\
2  &  251421  & 49.3 &  343268  & 319198 & 139981  & 50.5 &  219284
& 180373 \\
3  &  113103  & 41.4 &  153318  & 141990 & 67755  & 43.0 &  88476 &
82390 \\
4  &  48556  & 32.7 &  74432  & 66240 & 31190  & 34.5 &  45360 &
39462 \\
5  &  19321  & 22.0 &  40334  & 32237 & 13517  & 28.2 &  21982 &
19692 \\
6  &  9178  & 15.1 &  25304  & 16500 & 5976  & 20.4  &  12195 &
10144 \\
7  &  4218  & 11.1 &  15110  & 8508 & 3076  & 14.5 &  8431 &
5405 \\
8  &  1927  & 10.3 &  6925  & 4640 &  1165  & 9.77 &  4401 &
2998 \\
$\ge9$  &   955  & 2.58 &  11978  & 6053 &  223  & 2.70 &   2678 &
 4024 \\
\hline
\end{tabular}
\label{result_jpsi_to_all_pi0}
\end{footnotesize}
\end{center}
\end{table}

\end{document}